\documentstyle[12pt,epsfig]{article}
        \newdimen\eqskip
        \newdimen\txtskip
        \baselineskip=\txtskip
\begin{document}

  \newcommand{\ccaption}[2]{
    \begin{center}
    \parbox{0.85\textwidth}{
      \caption[#1]{\small\em{{#2}}}
      }                     
    \end{center}
    }
\newcommand{\BS}{\bigskip}
% MATH SYMBOLS
\def    \be             {\begin{equation}}
\def    \ee             {\end{equation}}
\def    \ba             {\begin{eqnarray}}
\def    \ea             {\end{eqnarray}}
\def    \nn             {\nonumber}
\def    \=              {\;=\;}
\def    \frac           #1#2{{#1 \over #2}}
\def    \ret            {\\[\eqskip]}
\def    \ie             {{\em i.e.\/} }
\def    \eg             {{\em e.g.\/} }
\def    \lsim           {\raisebox{-3pt}{$\>\stackrel{>}{\scriptstyle\sim}\>$}}
\def    \gsim           {\raisebox{-3pt}{$\>\stackrel{>}{\scriptstyle\sim}\>$}}
\def    \bentarrow      {\:\raisebox{1.1ex}{\rlap{$\vert$}}\!\rightarrow}
\def    \rd             {{\mathrm d}}    
\def    \Im             {{\mathrm{Im}}}  
\def    \bra#1          {\mbox{$\langle #1 |$}}
\def    \ket#1          {\mbox{$| #1 \rangle$}}
\newcommand\sss{\scriptscriptstyle}
\newcommand     \ptmin     {\ifmmode p_{\scriptscriptstyle T}^{\sss min} \else
$p_{\scriptscriptstyle T}^{\sss min}$ \fi}
\newcommand\muf{\mu_{\sss F}}
\newcommand\mur{\mu_{\sss R}}
\newcommand\muo{\mu_0}
\newcommand     \muF            {\mu_{\rm F}}
\newcommand     \muR            {\mu_{\rm R}}
\def \et   {\mbox{$E_{\scriptscriptstyle T}$}}
\def    \mb             {\mbox{$m_b$}}
\newcommand     \dphi           {\mbox{$\Delta \phi$}}

% UNITS                 
\def    \kev            {\mbox{$\mathrm{keV}$}}
\def    \mev            {\mbox{$\mathrm{MeV}$}}
\def    \gev            {\mbox{$\mathrm{GeV}$}}

% KINEMATICAL VARIABLES 

\def    \mq             {\mbox{$m_Q$}}  
\def    \mqq            {\mbox{$m_{Q\bar Q}$}}
\def    \mqqsq          {\mbox{$m^2_{Q\bar Q}$}}
\def    \pt             {\mbox{$p_T$}}
\def    \ptsq           {\mbox{$p^2_T$}}

% QCD PARAMETERS                                      
\newcommand     \lambdamsb     {\ifmmode
\Lambda_5^{\rm \scriptscriptstyle \overline{MS}} \else
$\Lambda_5^{\rm \scriptscriptstyle \overline{MS}}$ \fi}
\newcommand     \Lambdamsb      \lambdamsb
\newcommand     \LambdaQCD     {\ifmmode
\Lambda_{\rm \scriptscriptstyle QCD} \else
$\Lambda_{\rm \scriptscriptstyle QCD}$ \fi}
\newcommand     \MSB            {\ifmmode {\overline{\rm MS}} \else 
                                 $\overline{\rm MS}$  \fi}
\def    \muf            {\mbox{$\mu_{\rm F}$}}
\def    \mufsq          {\mbox{$\mu^2_{\rm F}$}}
\def    \mur            {{\mbox{$\mu$}}}
\def    \mursq          {\mbox{$\mu^2$}}
\def    \mul            {{\mu_\Lambda}}
\def    \mulsq          {\mbox{$\mu^2_\Lambda$}}
     
\def    \as             {\mbox{$\alpha_s$}}
\def    \asb            {\mbox{$\alpha_s^{(b)}$}}
\def    \assq           {\mbox{$\alpha_s^2$}}
\def    \ascube         {\mbox{$\alpha_s^3$}}
\def    \asfour         {\mbox{$\alpha_s^4$}}
\def    \asfive         {\mbox{$\alpha_s^5$}}

\def    \eps            {\ifmmode \epsilon \else $\epsilon$ \fi}
\def    \epsbar         {\ifmmode \bar\epsilon \else $\bar\epsilon$ \fi}
\def    \epsir          {\ifmmode \epsilon_{\rm IR} \else $\epsilon_{\rm IR}$ \fi}
\def    \epsuv          {\ifmmode \epsilon_{\rm UV} \else $\epsilon_{\rm UV}$ \fi}

% VARIOUS NOTATIONS
\def    \m              {\mbox{${\cal{M}}$}}
\def    \mbar           {\mbox{${\overline{\cal{M}}}$}}
\def    \mborn          {\mbox{${\cal{M}}_{\rm Born}$}}
\def    \sborn          {\mbox{$\sigma_{\rm Born}$}}
\def    \sborno         {\mbox{$\sigma^0_{\rm Born}$}}
\def    \sborno         {\mbox{$\sigma_0$}}
%%%%%%%%%%%%%%%%%%%%%%%%%%%%%%%%%%%%%%%%%%%%%%%%%%%%%%%%%%%%%%%%%%%
\def \oacube {\mbox{$ O(\alpha_s^3)$}}
\def \oatwo {\mbox{$ O(\alpha_s^2)$}}
\def \oas   {\mbox{$ O(\alpha_s)$}}
\def \QQ {Q \overline Q}
\def \Qb{\overline{Q}}
\def \qq {\mbox{$q \overline q$}}
\def\lqcd{\mbox{$\Lambda_{QCD}$}}
\def \cf {{\ifmmode C_F{}\else $C_F$ \fi}}
\def \ca {{\ifmmode C_A {}\else $C_A$ \fi}}
\def \caf {\mbox{$ C_F-\frac{1}{2}C_A $}}  
\def \tf {{\ifmmode T_F {}\else $T_F$ \fi}}
\def \nf {{\ifmmode n_f {}\else $n_f$ \fi}}
\def \nc {{\ifmmode N_c {}\else $N_c$ \fi}}
\def \da {{\ifmmode D_A {}\else $D_A$ \fi}}
\def \Bf {{\ifmmode B_F {}\else $B_F$ \fi}}
\def \df {{\ifmmode D_F {}\else $D_F$ \fi}}
                                           
\def\der{\mbox{$\stackrel{\leftrightarrow}{\bf D}$}}
\def\nder{\mbox{$\stackrel{\leftrightarrow}{D}$}}
\def\asopi{\mbox{$\frac{\as}{\pi}$}}
\def\aem{\mbox{$\alpha_{{\mathrm\tiny EM}}$}}
\def\muf{\mbox{$\mu_{\mathrm F}$}}
\def\mur{\mbox{$\mu_R$}}
\def\slash#1{{#1\!\!\!/}}

%%%%%%%%%%%%%%%%%%%%%%%%%%%%%%%%%%%%%%%%%%%%%%%%%%%%%%%%%%%%%%%%%%%%%%
\begin{titlepage}
\nopagebreak
{\flushright{
        \begin{minipage}{5cm}
        CERN-TH/97-328\\
        hep-ph/9711337\\
        \end{minipage}        }

}
\vfill
\begin{center}
{\LARGE { \bf \sc Two Lectures on \\[0.5cm] Heavy Quark Production in
\\[0.5cm] Hadronic Collisions \footnote{Presented at the International School
of Physics ``E. Fermi'', Course CXXXVII, {\em Heavy flavour physics: a probe of
Nature's grand design}.} }}
\vfill                                                       
\vskip .5cm
Michelangelo L. MANGANO\footnote{On leave of absence from                    
    INFN, Pisa, Italy}\\
{CERN, TH Division, Geneva, Switzerland} \\
{\tt mlm@vxcern.cern.ch}\\
\end{center}
\nopagebreak
\vfill
%\vskip 3cm
\begin{abstract} 
These lectures present a pedagogical introduction to the physics of
heavy-flavour production in hadronic collisions. The first lecture gives the
theoretical background, with a discussion of leading-order calculations and of
the effects of next-to-leading-order corrections. The origin and implications
of the large logarithmic corrections appearing at this order are presented in
an elementary way. The second lecture provides a survey of current experimental
data on charm and bottom production, and describes their comparison with
theoretical predictions. We emphasize the role played by some non-perturbative
effects in the determination of charm distributions, and study the theoretical
systematic uncertainties which affect our predictions.
\end{abstract}                                                
\vskip 1cm
CERN-TH/97-328\hfill \\
\today \hfill
\vfill 
\end{titlepage}
\begin{center}
{\LARGE { \bf \sc Heavy Quark Production \\[0.3cm] in Hadronic Collisions}}
\vskip .5cm                                       
Michelangelo L. MANGANO\footnote{On leave of absence from 
    INFN, Pisa, Italy}\\
{CERN, TH Division, Geneva, Switzerland} \\
\verb+mlm@vxcern.cern.ch+\\
\end{center}
\vskip 0.5cm
\section{Introduction}
Most of this wonderful School was devoted to illustrating the key role that
heavy quarks occupy in helping physicists probe the structure of elementary
particles and their interactions at the most fundamental level. Progress in
understanding the main conceptual weaknesses of the Standard Model, such as the
origin of the family structure or the origin of CP violation, might well come
in the future from more accurate studies of bottom and top quark decays and
interactions. 
The present set of lectures, will address instead less mind-boggling aspects
of heavy-quark physics, namely the physics involved in their production. 
Nevertheless, as I hope these lectures will convince you of, there
is more to learn from a theoretical study of heavy-quark production than just
cross-sections and predictions for the numbers of $b$ quarks to be expected at
the LHC. 

First of all, heavy-quark production provides a benchmark process for the study
of perturbative QCD.  This is because the heavy-quark mass $m_Q$ defines the
scale at which the strong-interaction coupling constant $\as$ is evaluated, and
$m_Q\gg \Lambda_{QCD}$ implies that the production properties (at least at the
inclusive level) should be calculable within perturbation theory (PT).
The existence of heavy quarks with different masses                   
($m_c \simeq 1.5$~GeV, $m_b\simeq 5$~GeV,
$m_t\simeq 175$~GeV), allows us to probe perturbative QCD in regions of different
$Q^2$, where the relative impact of radiative corrections and non-perturbative
(NP) effects are very different.  For example, we know that NP
effects are very important in turning a $c$-quark into a charmed hadron (say a
$D^+$).  Were we able to fully control 
the charm production within PT, we                   
could then hope to disentangle the NP physics from the comparison of PT
with data.  I shall give some                          
examples of this in the second part of these lectures.

The presence of a heavy quark in the final state is also a flag for very                                                 
specific production mechanisms and selects specific classes of relevant
higher-order corrections. This can be useful either as a probe of the nucleon
structure, or as a test of our understanding of radiative corrections in QCD.
For example:   
\begin{itemize}
\item
$W$+charm final states probe the strange content of the
proton, since the leading production mechanism is $g s \to Wc$.
\item Inclusive $b$ production at high energy probes the gluon density of the
proton, since the leading process is $gg\to b\bar b$.
\item Associated production of $W$ and heavy-quark pairs is sensitive to
gluon-splitting processes, since the dominant production process is the
production of a $W$ and an off-shell gluon, which then {\em decays} to the
heavy-quark pair: $q \bar q^\prime \to W g^* \to W Q \overline{Q}$.
\end{itemize}

The top quark is the heaviest of the heavy quarks. The history of its discovery
and a thorough review of its first measured properties have been nicely covered
in Bellettini's lectures at this School. There you certainly learned how much
of a theoretical input will be required in
measuring  important properties such as the top mass or the CKM couplings. The
mass, for example, is measured in hadronic collisions by reconstructing
multi-jet final    states from the top decay.
Statistical errors
on $m_{\rm top}$ achievable at the LHC ($\ll$ 1~GeV) are 
overwhelmed by the systematic                            
uncertainties due to the description of top production and decay.  
In fact a theoretical knowledge of the structure of the top-decay
jets and their origin is needed to accurately fit
the measured mass distributions~\cite{Abe94}
and extract a mass parameter\footnote{This is             
similar to what happens in the case of the $Z$ mass measurement at LEP/SLC,
where the $Z$ mass is obtained from a fit to the $Z$ lineshape. The lineshape
is affected by higher-order QED corrections, mostly initial-state radiation.
These need to be evaluated theoretically before the $Z$ mass, which is a    
parameter in the fit, can be extracted from a comparison with the data.}.
A similar need for theoretical input is required when extracting the top mass
from the threshold behaviour in $e^+e^-$ annihilations, one of the goals of
future Next Linear Colliders~\cite{NLC}. 
The direct measurement of the CKM matrix element
$V_{tb}$ can be performed by detecting production of single quarks in hadron
collisions, via the processes $q\bar q^\prime \to W^* \to t \bar b$ and
$g q \to t \bar b q^\prime$ (where the exchange of a $t$-channel $W$ is
understood). In both these processes, in fact, the production rate is directly
proportional to $\vert V_{tb}\vert^2$. Once more, accurate studies of the
QCD aspect of the production dynamics need to be performed to make a meaningful
measurement of $\vert V_{tb}\vert$ possible. 
                                             
The study of $b$ cross-sections at the Tevatron~\cite{Bedeschi}
is helping us to improve                                       
the reliability of our estimates of
$b$ rates at the LHC, where $b$'s will be used to measure CP
violation~\cite{Stone} and to
probe possible $b$-meson rare decays, foreseen in several theories beyond the
Standard Model, as reviewed in this School by Masiero. 
The expected total cross-sections will define
overall rates, the kinematical distributions will determine 
triggering strategies and detector design.

Heavy quarks are also important signals of possible new physics, in addition to
providing large backgrounds to searches for new physics:
\begin{eqnarray*}
& \mbox{\bf Signals:} & \mbox{\bf Backgrounds:} \\
& H \rightarrow b \overline{b} & \quad\quad gg \to b \overline{b} \\
& WH \rightarrow b \overline{b} & \quad\quad q\bar q \to W b \overline{b} \\
& \tilde t \rightarrow b \chi^+, ~c \chi^0, ~t \chi^0 & \quad\quad 
p \bar p \rightarrow W b \overline{b} jj,\quad \overline{t} t  \\
& b' \rightarrow \gamma b, ~Z^0b & \quad\quad  p \bar p \to b \overline{b} \gamma \gamma \\
& \chi^+\chi^0_2 \rightarrow \mbox{ 3  lept's} & \quad\quad p \bar p \rightarrow t
\overline{t}  \\ 
& \cdots & \quad\quad  \cdots  \\
\end{eqnarray*}       
Understanding the characteristics of final states with $b$'s (and $b$-jets) 
is fundamental in order to learn how to isolate signals from backgrounds.
                  
As a final example of possible uses of heavy-quark production, let me mention 
the searches for quark-gluon plasma done in heavy-ion collisions. Among other
probes, final states with $J/\psi$ and, more
generally, DY pairs, will be used to study the properties
of matter at high density.  Charm pairs will provide both signals $(c \bar c
\rightarrow J/\psi)$ and backgrounds $(c \bar c \rightarrow \mu^+\mu^-)$.
To make robust
predictions on the properties of charm produced in high-energy nuclear
collisions at future machines (RHIC and LHC) it becomes important to test our
understanding in 
simpler production environments, such as those provided by current
fixed-target hadro- and photo-production experiments. Due to the smallness of
the charm mass, significant NP corrections are expected. As mentioned earlier,
this is therefore also an important place to study the interplay of
perturbative and non-perturbative effects, as will be discussed in the second
part of these lectures.

The structure of the lectures is as follows:
\begin{enumerate}
\item The theory of heavy-quark production:
  \begin{itemize}
  \item leading order
  \item total and differential cross-sections
  \item NLO corrections
  \end{itemize}
\item Phenomenology of heavy-quark production:
  \begin{itemize}
  \item fixed-target charm production
  \item interplay of perturbative and non-perturbative effects
  \item bottom production at the Tevatron
  \end{itemize}
\end{enumerate}

\section{Theory of heavy-quark production}
In this Section I will present an elementary introduction to the theory of
heavy-quark production in hadron-hadron
collisions. Production
of heavy quarks in $e^+e^-$ collisions is to leading-order (LO) an electroweak
process, and is therefore relatively straightforward to control. The
complications of QCD arise only at the level of radiative corrections, and
become relevant only when one is interested in high-precision studies (such as
the measurement of $R_b$ at LEP/SLC) or in particular aspects of the QCD
dynamics (such as the study of fragmentation functions, to which we will return
in the following). Photoproduction, which is of relevance for several
fixed-target experiments and for the HERA collider, can be described by a
sub-class of the processes appearing in hadronic collisions, so we will not
consider it in detail here (although we will discuss its phenomenology in the
second lecture). For a complete and more technical review of the
theory of heavy-quark production, as well as for a precious list of references
to the original papers, see for example Nason's work in ref.~\cite{Book1}.
Additional details and phenomenological applications can be found in the
excellent book by Ellis, Stirling and Webber~\cite{Ellis96}.

\subsection{Leading order}                        
\begin{figure}[t]
\centerline{\epsfig{file=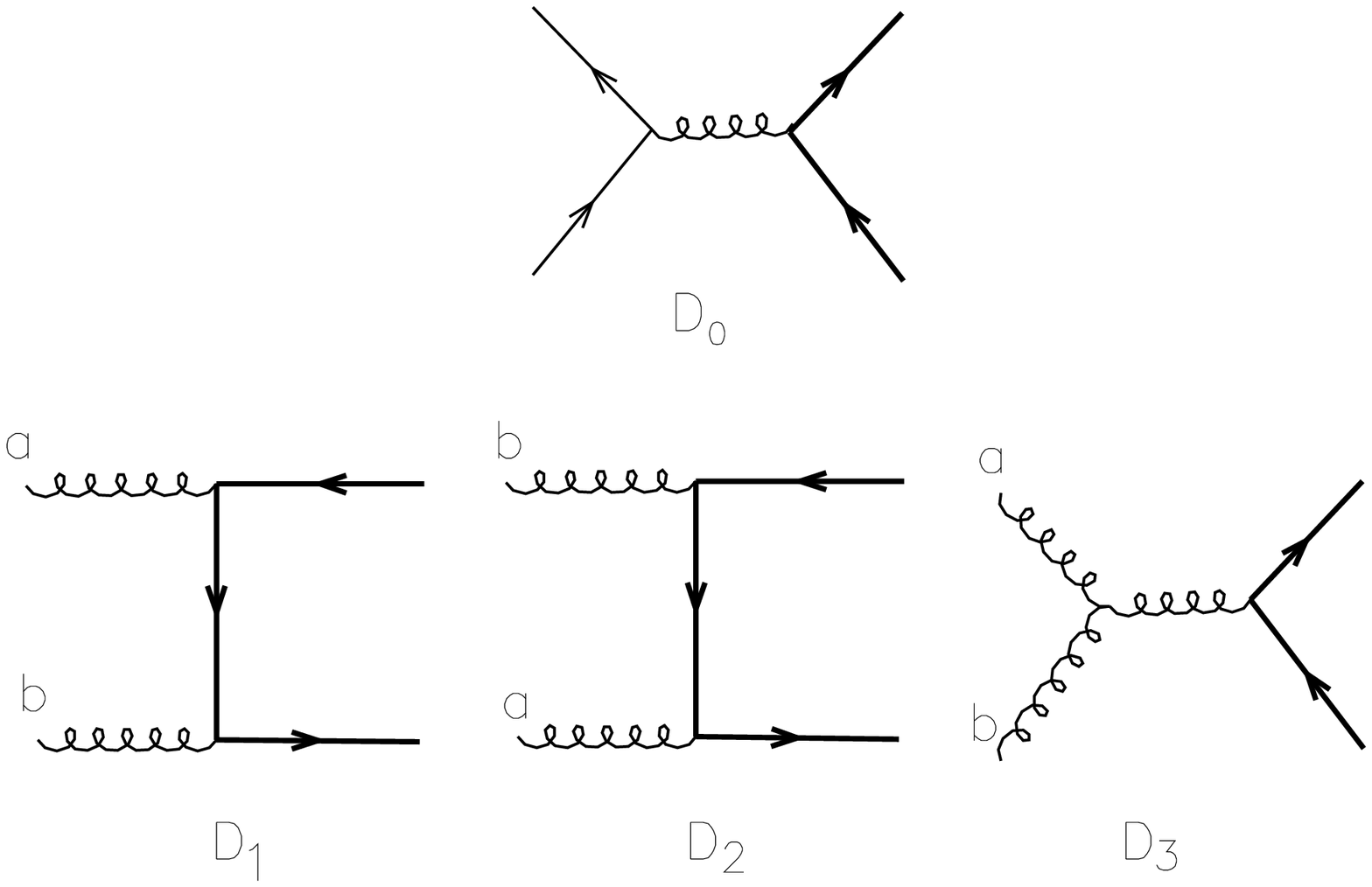,width=12cm,clip=}}
\ccaption{}{\label{fig:born} Leading-order diagrams for heavy-quark pair
production.}    
\end{figure}

Two processes are responsible for heavy-quark hadro-production at the LO in
perturbation theory:
\be
	q\bar q \to \QQ \quad \quad \mbox{and}\quad\quad gg\to \QQ \, .
\ee
The corresponding diagrams are shown if fig.~\ref{fig:born}.
Few simple comments are in order:
\begin{itemize}
\item In $q \bar q \rightarrow Q \Qb$, the $Q \Qb$ pair is always in a
colour-octet state.  In $gg \rightarrow Q \Qb$ 
both colour-singlet and octet are allowed.                              
\item The process   
$q \bar q \rightarrow Q\Qb$ is similar to $e^+e^- \rightarrow \mu^+\mu^-$.
The only difference is an overall colour factor
\be
\sum_a \lambda^a_{ij} \lambda^a_{kl} \equiv 
{1 \over 2}(\delta_{il}\delta_{jk} - {1 \over
N}~\delta_{ij}\delta_{kl}) \; ,
\ee                            
which when squared and summed over all possible initial and final colour states
gives a factor $\cf=(N^2-1)/2N$ ($\cf=4/3$ for $N=3$).
\item The total production cross-section for heavy quarks is
finite. In fact no poles can develop in the
intermediate propagators. This is obvious for the $s$-channel diagrams, since
$s>4m^2$. In the
case of the $t$-channel gluon exchange (say in diagram $D_1$ of the figure),
labeling the lower-line gluon and quark momenta as $p_1$ and $Q$, respectively, 
one obtains in the CM frame:
\be                         
(p_1-Q)^2-m^2 = -2p_1\cdot Q = - {\hat s \over 2}(1 - \beta\cos\theta)
\ee
with $\beta$ the heavy-quark velocity ($\beta\equiv 
\sqrt{1 - {4m^2 \over \hat s}}$) and $\theta$ the scattering angle.
Therefore the propagator never vanishes:
\be
2p_1 \cdot Q \ge {\hat s \over 2} (1-\beta) = {\hat s \over 2(1 + \beta)} 
(1-\beta^2)
= {2m^2 \over 1 + \beta} \geq m^2~.
\ee
\end{itemize}
Therefore $m^2$ is the minimum virtuality exchanged in the $t$-channel, and the
total cross-section is finite at LO in PT. This would not be the case for
massless quarks. The minimum transverse momentum transferred in
the $t$-channel sets the scale for the argument of the strong coupling constant
$\as$. Since this scale is of the order of the heavy-quark mass, and since this
is (by definition of heavy quark) much large than $\Lambda_{QCD}$, perturbative
calculations should be reliable. 
Once again, this could not be the case for light quarks: the {\em total}
production cross-section for {\em up} or {\em down} quarks in not calculable in
PT!

The evaluation of the matrix elements can be done using standard Feynman
diagram techniques. The results are:
\ba  \label{eq:qqlo}
{1 \over g^4} \overline{\Sigma}
\left|M(q \bar q \rightarrow Q \Qb)\right|^2 &=& {V \over 2N^2}
(\tau^2_1 + \tau^2_2 + {\rho \over 2})  \\
\label{eq:gglo}
{1 \over g^4} \overline{\Sigma}\left|M(gg \rightarrow Q \Qb)\right|^2 &=&
{1 \over 2VN}\left({V \over \tau_1\tau_2} - 2N^2\right)   
\left( \tau^2_1 + \tau^2_2 + \rho - {\rho^2 \over 4\tau_1\tau_2}\right)
\ea                                                                   
where:
$V = N^2-1$ is the dimension of the $SU(N)$ gauge group, i.e. the number of
gluons, and
\be
\tau_{1,2} = 2 \; {p_{1,2} \cdot Q \over \hat s} 
= {1 \mp \beta \cos \Theta \over 2}~, \quad\quad \rho = {4m^2 \over \hat s}
,\quad\quad \hat{s}=(p_1+p_2)^2
\ee                            
\underline{Exercise~1:} show that in the limit $\tau_1\tau_2 \to 0$ (which can
be achieved at very high energy when $\beta\to 1$) the $gg\to \QQ$ matrix
element squared goes at most like $1/(\tau_1\tau_2)$, and not like 
$1/(\tau_1\tau_2)^2$, as the eq.~(\ref{eq:gglo}) naively indicates.
%$$\tau_1 + \tau_2 = 1~,~~~\tau_1\tau_2 = {1 - \beta^2\cos^2\theta \over 4} \geq {1 - \beta^2
%\over 4} = {\rho \over 4}$$
%$\Rightarrow {\rho^2 \over 4\tau_1\tau_2} \leq \rho$ and therefore the most singular term
%in the $(gg \rightarrow Q \Qb)$ process goes like ${1 \over \tau_1\tau_2}$, not like ${1
%\over (\tau_1\tau_2)^2}$.
\\
\underline{Exercise~2:} show that the term inside the first parenthesis of
eq.~(\ref{eq:gglo}) is always positive.
\\[0.3cm]
%$$
%\tau_1\tau_2 \leq {1 \over 4} \Rightarrow {V \over \tau_1\tau_2} - 2N^2 \geq 4V-2N^2 = 2N^2
%- 2 = 2(N^2-1) \geq 0
%$$
 
The total partonic cross-sections are obtained by integrating over the 
two-body phase space:
\be
d\phi_{(2)} \equiv {1 \over 2\hat s}~{d^2Q \over (2\pi)^32Q^0}~{d^3
\overline{Q}\over                            
(2\pi)^32\overline{Q}^0}~(2\pi)^4 \delta^4(P_{in}- P_{out}) 
\= {\pi \over 2\hat s} \left( {1 \over 4\pi}\right)^2 \beta~d(\cos\theta)
\ee                                                                  
giving the following results:
\ba
\hat\sigma(q \bar q \rightarrow Q \overline{Q}) &=& 
{\alpha^2_s \over m^2}~\left({V \over
N^2}\right)~{\pi\beta \over 24} \rho(2 + \rho) 
\buildrel{\hat s \rightarrow \infty}
\over{\rightarrow} {1 \over \hat s}
\\
\hat \sigma(gg \rightarrow Q \overline{Q}) 
&=& {\alpha^2_s \over m^2} \left( {1 \over NV} \right)
{\pi \beta \over 24} \rho
\left[ 3{\cal L}(\beta)\left(\rho^2 + 2V(\rho+1)\right) \right. \nn \\
 && \left. + 2(V-2)(1+\rho) +                                              
\rho(6\rho-N^2)\right] 
\buildrel{\hat s \rightarrow \infty}\over{\rightarrow} {1 \over \hat{s}}
{\cal L}(\beta)                                                        
\ea
where
\be
{\cal L}(\beta) = {1 \over \beta} \log \left({1 + \beta \over 1 - \beta} \right) - 2
\ee
\underline{Comments}:
\begin{itemize}
\item At large $\hat s$ the $q \bar q$ rate vanishes more quickly.
\item 
Notice that the threshold suppression in the $q \bar q \rightarrow Q \Qb$
process goes like
\be
\beta(2 + \rho) = \sqrt{1-\rho} \; (2 + \rho) \simeq 2(1 - {\rho \over 2})(1 +
{\rho \over 2}) = 2(1 - 4{m^4 \over \hat s^2})
\ee             
As a result threshold effects vanish very quickly as soon as $\hat s > 4m^2$.
This {\em suppressed suppression} is related to the
spin-1/2 of quarks.  For scalar particles (e.g., squarks) the phase-space
suppression is stronger ($\sim\beta^3$).
\end{itemize}                                     
This is important for top production, which for the
measured value of the top mass is dominated at the Tevatron by $q \bar q$
annihilation close to the kinematic threshold.              
 Including the reduction in rate due
to the fewer degrees of freedom of a scalar particle relative to a spin-$1/2$
one, the squark cross-section turns out to be 
approximately $1/10$ of that of a quark with the same mass, 
if the dominant production mechanism is $q \bar q$ annihilation.  
\\[0.5cm]
\underline{\bf Examples:}
\begin{itemize}
\item The relative productions rates for heavy quarks of mass $m_1$ and $m_2$
behave, at high energy, like:
\be
{\hat \sigma(gg \rightarrow Q_1 \Qb_1) \over \hat\sigma(gg \rightarrow Q_2
\Qb_2)} \buildrel{s \rightarrow \infty} \over \rightarrow 1 -
{\log(m^2_1/m^2_2) \over \log(s/m^2_2)}
\ee
For example,
\be
{\sigma(gg \rightarrow b \bar b) \over
\sigma(gg \rightarrow c \bar c)} \sim 0.7 \quad \quad \mbox{at}
\quad \sqrt{\hat s} = 100~{\rm GeV}                 
\ee   
\item For the $q\bar q$ process we have instead:
\be
{\hat\sigma(q \bar q \rightarrow Q_1 \Qb_1) 
\over \hat \sigma(q \bar q \rightarrow
Q_2\Qb_2)} \rightarrow 1 - O(m^4_1/\hat{s}^2)
\ee
with
\be
{\sigma(q\bar q \rightarrow b \bar b) \over \sigma(q \bar q
\rightarrow c \bar c)} \sim 0.99\quad \quad \mbox{at}
\quad \sqrt{\hat s} = 100~{\rm GeV}                 
\ee
\end{itemize}
In general, we are interested in {\em hadronic} rather than
{\em partonic} rates.  The obtain these we need to convolute the partonic
cross-sections with the parton densities in the hadron.
It is then useful to parametrize the final
state in terms of the transverse momentum ($p_{\perp}$) and 
of the rapidities ($y$) of the produced quarks. The resulting phase-space,
including the integration over the momentum fractions of the hadrons
carried by the initial-state partons $x_{1,2}$, is then:        
\be                                           
d\phi_{p \bar p} = {1 \over 2\hat s}~dx_1dx_2 {d^3Q \over (2\pi)^32q^0}~{d^3\Qb \over
(2\pi)^32\Qb^0}~(2\pi)^4 \delta^4 (P_{\rm in}-P_{\rm out})
\ee
Use:
\ba
&&
dx_1dx_2 ~\delta (E_{\rm in}-E_{\rm out})\delta(P^z_{\rm in}-P^z_{\rm out}) =
{1 \over 2E_{\rm beam}^2} = {2 \over S_{\rm had}}
\\
&&
{dQ_z \over Q^0} = dy~,~~y = {1 \over 2}\log {Q^0+Q^z \over Q^0-Q^z}
\quad (\rightarrow \eta=-\log(\tan\frac{\theta}{2})~~~{\rm for}~~ m = 0)
\ea                                                 
to get:
\be
d\phi_{p\bar p} 
= {\pi x_1x_2 \over \hat s^2} \left( 1 \over 4\pi \right)^2 dy d\bar y dp^2_T
\ee
It is easy to rewrite $\hat s$ in terms of $y, \bar y$ and of the transverse
mass $m_T^2=p_\perp^2+m^2$ using:
\be                        
\left\{
\matrix{ 
Q_0 &=& m_T ~\cosh y\cr
Q_z &=& m_T~\sinh y}  \; . \right.
\ee                        
This gives: $\hat s = 2m^2_T \left\{ 1 + \cosh(y - \bar y)\right\}$ and
finally:                                                               
\be
{d\sigma \over dyd\bar y dp^2_T} ={\pi \over 4m^4_T}~{1 \over [1+\cosh (y-\bar
y)]^2} \frac{1}{(4\pi^2)} \times 
\sum_{i,j}x_1f_i(x_1)x_2f_j(x_2) \overline{\sum}\vert M(ij\to
Q\Qb)\vert ^2 
\ee
For a fixed value of $p_T$ (and therefore of $m_T$), the rate is heavily
suppressed when $\vert y - \bar y\vert$ becomes large.  Therefore $Q$ and $\Qb$
tend to be produced with the same rapidity.  This has important consequences
for the design of detectors aimed at collecting large numbers of $b \bar b$
pairs, as required to perform CP studies. In fact, it implies that once a
detector has registered the $b$ quark, there is a large probability that
the $\bar b$ will be nearby in rapidity. This is fundamental
for detectors covering the forward region only, since it shows that there is no
need to build a backward spectrometer as well to efficiently collect the
heavy-quark {\em pair}. The gain in acceptance for the pair would only be a
factor of 2, and not 4 as naively expected in absence of such strong rapidity
correlations!
\subsection{Some results}
\begin{figure}[t]
\centerline{\epsfig{file=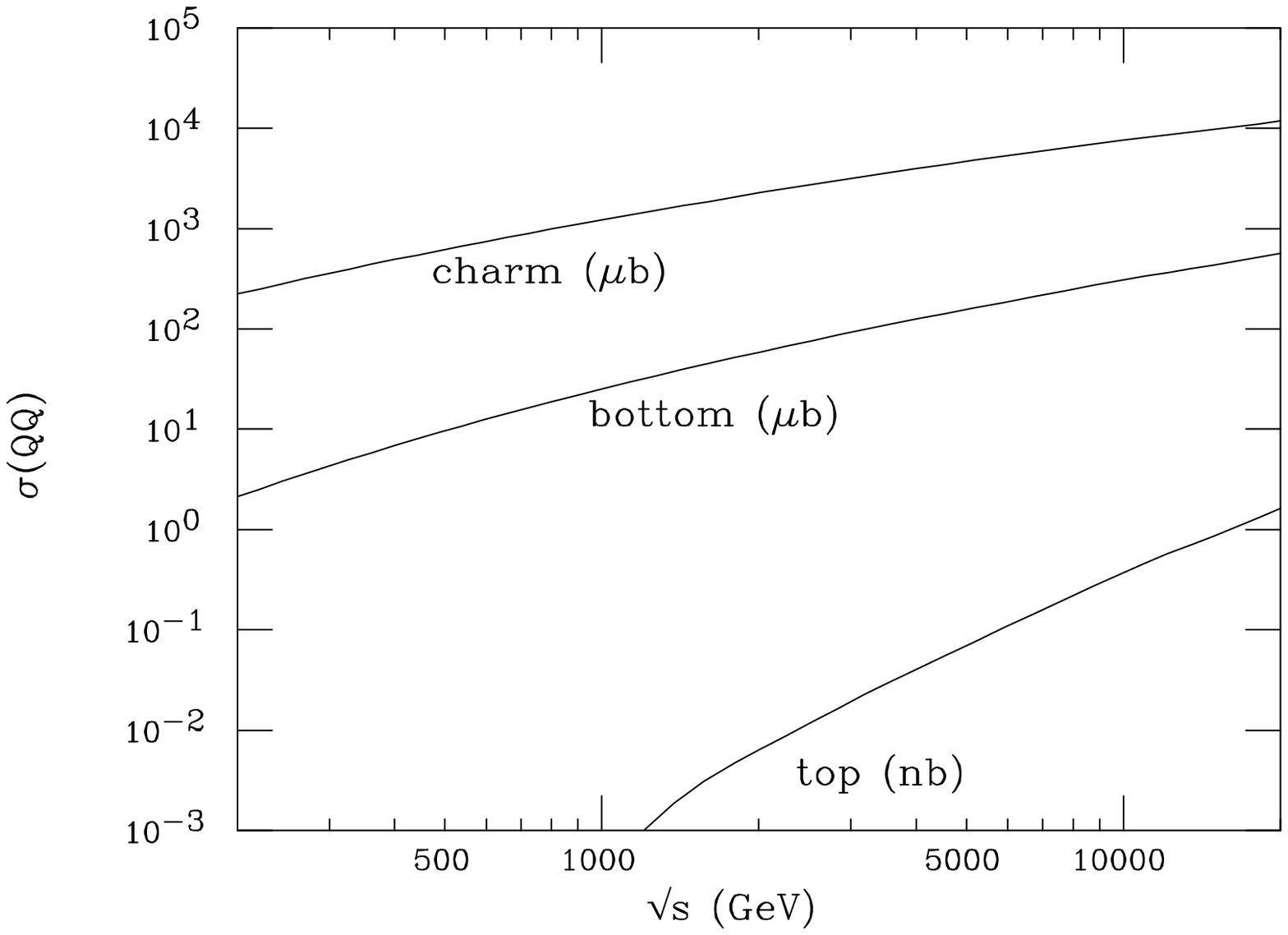,width=12cm,clip=}}
\ccaption{}{\label{fig:cbttot} Total production cross-sections for charm,
bottom and top quark pairs, in $p \bar p$ collisions.}
\end{figure}
We discuss here some simple applications of the cross-section formulae obtained
in the previous section. They can be convoluted numerically with standard
parametrizations of the parton densities inside the proton, to obtain hadronic
cross-setions at different energies and for different quark masses.
\begin{figure}[t]
\centerline{\epsfig{file=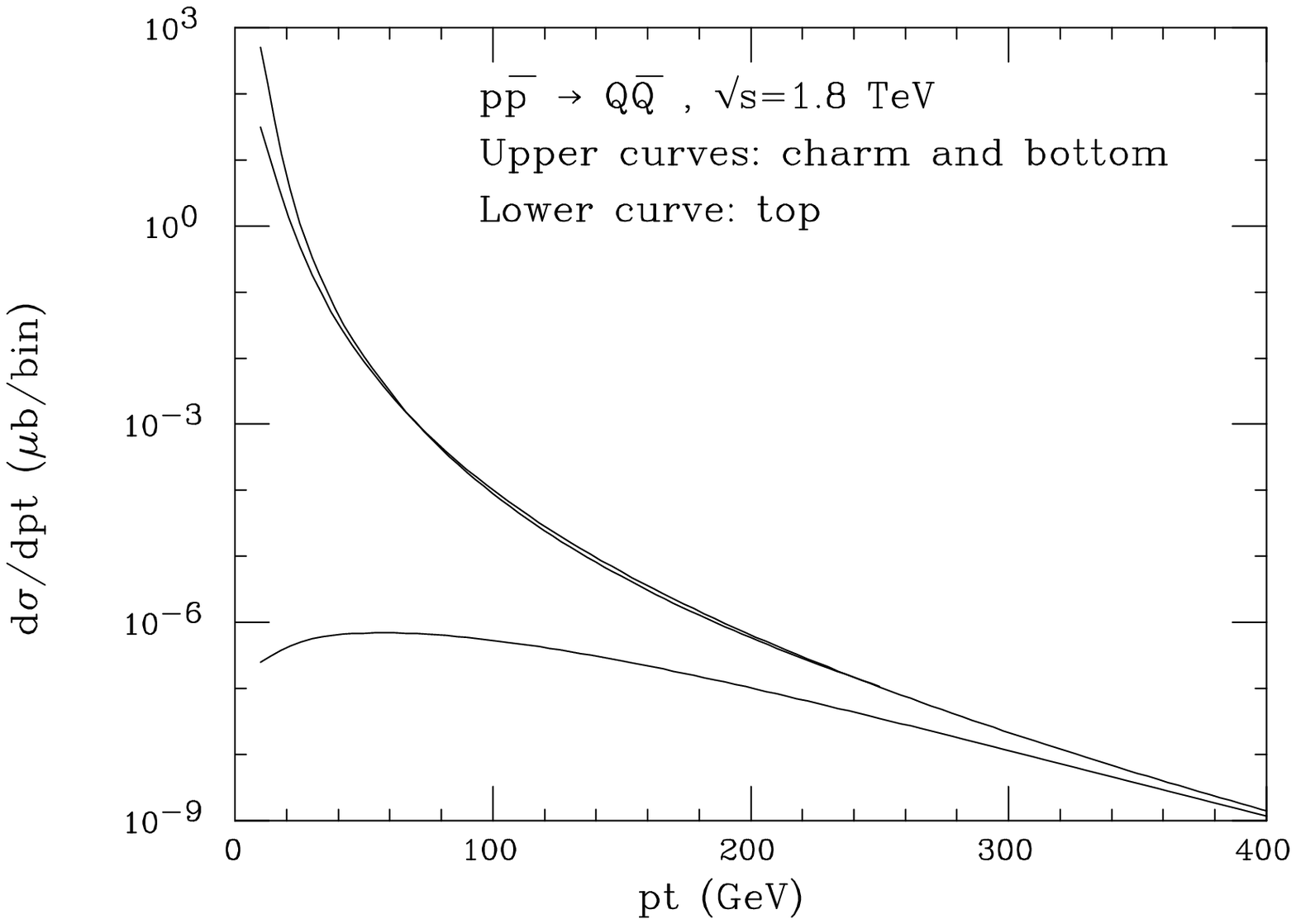,height=0.35\textheight,clip=}}
\ccaption{}{\label{fig:alllow} Inclusive $\pt$ distributions      
for charm,
bottom and top quarks pairs, in $p \bar p$ collisions at $\sqrt{S}=1.8$~TeV.}
\end{figure}
\begin{figure}
\centerline{\epsfig{file=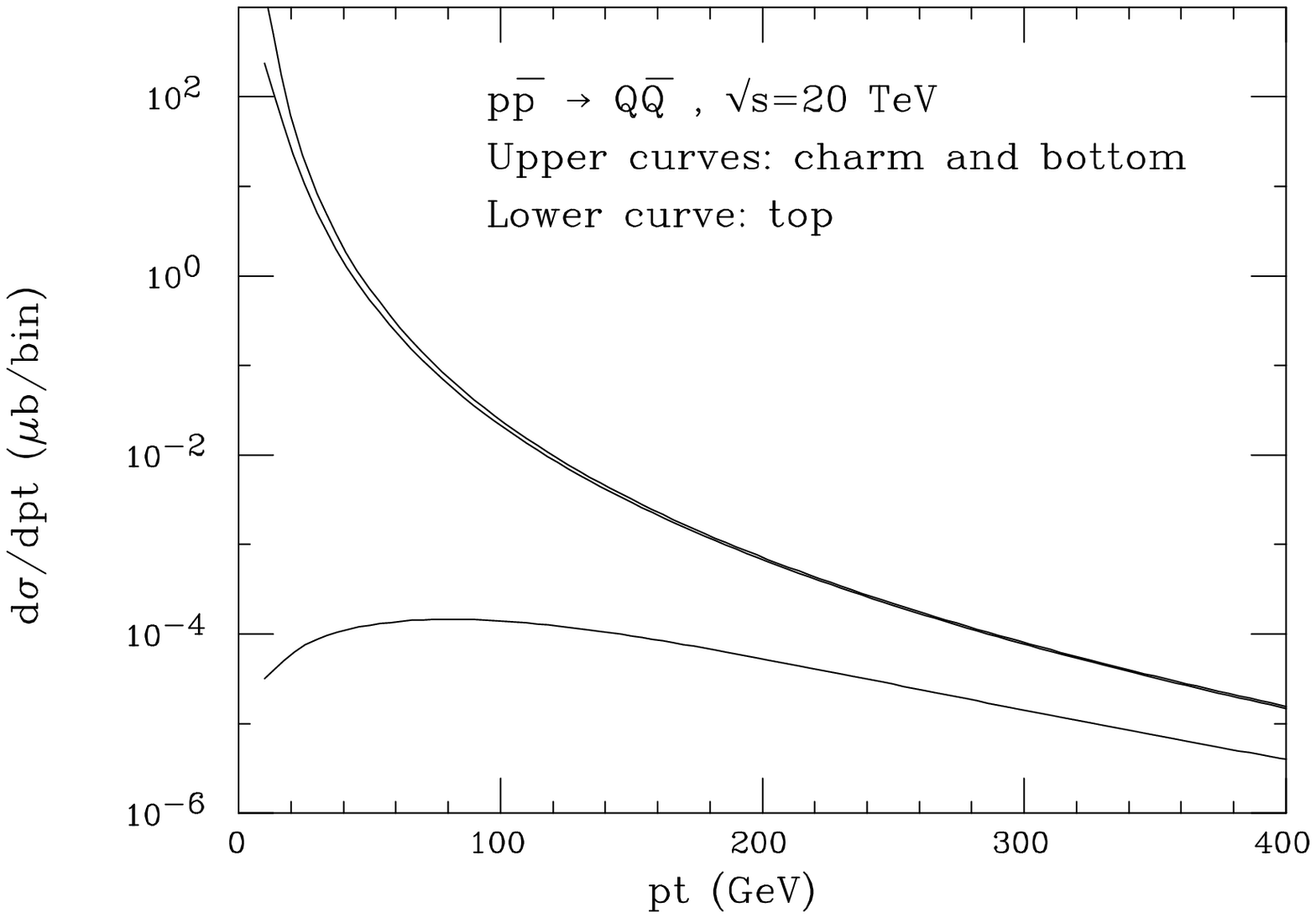,height=0.35\textheight,clip=}}
\ccaption{}{\label{fig:allhigh} Inclusive $\pt$ distributions      
for charm,                    
bottom and top quarks pairs, in $p \bar p$ collisions at $\sqrt{S}=20$~TeV.}
\end{figure}                                                         
In fig.~\ref{fig:cbttot} I show the total production cross-sections in $p\bar
p$ collisions, as a function of the hadronic center of mass energy $\sqrt{S}$,
for charm, bottom and top quark pairs. 
Notice the different units used in the
case of the top quark relative to those used for charm and bottom! Top cross
sections are suppressed by several orders of magnitude even at the highest
energies reachable in the foreseeable future. When the {\em partonic} CM energy
grows, however, the suppression due to the large mass becomes less and less
important, as discussed in the previous section. 
This can be seen in
fig.~\ref{fig:alllow}, which shows the transverse momentum distributions at
$\sqrt{S}=1.8$~TeV. Notice that the charm and bottom rates 
become almost equal as soon as $\pt\gsim 40$~GeV (which means $\sqrt{\hat
s}\gsim 80$~GeV). In the case of the top, we need to go out to $\pt>400$~GeV.
It is instructive to see how this picture changes at higher
hadronic energies. This is shown in fig.~\ref{fig:allhigh} for
$\sqrt{S}=20$~TeV. Notice that at $\pt=400$~GeV there is still a large
difference between the $c,b$ rates and the top rate. This is because, contrary
to the case of 1.8~TeV, at the LHC the
production rate is dominated by $gg$ annihilation. As we showed earlier,
the mass effects vanish more slowly for $gg$ processes than for the $q \bar q$
annihilation process.
More details on the comparison of the predicted production rates with the
available experimental data will be given in the second lecture.
For the time being, we shall move on to the discussion of the corrections
beyond the Born level. 

\subsection{Next-to-leading order corrections}
Next-to-leading-order (NLO) corrections come from two sources of
$O(\alpha^3_s)$ diagrams:
\\           
\centerline{\epsfig{file=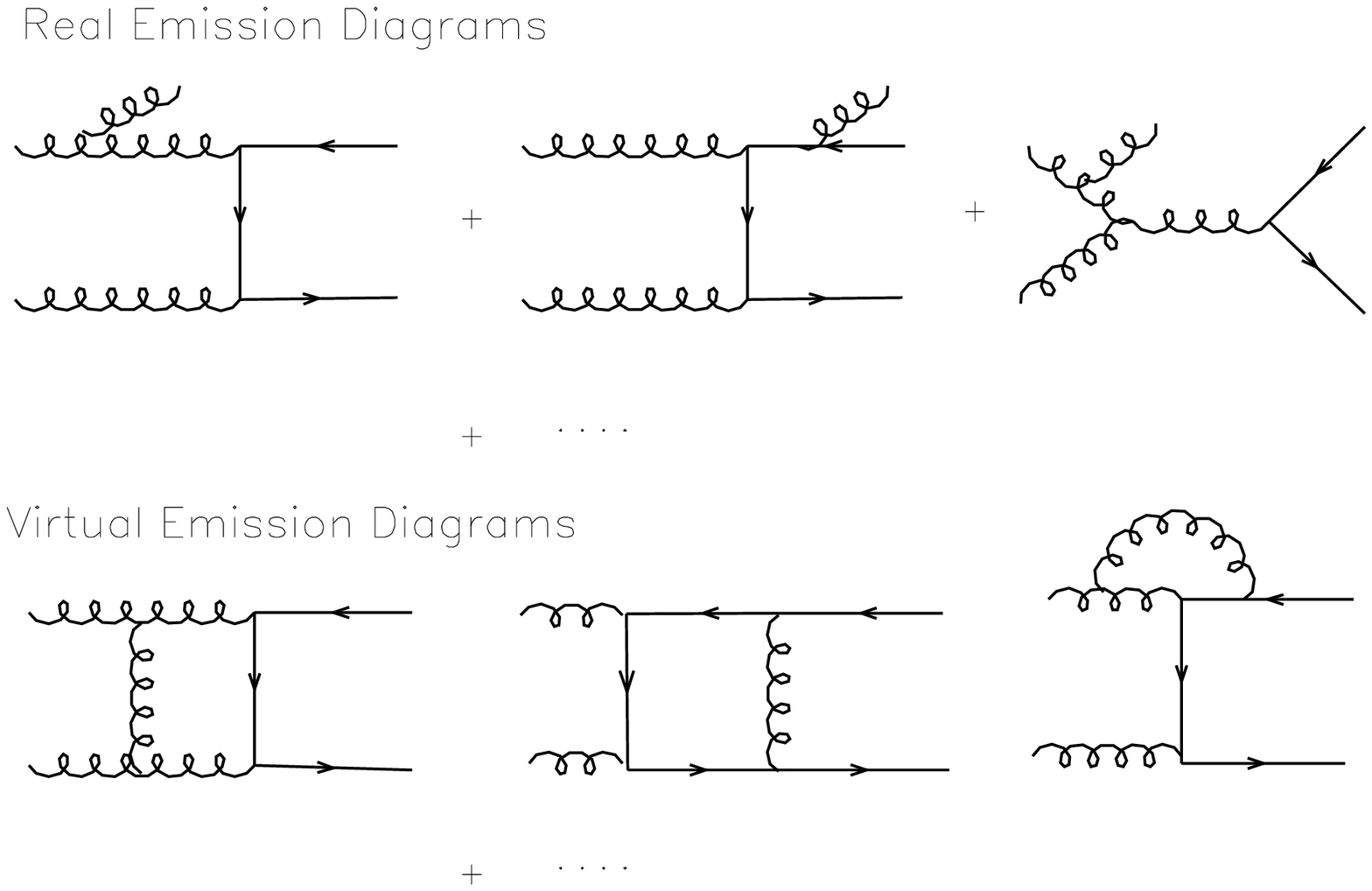,height=0.35\textheight,clip=}}
\\                                               
In the first case,
the corrections come from the square of the real emission matrix
elements. In the second case, 
from the interference of the virtual matrix elements (of
$O(g^4)$) with the Born level ones (of $O(g^2)$).
Ultraviolet divergences in the virtual diagrams are removed by the
renormalization process.  Infrared and collinear divergences
which appear both in the virtual diagrams and in the integration over the
emitted parton in the real-emission processes cancel each other,  or are
absorbed in the initial-state parton densities.
The complete calculations of NLO corrections to the production of heavy-quark
pairs in hadro- and in photo-production were done in
ref.~\cite{Nason88,Beenakker89} (total hadroproduction cross-sections),
ref.~\cite{Nason89,Beenakker91} (one-particle inclusive distributions in
hadroproduction), 
ref.~\cite{Ellis89,Smith92} (total and one-particle inclusive 
distributions in photoproduction),
ref.~\cite{Mangano92} (two-particle inclusive distributions
in hadroproduction) and ref.~\cite{Frixione94a} (two-particle inclusive
distributions in photoproduction).
The explicit results are too complicated for me to reproduce them here. So I
will limit myself to a qualitative discussion only.

\begin{figure}[t]
\centerline{    
      \epsfig{figure=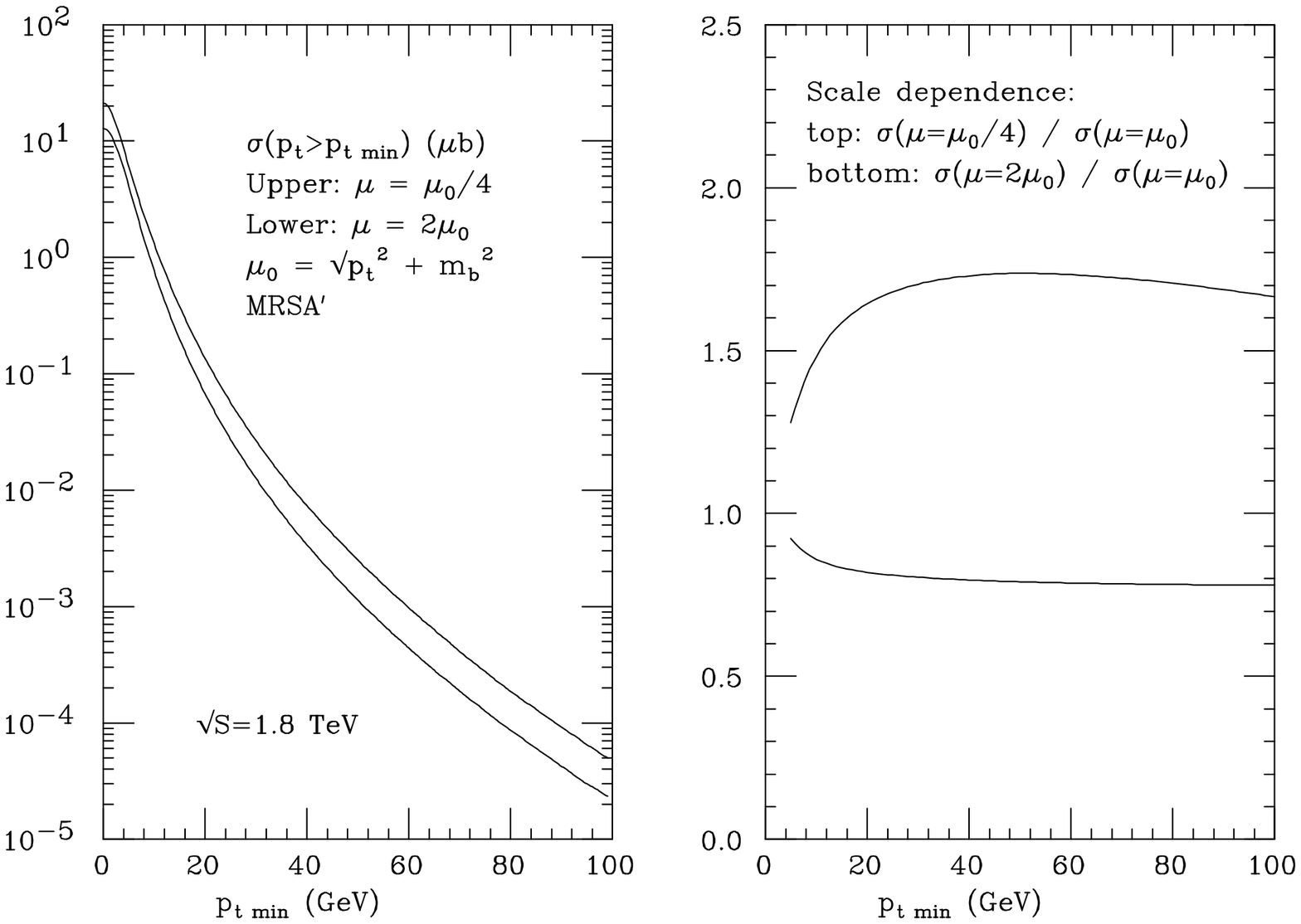,height=0.4\textheight,clip=}}
\ccaption{}{\label{fig:scale} Scale dependence of the 
inclusive $\pt$ distributions      
for bottom quarks, in $p \bar p$ collisions at $\sqrt{S}=1.8$~TeV.
$\muR=\muF=\mu$.}
\end{figure}

NLO corrections should improve our knowledge of the production cross-section.
In particular, uncertainties related to the renormalization ($\muR$)
and factorization ($\muF$) scale dependence                          
should be reduced\footnote{The renormalization scale $\muR$ is the energy scale
used in the evaluation of $\as$. The factorization scale $\muF$ is the scale
used in the evolution of the parton densities.}.              
There is evidence however that the
NLO is not sufficient to get accurate estimates, since a large
scale dependence is still present. 
This is shown is fig.~\ref{fig:scale}, which shows the scale
dependence of the inclusive $\pt$ distribution of $b$ quarks at the Tevatron.
Large scale dependence is a symptom of large NNLO corrections
A full NNLO calculation goes beyond our ability. So, we are stuck.

We can however make some progress in this direction by trying to understand
what are the sources of these large NNLO corrections. We can classify them as
follows:
\begin{itemize}                                       
\item ``Accidentals'' (e.g. large numerical coefficients, $\pi^2$'s, etc.,
     which should not affect the scale dependence).
\item Large logarithms due to the coexistence of different scales. E.g.:
\[                                                        
    \log \frac{S}{m^2} \quad , \quad \log \frac{p_T^2}{m^2} 
\]                                                      
\end{itemize}
The origin of these logarithms is known, and will be discussed in the 
remaining of this lecture. They are universal, namely they are
associated to a universal behaviour of the cross-section in some particular
kinematic configurations. Because of this, their structure can be predicted at
all orders of PT, and in principle they can be resummed. One then hopes that
these resummed large logarithms capture all the essential features of the
higher-order corrections, and help to improve the
behaviour of the perturbative predictions, without the need of a complete NNLO
(or even higher-order) calculation. 
                                    
Before we start analysing the sources of these logarithms in detail, 
some general comments are in order:
\begin{itemize}
\item[i)]
No collinear singularities appear when gluons are emitted from the final state
heavy quarks, since they are screened by the quark mass.
Therefore, contrary to the case of a light parton,
$d\sigma/dp_T$ for a heavy quark
is a well-defined quantity in NLO.  For light partons, one would
encounter a collinear singularity and would have to introduce a fragmentation
function, usually not calculable from first principles.
\item[ii)]                                             
At large $p_T$, nevertheless, 
large $\log(p_T/m)$ factors  appear, signalling the
increased probability of collinear gluon emission.   At large $p_T$, the
massive quark looks in fact more and more like a massless particle.  These
logarithms  can be resummed using the fragmentation function formalism. 
Contrary to the case of light quarks, however, the fragmentation function for a
heavy quark can be calculated from first principles~\cite{Azimov82,Mele91}. 
\item[iii)]                                                                
New processes appear at NLO 
which drastically change either the $\hat s$ dependence of the
cross-sections, or the kinematical distributions. 
\end{itemize}                   
In the following few subsections I will present some examples of where these
large corrections arise, and how they can be accounted for. The discussion will
be rather qualitative, and is just meant to give an idea of the main issues and
to introduce a nomenclature which is often used in the literature and heard in
seminars. I hope these examples will help you demistifying some possibly 
esoteric language.
For more detailed treatments, look at the literature quoted.
      
\subsubsection{$t$-channel gluon exchange diagrams}
Diagrams with a gluon exchanged in the $t$-channel alter the 
high-energy behaviour of the total cross-section~\cite{Nason88}.
Consider for example the following diagram:     
\\[0.5cm]
\centerline{
      \epsfig{figure=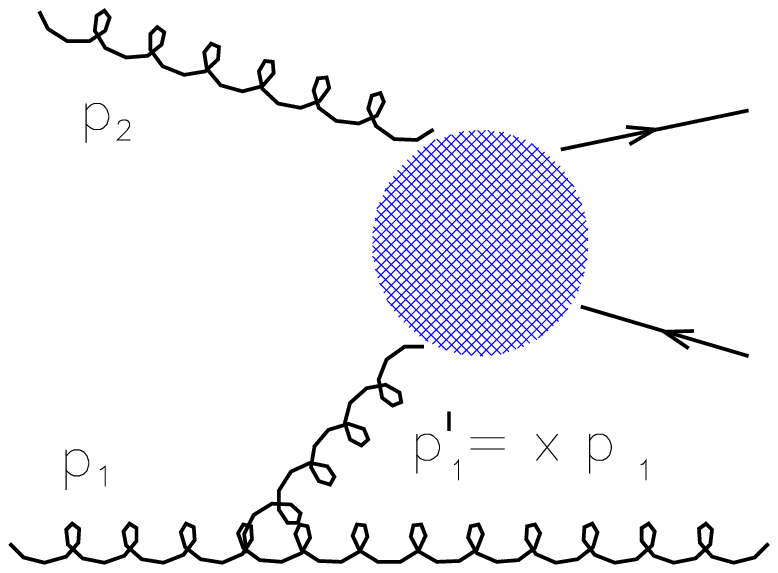,height=0.2\textheight,clip=}}
\\[0.5cm]                                       
Let us call $\hat s$ the incoming total energy squared, and $s^\prime$ the
energy squared of the hard subprocess involving gluon 2 and 1$^\prime$.
Let us concentrate on the phase-space region where $s^\prime \ll \hat{s}$.
Since the probability to find the gluon 1$^\prime$ with momentum fraction $x$ 
inside gluon 1 is proportional to 
$\as/x$, the approximate contribution from this process to the heavy-quark 
cross-section is
\be
\sigma_{12 \rightarrow Q\overline{Q}X} \sim \int^1_{4m^2/\hat s} dx
{\alpha_s \over x}~ \sigma (1^\prime 2 \rightarrow Q \overline{Q})
\ee                                   
Using
\begin{eqnarray*}
\sigma_{1^\prime 2 \rightarrow Q\Qb} &\sim& {\alpha^2_s \over s^\prime} = 
{\alpha^2_s \over
x\hat s} \\       
&\Rightarrow&
\sigma_{12 \rightarrow Q\Qb X} \sim \int^1_{4m^2/\hat s} \alpha_s~{dx \over
x^2}~{\alpha^2_s \over \hat s} = {\alpha^3_s \over \hat s} \left( {\hat s \over 4m^2} -1
\right) \rightarrow \left( {\alpha_s \over 4m^2} \right) \alpha^2_s
\end{eqnarray*} 
Therefore,
\be
\frac{\hat\sigma_{NLO}(gg\rightarrow Q \overline{Q}g)   }
     {\hat\sigma_{LO} (gg\rightarrow Q \overline{Q})    }  
  \;   \stackrel{\hat s \rightarrow \infty}{\rightarrow}     \;
           \as{\hat s \over m^2}  
\ee
At very high partonic energies, the NLO process will dominate over the LO one!
Of course production at large energy is suppressed in the total hadronic
cross-section
by the convolution with the steeply falling parton densities. To quantify the
impact of these effects, we can take for example
a gluon density inside the proton with the form:
$$                                              
f(x) \sim {{\cal A} \over x^{1+\delta}}
$$                             
with $\delta < 1$
It is then easy to show (Exercise!) that
\[                                       
{\sigma_{\rm NLO}(p \bar p \rightarrow Q \overline{Q}) \over \sigma_{\rm LO} (p
\bar p \rightarrow Q \overline{Q})} \sim
\left\{                                 
\matrix{
\alpha_s \log {S\over 4m^2} &{\rm if~} \delta \log {S \over 4m^2} \ll 1\cr \cr
\alpha_s {1 + \delta \over \delta} &{\rm if~} \delta \log {S \over 4m^2} \gg 1
}\right.                    
\]                                        
Therefore, for small $\delta$ the NLO correction becomes larger than the LO term
as soon as $\alpha_s \log S/4m^2 \gsim~1$.  At the Tevatron (and for $m = m_b
\sim$ 4.5~GeV), this number is indeed of order 1!
                                                 
These large logarithms are known as ``$\log {1 \over x}$" terms, where $x \sim
\sqrt{4m^2/S}$ is the average value of the hadron energy fraction
needed to produce the pair. 
These $\log 1/x$ terms appear at all
orders of PT, but fortunately 
techniques are known to resum them~\cite{Catani91,Collins91}.
Notice, however, that they would not be present if the gluon
density were steeper than $1/x$.  For example,  $\delta = 0.5$ would give only
a moderate NLO contribution.  The apparent paradox of this different behaviour
for different values of $\delta$ is resolved by the observation that these
large $\log 1/x$ terms can be resummed in a universal fashion and absorbed in
an improved evolution equation for the gluon density (the BFKL evolution
equation).  As a result of this evolution, a gluon density with $\delta = 0$
immediately develops a $\delta > 0$.  One can therefore work with a BFKL
resummed gluon density, and forget about large $\log(1/x)$ corrections to the
heavy-quark production cross-section.

More accurate estimates of the effect of $\log(1/x)$ resummation for $b$
productionat the Tevatron can be found in~\cite{Collins91}, where it was
concluded that corrections of up to $30$\% in addition to the NLO calculation
can be expected. 

\subsubsection{Gluon-splitting contributions}
The so-called gluon-splitting contributions come from diagrams of the following
type:
\\[0.5cm]
\centerline{
      \epsfig{figure=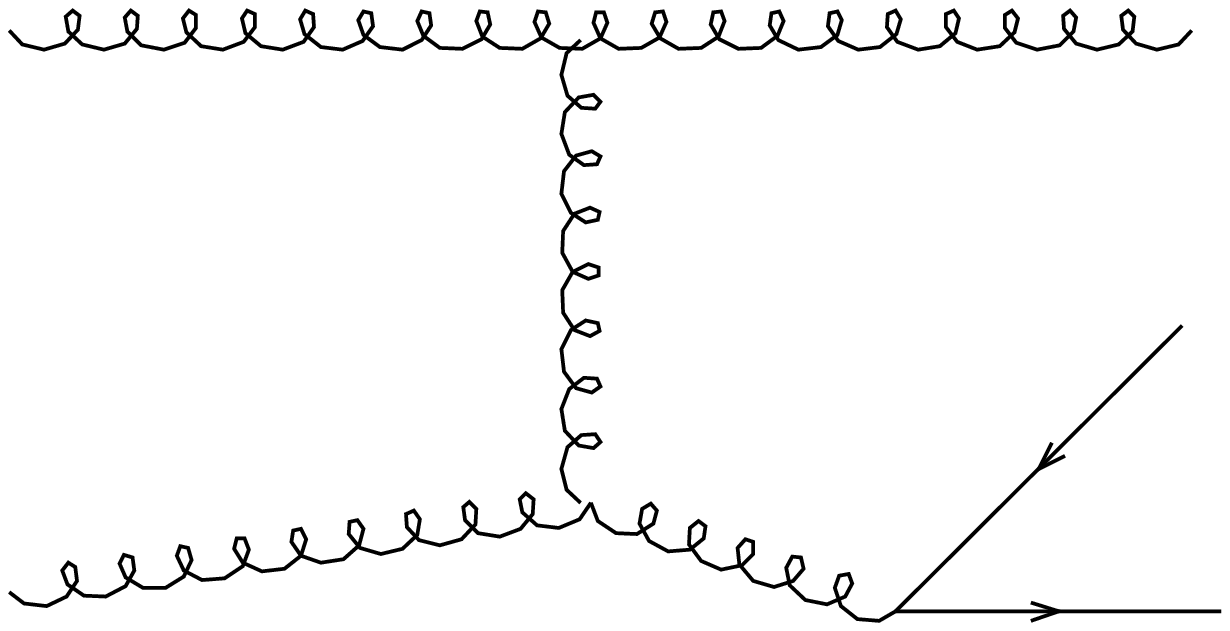,height=0.15\textheight,clip=}}
\\[0.5cm]                                       
In this case we should focus on the region of phase-space where the $t$-channel
gluon is very hard and far off-shell, while the {\em splitting} gluon has a
virtuality much smaller than its transverse momentum.
These diagrams then give rise to potentially large logarithms.  We can use a
fragmentation-function language to estimate the contribution of these processes
to heavy-quark production. One factorizes the process into a $gg\to gg$ (or
$qg \to qg$) scattering, with one final-state gluon off-shell, followed 
by the evolution of the off-shell gluon into a $\QQ$ pair. Higher-order
corrections, such as the emission of multiple gluons from the final state
gluon, can be included by convoluting the light-parton cross-section with the
$g\to Q$ fragmentation function. This approach is similar to what done in the
case of inclusive light-hadron production inside a jet, with the difference
that the $g\to Q$ fragmentation function can be calculated from first
principles within PT~\cite{Mele91}.                                       
This will account both for the increase
in probability that the gluon will turn into a heavy-quark pair (due to the
larger number of gluons present when one considers the multi-gluon emission),
and for the slow-down of the heavy-quark momentum due to the loss of energy in
the multi-gluon radiation process. 
The fragmentation function for a heavy quark inside a gluon
jet with transverse energy $p_\perp$ obeys the evolution equation:
\be \label{eq:gtoQ}
{dD(x,q^2) \over d\log q^2} = 
{\alpha_s \over 2\pi} \int^1_x {dz \over z} G(z,q^2,\pt^2) P_{Qg}({x\over z})
\ee
where $G(z,q^2,\pt^2)$ is the density of gluons of virtuality $q^2$ carrying a
momentum fraction $z$ inside 
a gluon of maximal virtuality $\pt^2$, and $P_{Qg}(x)$ 
is the Altarelli-Parisi kernel given by           
$$                         
P_{Qg} = {1 \over 2} [z^2 + (1-z)^2]
$$                              
The heavy-quark multiplicity  (which, for small average multiplicity, coincides
with the probability to find a quark in the gluon jet) is given by the first
moment of the fragmentation function\footnote{The $n$-th moment of a function
$f(x)$ is defined by  $\int_0^1 dx \; x^{n-1} f(x)$.}:
\be
N \equiv \int^1_0 dz D(z) 
\ee
which obeys the equation:
$$
{dN \over d \log q^2} = {\alpha_s \over 2\pi} G_{(1)} P^{(1)}_{Qg}
$$
with $G_{(1)}$ and $P^{(1)}_{Qg}$ first moments of $G(x)$ and
$P_{Qg}(x)$. 
$G_{(1)}(q^2,\pt^2)$ is the number of gluons with virtuality $q^2$ in the jet,
and
$$                                                       
P^{(1)}_{Qg} = \int^1_0 dz~{1 \over 2} (z^2 + (1 - z)^2) = {1 \over 3}~.
$$
At the first order, neglecting the $q^2$-dependence of the gluon multiplicity 
and therefore assuming $G_{(1)}\equiv 1$, we get          
\be \label{eq:gsplit}                   
N(Q \bar Q) \sim {\alpha_s \over 6\pi} \log \left( {p^2_T \over m^2} \right)
\ee
where $p^2_T$ and $m^2$ are the maximum and minimum gluon virtuality. In the
case of light quarks, $m=0$ and this calculation would not be possible. A
collinear divergence would develop, associated to the vanishing minimum gluon
virtuality, and one would need to introduce a non-perturbative fragmentation
function. Once more, this shows the advantage of working with heavy quarks!

The logarithm in eq.~(\ref{eq:gsplit}) signals the growth in probability to
find a heavy quark inside a large-$\pt$ gluon. It is yet one more potentially
large logarithm that was not present at leading order, and which appears for
the first time at NLO. Similar large logarithms will appear at higher orders of
perturbation theory (associated to processes where more gluons are radiated in
the final state). These logarithms are resummed by solving exactly the
evolution equation for the $g\to Q$ fragmentation function,
eq.~(\ref{eq:gtoQ}). 

\subsubsection{Flavour-excitation processes}
An analogous class of NLO corrections is given by the so-called
{\em flavour-excitation} diagrams.  These can be thought of as initial-state
gluon-splitting processes:                                             
\\[0.5cm]
\centerline{
      \epsfig{figure=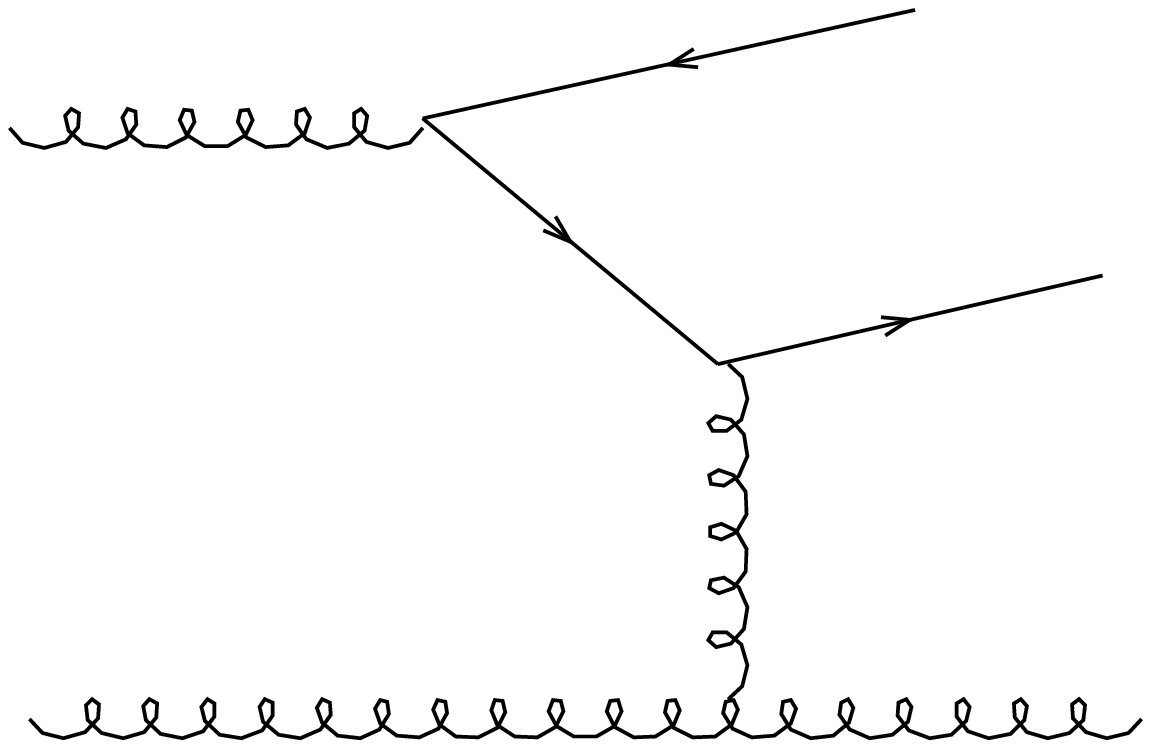,height=0.15\textheight,clip=}}
\\[0.5cm]                                       
The relevant region of phase-space in this case is the one with the heavy-quark
propagator close to the mass shell.
These processes are analogous to the singlet contribution to $F_2$ in DIS.
It is known that these effects can be reabsorbed in the AP evolution of the
parton densities~\cite{Collins86}.  
In our case, this is equivalent to defining a heavy-quark
density inside a proton by:
\be                                                 
\left\{
\matrix{
{dD_Q(x,q^2) \over d \log q^2} &=& {\alpha_s \over 2\pi} \int^1_x {dz \over z}
G(z,q^2)
P_{Qg} \left( {x \over z}\right) \cr\cr
D_Q(x,m^2) & = & 0.}\right.
\ee
It is easy to get a crude estimate of the ``heavy-quark density" by assuming a
simple functional form for $G(x)$, for example, $G(x) = A/x$.  Then:
\begin{eqnarray*}
{dD_Q(x) \over d \log q^2} &=& {\alpha_s \over 2\pi} \int^1_x {dz \over
z}~G\left({ x \over z}
\right)~P_{Qg}(z) \\  
&=& {\alpha_s \over 2\pi} \int^1_x {dz \over z}~{A \over x/z}~{1 \over 2}
\left[ z^2 +                                                             
(1-z)^2 \right] \\
&\simeq& {\alpha_s \over 6\pi} G(x)
\end{eqnarray*}                   
Assuming $G(x,q^2)$ slowly varying in the $q^2 \gsim m^2$ region:
\[ \begin{tabular}{|c|}\hline  \\
$ D_Q(x,q^2) \simeq {\alpha_s \over 6\pi} \log \left( {q^2 \over m^2} \right)
G(x,q^2)  $ \cr \cr \hline
\end{tabular}
\]
Since the NLO corrections contain these ``structure function-like'' piece, it
would be incorrect to add to LO processes contributions from $gQ \rightarrow
gQ$.  In fact, this would imply a double-counting of the NLO flavour-excitation
diagrams.  Of course, the use of the $gQ \rightarrow gQ$ process allows one to
capture higher-order effects included in the AP evolution
equations, which are usually solved to generate the heavy-quark densities found
in standard Fortran libraries of parton density functions.  
The use of the NLO flavour-excitation diagrams
reproduces instead more faithfully the exact kinematics and correlations of the
flavour-creation process in the region next to the threshold.
                                                             
\begin{figure}
\centerline{\epsfig{file=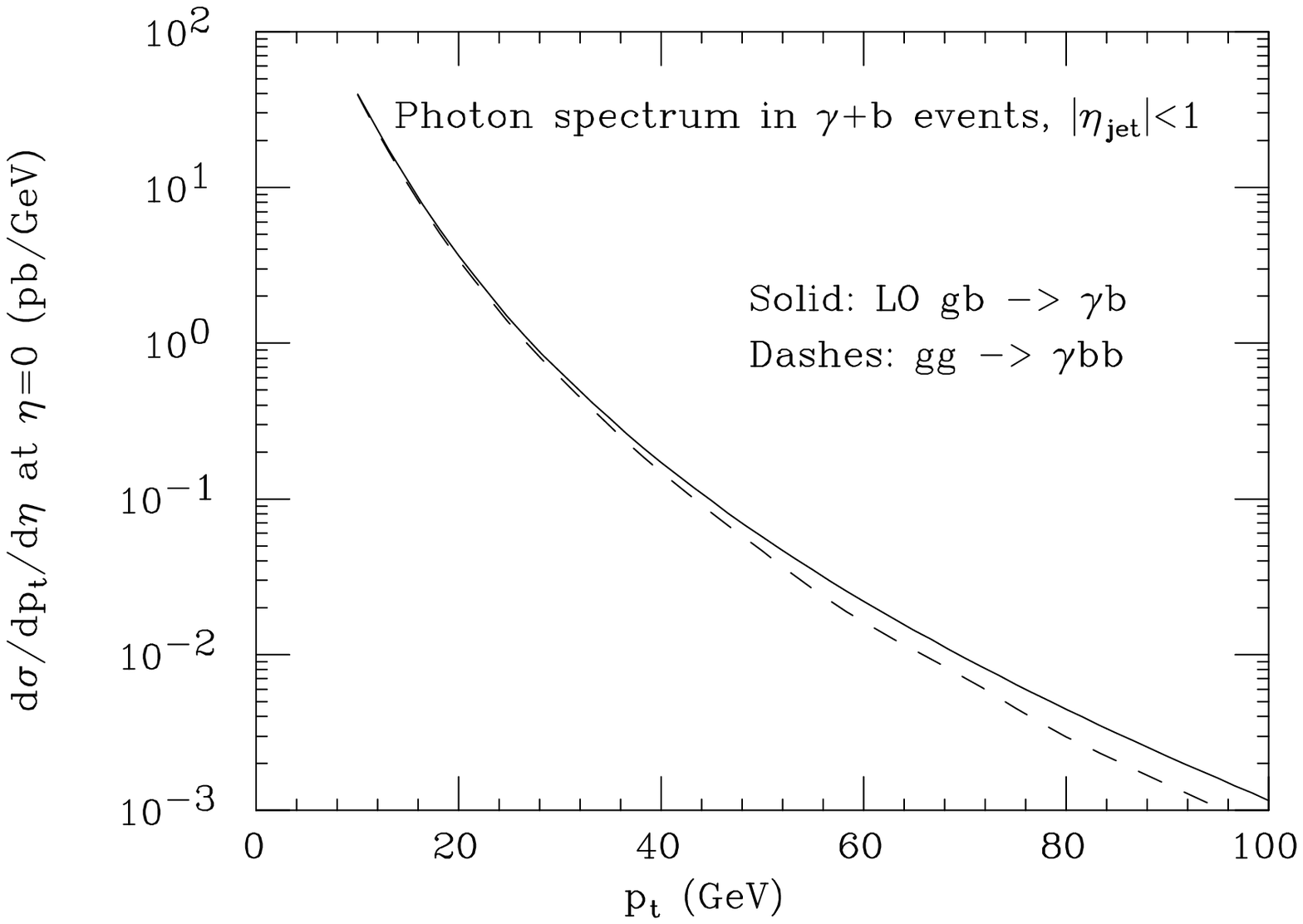,height=0.35\textheight,clip=}}
\ccaption{}{\label{fig:phobot1} The $\pt$ distributions for central photons
produced in asociation with a $b$ quark, 
in $p \bar p$ collisions at $\sqrt{S}=1.8$~TeV. The results of the two 
different calculations discussed in the text are displayed.}
\end{figure}

\begin{figure}
\centerline{\epsfig{file=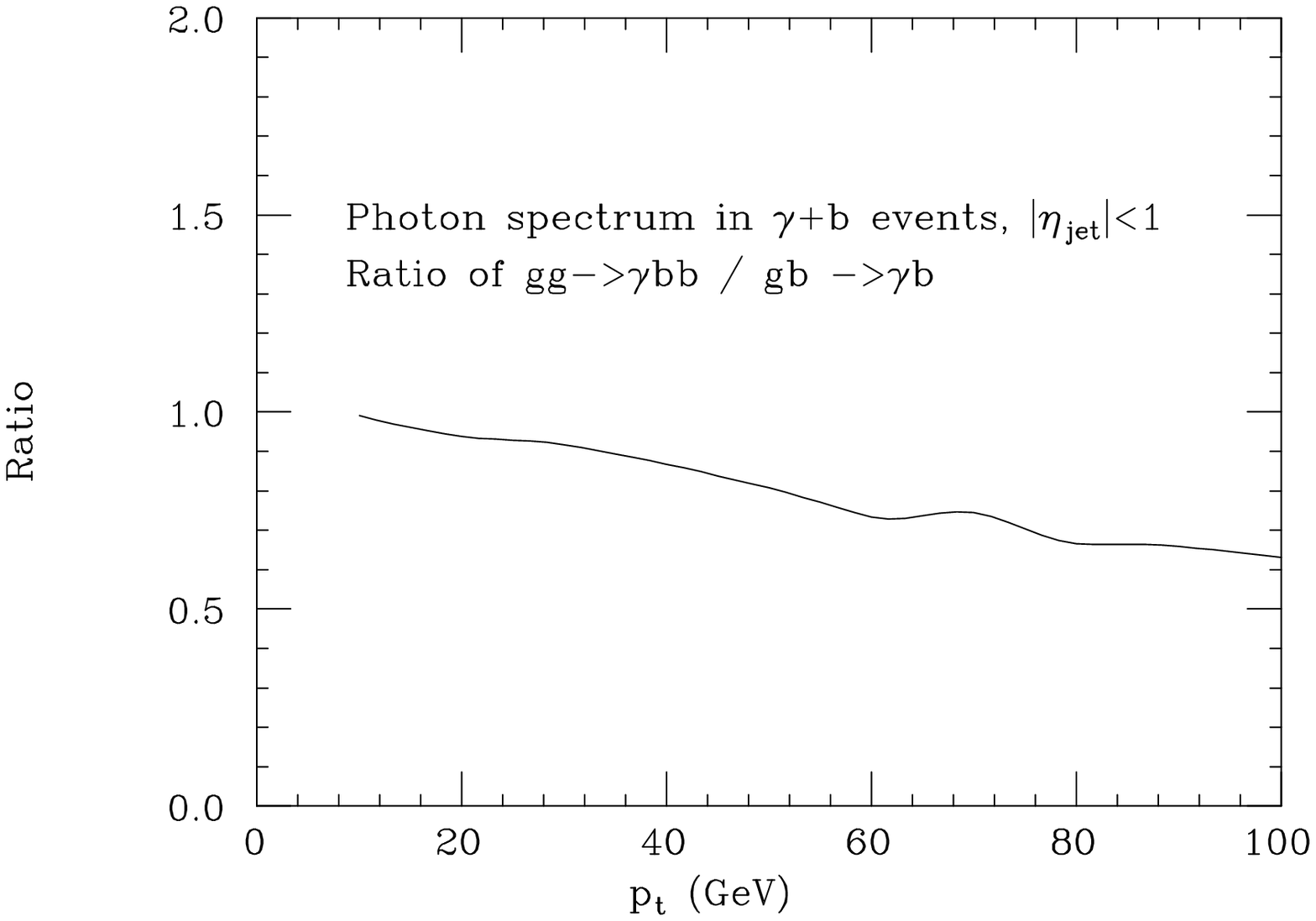,height=0.35\textheight,clip=}}
\ccaption{}{\label{fig:phobot2} The ratio between the two curves shown in the
previous figure.}
\end{figure}

The example I will give here describes the comparison between the two
approaches when applied to the simpler case of associated production of a
$\gamma$ and heavy quarks. In this case I compare the results of the two
calculations of the photon $p_\perp$ spectrum obtained by using either of the
these two processes:
\\[0.5cm]                            
\centerline{
      \epsfig{figure=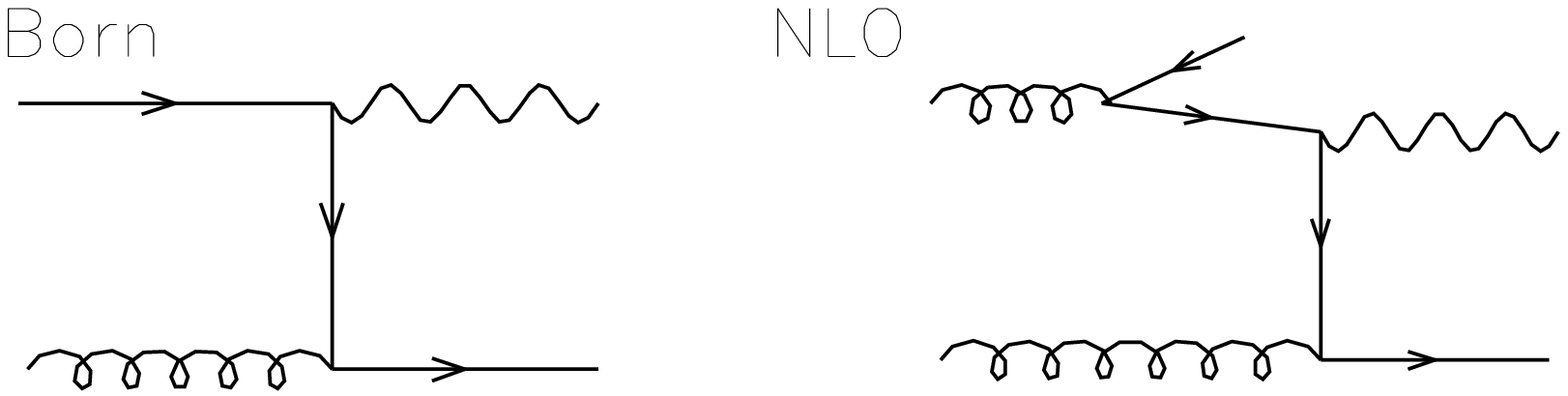,height=0.15\textheight,clip=}}
\\[0.5cm]                                       
The calculation done using the LO approach, which uses as an input the
$b$ structure function taken from a given set of parton distribution
functions (CTEQ1M, in this case), is shown in fig.~\ref{fig:phobot1} as a solid
line. 
The calculation done by evaluating the ${\cal O}(\as^2\aem)$ matrix
element is shown by the dashed line. The ratio of the two distributions is
given in fig.~\ref{fig:phobot2}. Notice that the two calculations agree
remarkably well down to low values of $\pt$, indicating that even the structure
function approach can be used to reproduce rather accurately the onset of the
mass threshold (this is because we are looking at a rather inclusive quantity.
Were we looking at more exclusive quantities, or at correlations between the
$b$ and the $\bar b$ in the final state, the LO approach would not even be
usable: in this approach the antiquark is always integrated over and we
cannot keep track of where it went!)
At large $\pt$ some differences between the two calculations
emerge,
due to the absence of higher-order terms in the NLO calculation. These
higher-order terms are included in the structure-function approach, since the 
$b$ structure function itself has been evolved to large $\pt$ using the exact
Altarelli-Parisi evolution equation. 
The ideal calculation, therefore, would merge the two approaches by smoothly
interpolating betwen the two regimes. This was done in
refs.~\cite{Cacciari94,Scalise96},
and the results will be discussed in the next lecture.

\subsection{Heavy-quark fragmentation}
At the end of the perturbative evolution, one still expects hadronization
effects to alter the spectrum of the heavy quark while it transforms into a
hadron. These effects are usually described in terms of a non-perturbative
fragmentation function. The fundations of this approach lie in a {\it
factorization theorem}:
\[
   \frac{d\sigma}{dP_T} \;=\; \int_{\frac{P_T}{E_{beam}}}^1 \; \frac{dz}{z} 
   \; F(z)                                                        
   \; \frac{d\sigma}{dp_t}(p_t=P_T/z)
\]                    
where $P_T$ is the hadron momentum, and 
$d\sigma/dp_t$  includes the {\it perturbative} part of
the heavy-quark fragmentation function, namely the resummation of the $\log
(\pt/m)$ terms~\cite{Mele91}. The factorization theorem ensures that, up
to corrections of order $\Lambda/P_T$, with $\Lambda$ a scale typical of the
hadronization phenomena, the function $F(z)$ for a given heavy flavour
is universal.
Although non-perturbative in nature, some general features of fragmentation
functions can be extracted from first principles. For example, one can
establish a relation between the effects of hadronization on heavy quarks
of different masses:
\be
     1-\bar{x}_B = \left(\frac{m_c}{m_b}\right)  ( 1-\bar{x}_D)
\ee
where $\bar{x}$ is any parameter such as
$\langle x\rangle$ (the average value of $x$) or $x_{\rm max}$ (the value of
$x$ at which the fragmentation function peaks). 
This follows easily from phase-space considerations. One can idealize the
fragmentation process with a transition of a slightly off-shell heavy quark
(possibly in the field of  nearby soft gluons) decaying to a heavy hadron $H$ 
plus  a light system $\Pi$ (e.g. pions):
\\[0.5cm]                    
\centerline{
      \epsfig{figure=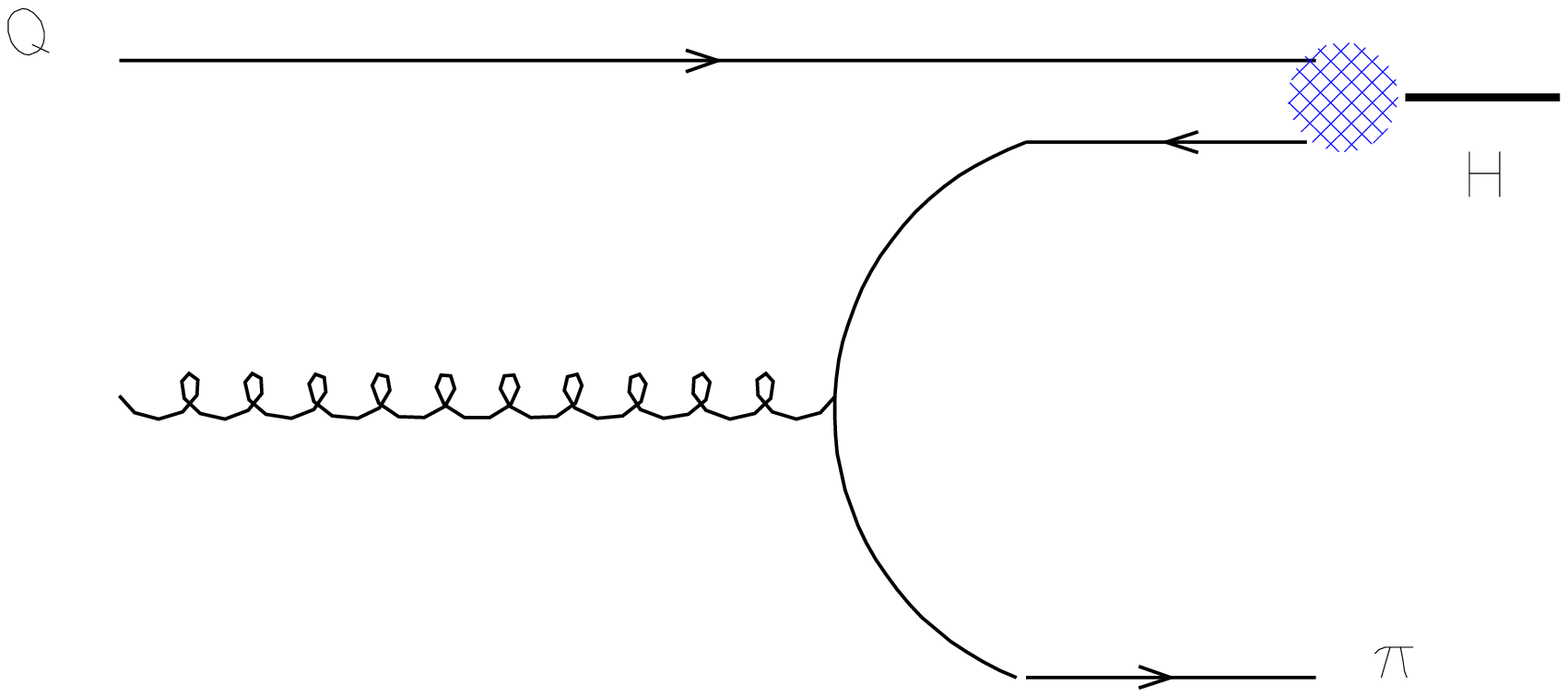,height=0.15\textheight,clip=}}
\\[0.5cm]                                       
In this case:
\be
      P_H \; : \; P_{\Pi} \; \equiv \;  
      z_H \; : \; z_{\Pi} \; = \;  M_H \; : \; M_{\Pi} 
\ee
and, as a result,
\be
   1-z \sim 1-\frac{M_H}{M_H+M_{\Pi}} \sim \frac{M_{\Pi}}{M_H}
\ee
A typical example of parametrizations for non-perturbative fragmentation
functions is the {\em Peterson, Schlatter, Schmitt {\em and} Zerwas}
form~\cite{Peterson83}: 
\be                  
    \frac{dN}{dz} \propto \frac{1}{x \left[ 1-1/x-\epsilon/(1-x)\right]^2}
\ee                                                                      
Here $\epsilon$ is the only parameter describing the
non-perturbative part, and scales as follows:
\be                               
    \epsilon \sim \frac{m_0^2}{m_Q^2} \quad \quad
    \langle 1-z\rangle \sim \sqrt{\epsilon}      
\ee                         
Another interesting form was proposed by {\em Nason {\em and}
Colangelo}~\cite{Colangelo92}:
\be                                                                      
    \frac{dN}{dz} \propto (1-z)^\alpha \; z^\beta
\ee
In this case the relation between the values of $\alpha$ and $\beta$ for
different values of the heavy-quark mass (say charm and bottom hadrons)
can be extracted by imposing the two relations:
\ba         \label{eq:alpha}
  1 -\langle x_B\rangle &=& \frac{m_c}{m_b} (1 -\langle x_D\rangle )
\\          \label{eq:beta}                                
  1 - \hat{x}_B&=& \frac{m_c}{m_b} (  1 - \hat{x}_D) \; ,
\ea                                                      
which can be easily solved analitically (Exercise!) in terms of $\alpha$ and
$\beta$.
The values of the parameters characterizing the non-perturbative fragmentation
functions can be extracted by fitting some set of data. The factorization
theorem will then allow us to use the same parameters to predict cross-sections
for other data. The best place to extract information on heavy-quark
fragmentation is $e^+e^-$ collisions, since there at least the energy of the
primary heavy quark before the perturbative and non-perturbative fragmentation
processes occur is well determined, being equal to the beam energy.
Recent fits
to charm production data from the old ARGUS experiment~\cite{ARGUS} at DESY
($\sqrt{s}=10.6$~GeV)                                               
and the more recent OPAL experiment~\cite{Akers95} at 
LEP ($\sqrt{s}=91.2$~GeV) have been
performed in ref.~\cite{Cacciari97}. 
\begin{figure}                    
\centerline{\epsfig{figure=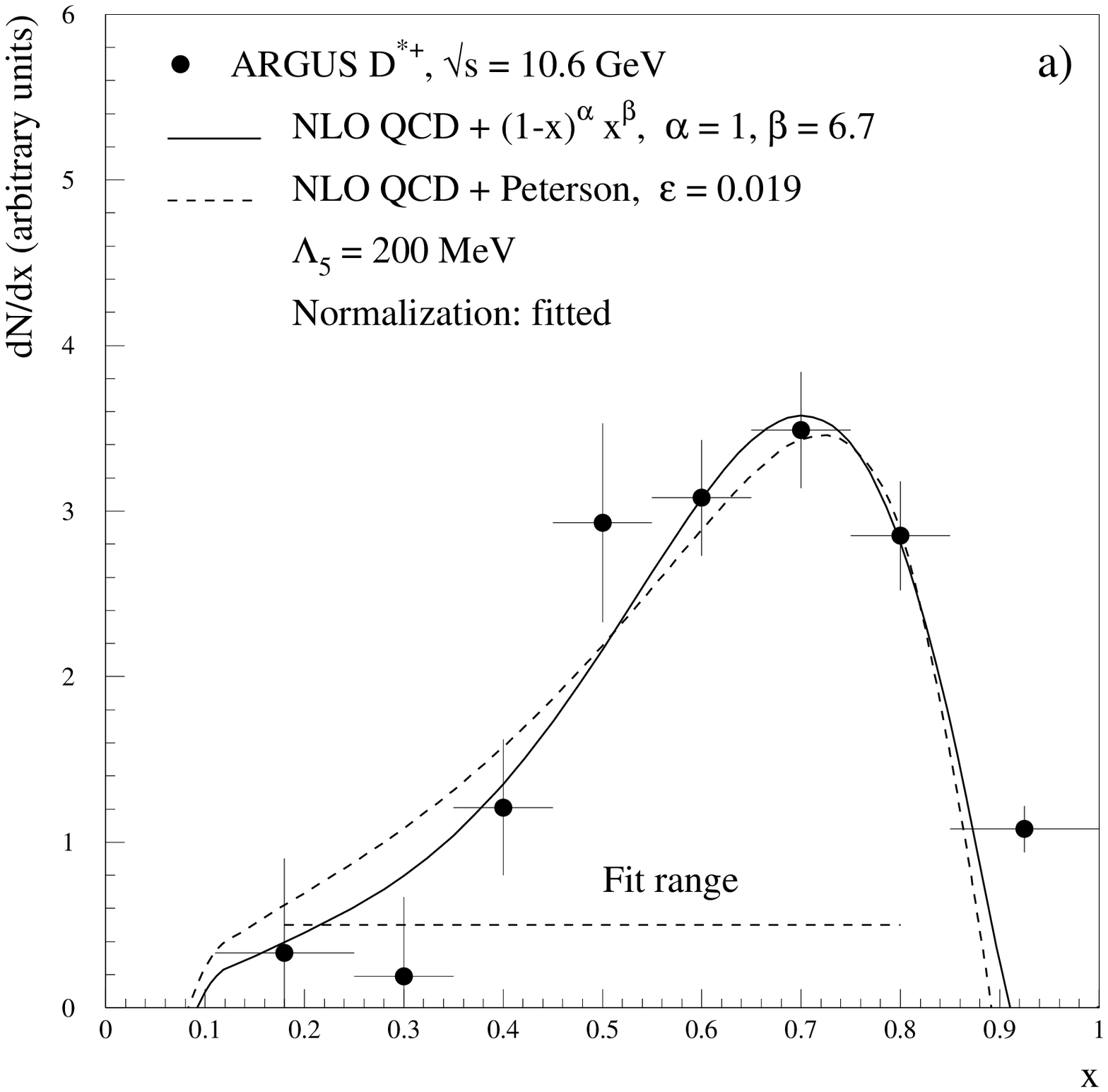,width=0.4\textwidth,clip=}\hfil
            \epsfig{figure=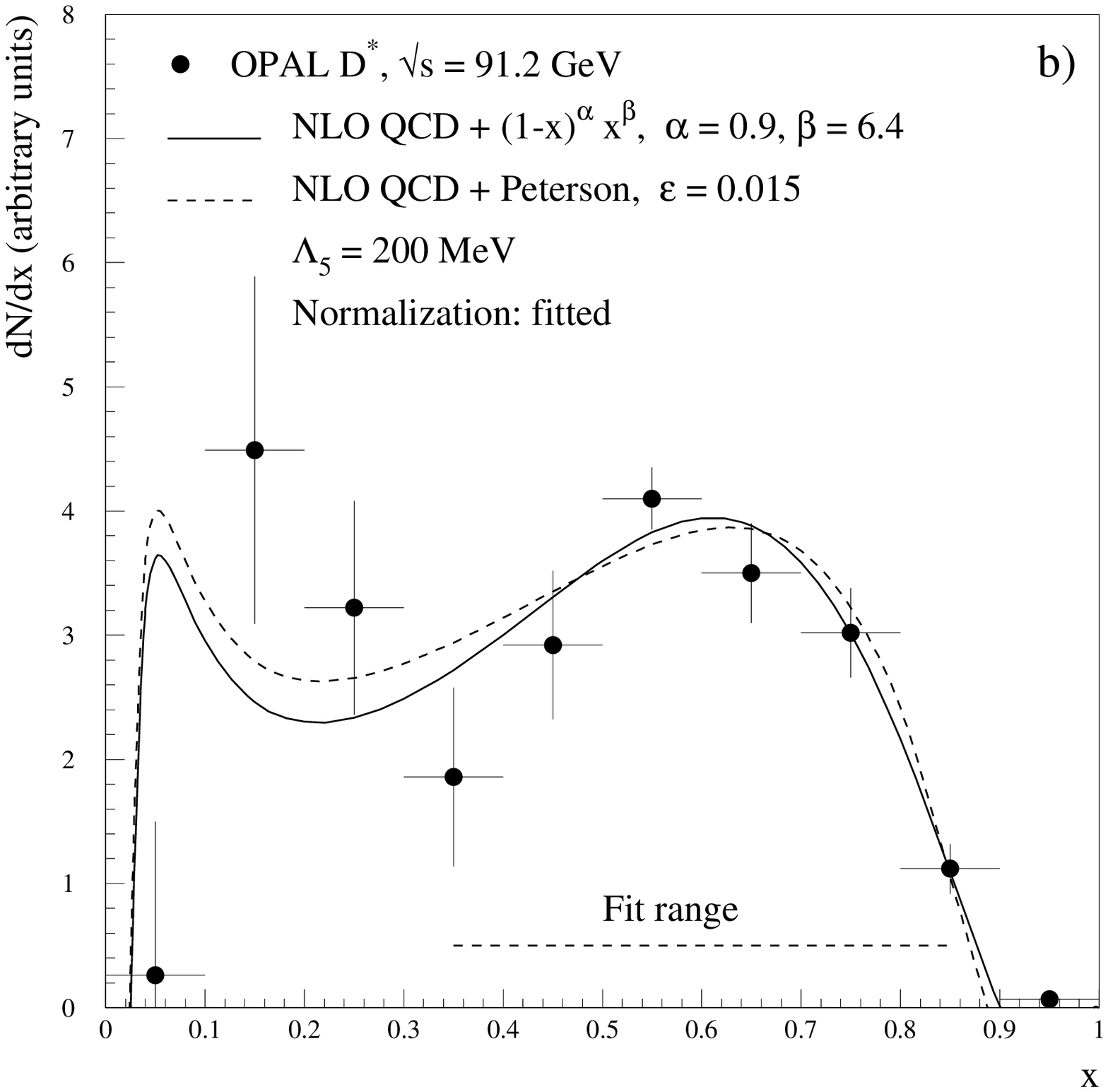,height=0.4\textwidth,clip=}}    
\ccaption{}{\label{fig:argus}
  Fits to the charm fragmentation function measured at ARGUS (left) and OPAL
(right).}                                                                   
\end{figure}
The results are shown in
fig.~\ref{fig:argus}. The parameters extracted from the fits to the data of the
two experiments, using different values of $\Lambda_{QCD}$, are contained
in the following tables (for ARGUS and OPAL, respectively):
\\[0.5cm]                                                 
         \centerline{
            \hfill
            \begin{tabular}{ll}   \hline
              $\Lambda_5 $=200~{\rm MeV} & 300~{\rm MeV}\\ \hline
              $\epsilon_c$=0.019 & 0.011 \\                      
              $\alpha=1 $&  - \\
              $\beta=6.7 $&  -  
            \end{tabular} \hfill
            \begin{tabular}{ll}    \hline
              $\Lambda_5$=200~{\rm MeV} & 300~{\rm MeV} \\ \hline
              $\epsilon_c$=0.015 & 0.008 \\
              $\alpha=0.9 $& - \\           
              $\beta=6.4  $& -
            \end{tabular}                     
                        \hfill} 
\\[0.5cm]
Few comments should be made:
\begin{itemize}
\item
  There is a very good agreement between the value of the parameters extracted
from the two experiments. This confirms the validity of the factorization
theorem.
\item 
  The value of the parameters depends strongly on the input value of
$\Lambda_{QCD}$. This is reasonable, since the perturbative part of the
heavy-quark evolution will depend on the strength of the perturbative coupling
constant $\as$. A larger value of $\as$ implies a larger amount of energy
radiated off in the form of gluons. This leaves less energy to the quark before
it hadronizes, and calls for a harder non-perturbative fragmentation to
reproduce the data than if a small value of $\as$ were used to start with. 
This correlation is faithfully reproduced by the fits shown above.
Notice also that the value of the Peterson parameter $\epsilon_c$
which is usually obtained by using the LO result for the 
perturbative part of the fragmentation function
($\epsilon_c=0.06$, \cite{Cacciari97,Chrin87}), 
is much larger than what extracted from these NLO fits ($\epsilon_c=0.01-0.02$).
\end{itemize}                                                                  

\begin{figure}
\centerline{\epsfig{figure=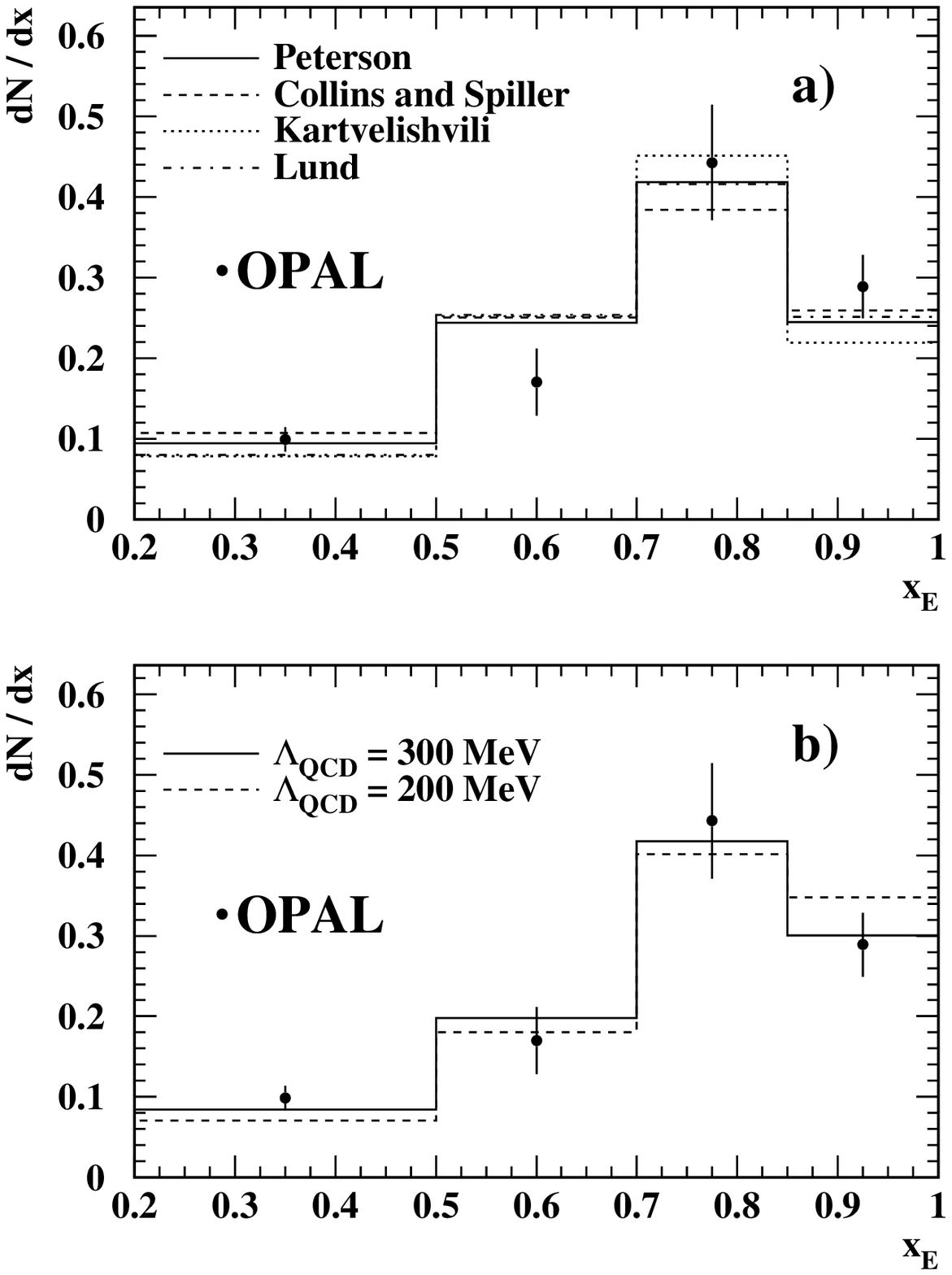,width=0.6\textwidth,clip=}}
\ccaption{}{\label{fig:opalbfrag}
   Comparison of OPAL $b$-fragmentation function data with 
{\em fits} to different
fragmentation function forms (upper figure), 
and with {\em predictions} based on the convolution of
the $b$-quark NLO fragmentation function and the Colangelo-Nason
non-perturbative fragmentation function, with parameters extrapolated from
fits to the charm fragmentation function (lower figure).}
\end{figure}

The relations given in eqs.~(\ref{eq:alpha}-\ref{eq:beta})
allow us to predict the value of the parameters
governing the non-perturbative
fragmentation of $b$ quarks. When this is done, one can predict the $b$-hadron
fragmentation function at LEP. The comparison of data~\cite{Alexander96}
and theory (using the                                                   
Colangelo and Nason fragmentation model, and the $\alpha$ and $\beta$ values
extracted from the charm fits) is shown in the bottom plot of
fig.~\ref{fig:opalbfrag}. Excellent agreement is seen, again a good test that we
are on the right track!
\section{Phenomenology of heavy-quark production}
The ultimate test of our ability to describe the heavy-quark production
processes rests on the comparison with actual data. In the remaining part of
these lectures I will present a survey of current experimental results, and how
they compare to theory. Unless otherwise indicated, {\em theory} will refer to
NLO calculations. A more complete review can be found in
ref.~\cite{Frixione97}.
\subsection{Fixed target production}
Heavy-flavour production has been studied extensively in fixed-target
experiments, with both hadron and photon beams. The typical
centre-of-mass energy is in the range 10-40~GeV, where the bottom
cross-section is rather small. Therefore most of the available data
are on charmed-hadron production.
Total cross-sections, single-inclusive
distributions, correlations between the quark and the antiquark have
been measured in both hadro- and photoproduction.  The theoretical
apparatus of perturbative QCD is in this case at its very limit of
applicability, because the charm mass is very close to
typical hadronic scales. Thus, effects of non-perturbative origin
will very likely play an important role. Conversely, it is hoped that
these effects may be better understood by studying charm production.
Although modern fixed-target experiments have considerably improved
the situation, many open problems remain in this field. All
experimental results are in qualitative agreement with perturbative
QCD calculations, thus supporting the ``hard'' nature of
charm-production phenomena. However, several quantitative deviations from
pure QCD are observed.  It is interesting to see whether simple models
of non-perturbative phenomena, such as fragmentation effects and intrinsic
transverse momenta, may be sufficient to   
explain these deviations. I will discuss these problems at length.
I will not discuss, instead, the subject of $x_F$-asymmetries. A discussion of
this issue can be found in ref.~\cite{Frixione97}, and recent relevant data
(and references to previous literature), in ref.~\cite{Aitala96}.

\begin{figure}[ht]
\begin{center}
\mbox{\psfig{file=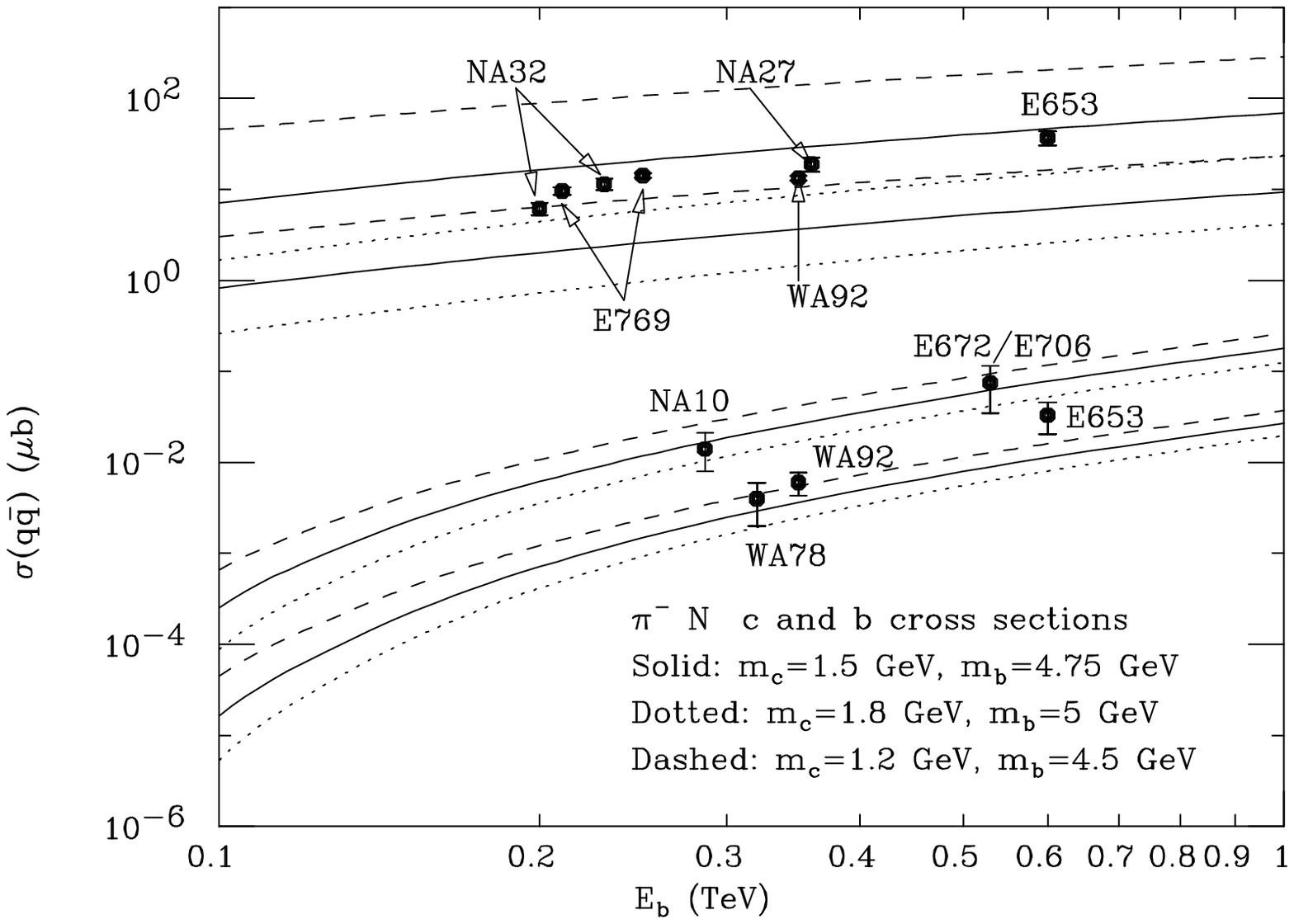,width=0.70\textwidth}}
\ccaption{}{\label{bcpion}
Pair cross-sections for $b$ and $c$ production in $\pi^-N$
collisions versus experimental results.
}
\end{center}
\end{figure}

\subsubsection{Total cross-sections}
\begin{figure}[htb]
\begin{center}
\mbox{\psfig{file=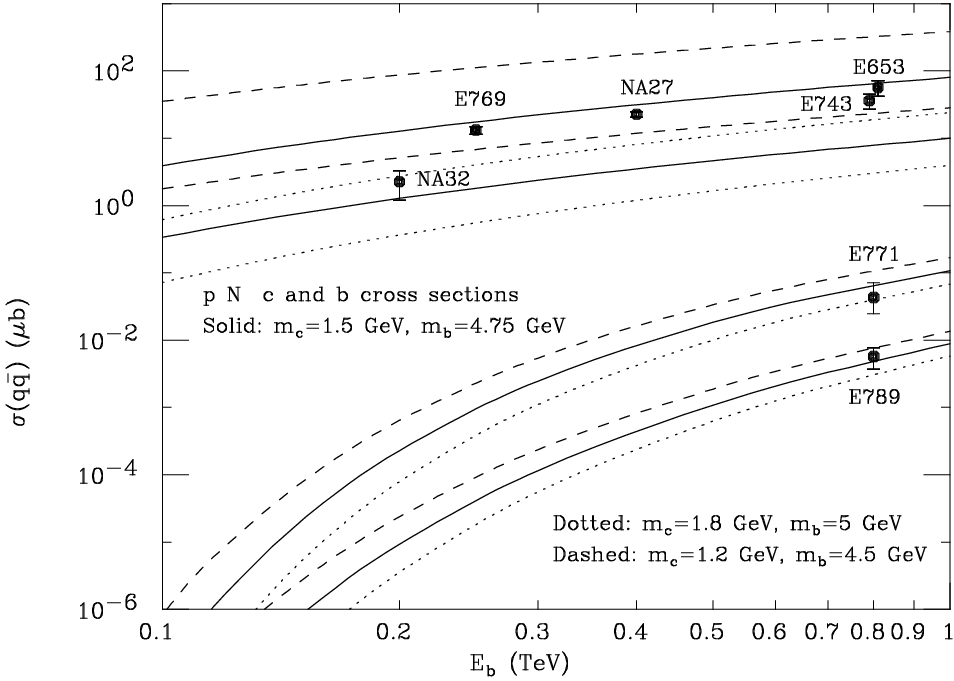,width=0.70\textwidth}}
\ccaption{}{\label{bcproton}
Pair cross-sections for $b$ and $c$ production in $p N$ collisions versus
experimental results.
}
\end{center}
\end{figure}

In fig.~\ref{bcpion} I plot the $c\bar{c}$ and $b\bar{b}$ cross-sections,
computed in QCD at NLO, as functions of the beam energy, for
$\pi^- N$ collisions.
The same quantities are shown in fig.~\ref{bcproton} for a proton beam.
The cross-sections are calculated using the parton distribution sets
of ref.~\cite{Martin95a} for the nucleon, and the central set SMRS2
\cite{Sutton92} for the pion. The default values of the charm and bottom masses
are 1.5 and 4.75 GeV respectively, and the default choices for the
factorization scale $\muf$ and the renormalization scale $\mur$ are
\be
\muf=2m_c,\quad \mur=m_c \quad\mbox{and}\quad  \muf=\mur=m_b
\ee                        
for charm and for bottom.

The bands in the figures are obtained as follows. We varied $\mur$ between half
and twice the central value. The factorization scale $\muf$ was also
varied between $m_b/2$ and $2m_b$ in the case of bottom,
while it was kept fixed
at $2m_c$ in the case of charm. This is because the adopted parametrizations of
parton densities are given for $Q^2$ larger than 5 GeV$^2$.
The bands shown in the figures are therefore
only an underestimate of the uncertainties involved in the computation of charm
production cross-sections. Considering independent variations
for the factorization and renormalization scales does not lead to a wider range
in the bottom cross-section for the energies shown in the figures. We also show
the effect of varying $m_c$ between $1.2$ GeV and $1.8$ GeV, and $m_b$ between
4.5 and 5 GeV.

%%%%%%%%%%%%%%%%%%%%%%%%%%%%%%%%%%%%%%%%%%%%%%%%%%%%%%%%%%%%%%%%%%%%%
%%%%%%%%%%%%%%%%%%%%%%%%%%%%%%%%%%%%%%%%%%%%%%%%%%%%%%%%%%%%%%%%%%%%%

The proton parton densities of ref.~\cite{Martin95a} are available for a wide
range of $\LambdaQCD$ values, corresponding to
$\as(m_{\rm \sss Z})$
values between $0.105$ and $0.130$. The bands shown in figs.~\ref{bcpion} and
\ref{bcproton} for bottom production are obtained by letting $\LambdaQCD$
vary in this range. In the case of charm, values of $\LambdaQCD$
corresponding to $\as(m_{\sss Z})$ above 0.115 induce values of $\as(m_c)$ too
large to be used in a perturbative expansion. For this reason, the upper bounds
on charm production cross-sections are obtained with $\as(m_{\sss Z})=0.115$.
We point out that, by varying $\LambdaQCD$, one is
forced to neglect the correlation between $\LambdaQCD$
and the pion parton densities, which were fitted in ref.~\cite{Sutton92} with
$\Lambdamsb=122$~MeV.

As can be seen, experimental results on total charm cross-sections
\cite{Aguilar85a,Alves96,Adamovich96,Kodama91,Ammar88,Aguilar88}
are in reasonable agreement with theoretical  
expectations, if the large theoretical uncertainties are taken into proper
account. We can see that the hadroproduction data are compatible with a value
of 1.5 GeV for the charm-quark mass. In the case of bottom 
production \cite{Bari91}, the
spread of the experimental data is almost as large as that of the theoretical
predictions. The two results in proton-production~\cite{Jansen95,Alexopoulos97}
differ at the level of                                                         
2.5$\sigma$. This is unfortunate, since the beam energy 
($E_p=800$~GeV) is almost the same of the forthcoming HERAB experiments. A more
consistent experimental estimate of the production cross-section would have
provided more reliable estimates of the $b$ rate at HERAB.
                                                          
%%%%%%%%%%%%%%%%%%%%%%%%%%%%%%%%%%%%%%%%%%%%%%%%%%%%%%%%%%%%%%%%%%%%%
\begin{figure}[htb]
\begin{center}
\mbox{\psfig{file=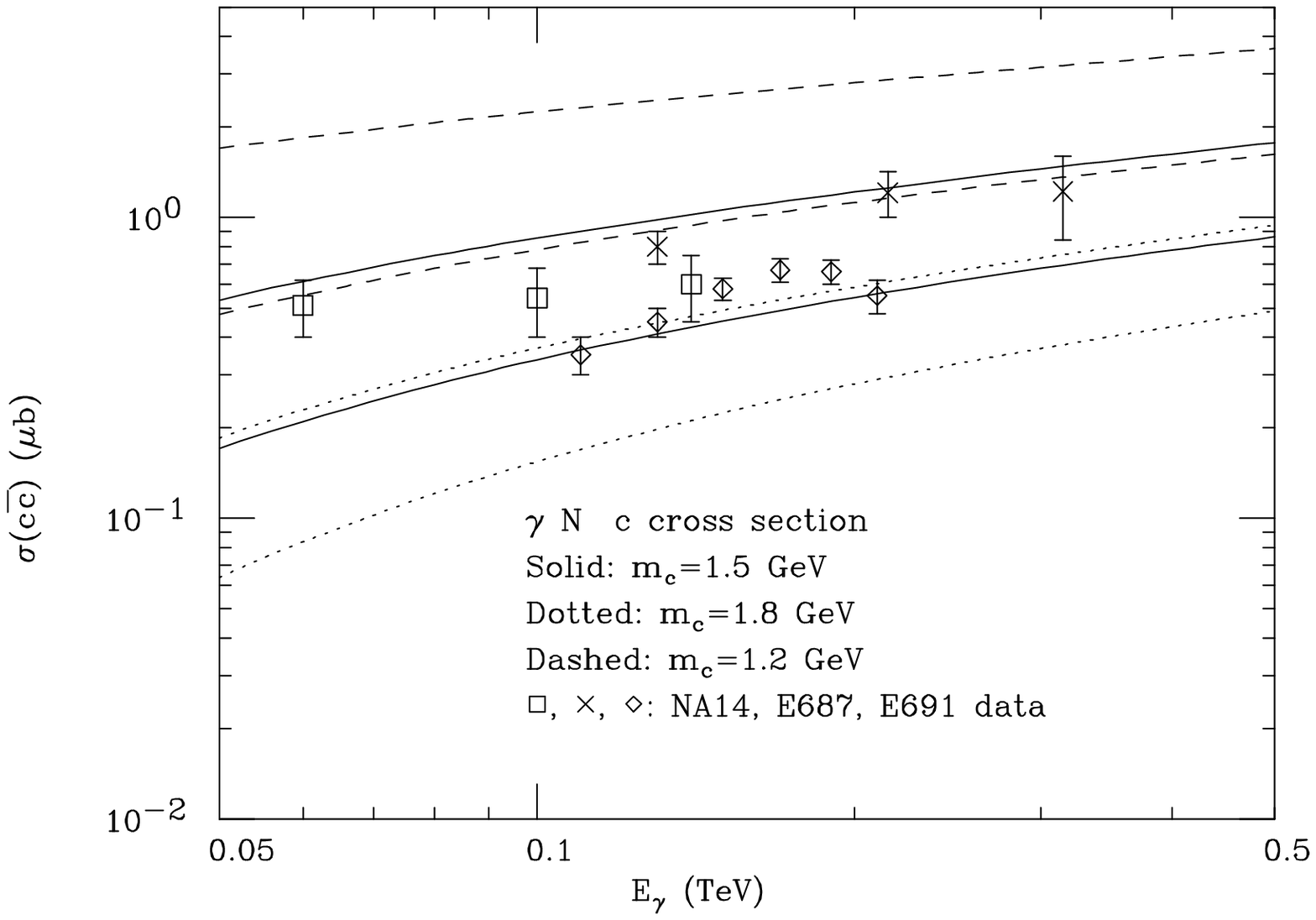,width=0.70\textwidth}}
\ccaption{}{\label{totgamma}
Pair cross-sections for $c$ production in $\gamma N$ collisions versus
experimental results.
}
\end{center}
\end{figure}
%%%%%%%%%%%%%%%%%%%%%%%%%%%%%%%%%%%%%%%%%%%%%%%%%%%%%%%%%%%%%%%%%%%%%
Total cross-sections for charm production have also been measured in
photoproduction experiments. In fig.~\ref{totgamma} the relevant experimental
results of refs.~\cite{Alvarez93,Anjos89,Anjos90,Bellini94} are shown in
comparison with NLO QCD predictions.
As can be seen, the theoretical uncertainties are smaller in this
case than in the hadroproduction case. Again, a charm mass of 1.5~GeV
is compatible with photoproduction data. It should be stressed however
that some of the experimental results are incompatible        
with one another. Until these discrepancies are resolved, it will not
be possible to use the data to constrain physical parameters.
For example, while the E687 data are inconsistent with a charm mass of
1.8~GeV, this mass value cannot be excluded because of the E691 data.
\subsubsection{Single-inclusive distributions}
\label{single}
While total cross-section measurements are certainly important,  it is
equally important to explore in more detail the production properties of heavy
quarks. In addition to the theoretical interest, this is relevant to
guarantee that some of the experimental systematics (such as the detector
acceptances and efficiencies) are under control. When experiments measure total
cross-sections, they often do so by probing a fraction of the full available
phase-space, and correcting later for the remaining part of phase-space.
This is done by using a theoretical modeling of the differential distributions. 
Checking that these agree with the data is therefore a fundamental test. 

Many experiments have measured       
single-inclusive $x_{\sss F}$ and $\pt$ distributions for charmed hadrons
in $\pi N$ collisions \cite{Aguilar85a,
Aoki88,Alves92,Adamovich92,Aoki92a} and in $pN$ collisions
\cite{Kodama91,Ammar88,Aguilar88,Adamovich92}.
Distributions are expected to be more
affected by non-perturbative phenomena than total cross-sections.
For example, an intrinsic transverse momentum of the incoming
partons, and the hadronization of the produced charm quarks, may
play an important r\^ole in this case. We will therefore try to
assess the impact of such phenomena by means of simple models.

%%%%%%%%%%%%%%%%%%%%%%%%%%%%%%%%%%%%%%%%%%%%%%%%%%%%%%%%%%%%%%%%%%%%
\begin{figure}
\centerline{\epsfig{figure=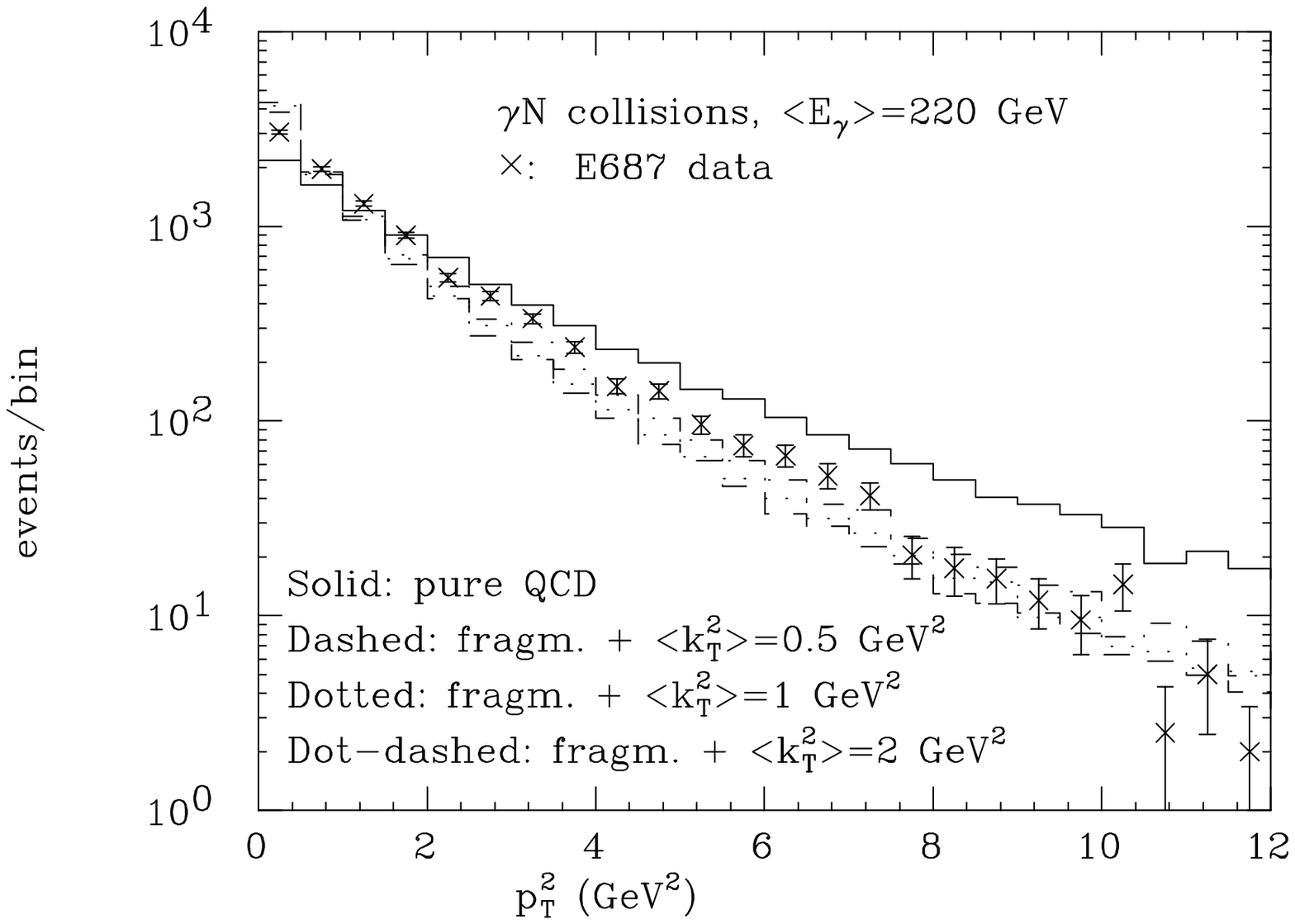,width=0.5\textwidth,clip=}
            \hspace{0.3cm}
            \epsfig{figure=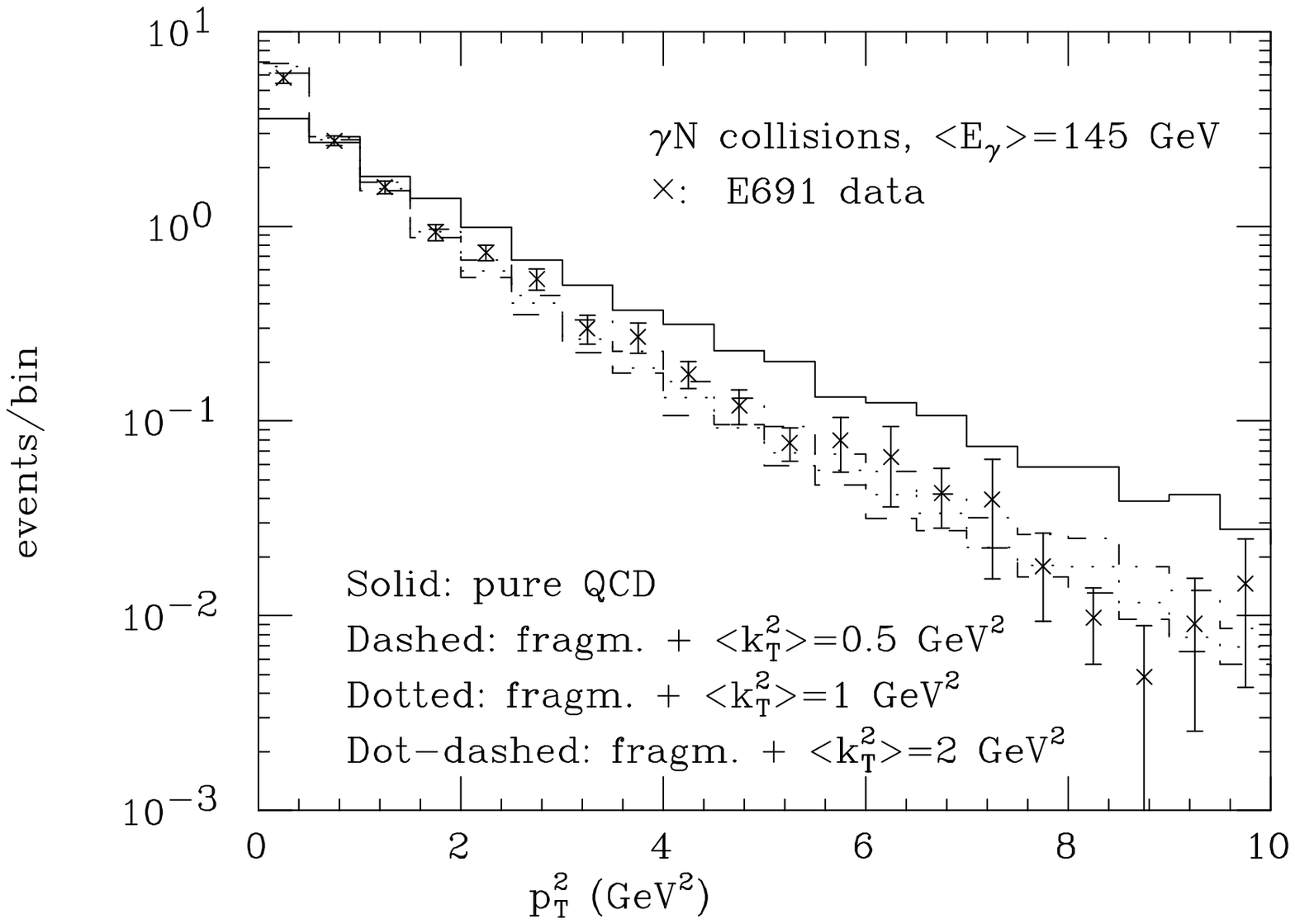,width=0.5\textwidth,clip=}}
\ccaption{}{ \label{phpt2}
Experimental $p_{\sss T}^2$ distribution compared to the NLO
QCD predictions ($m_c=1.5$~GeV), with and without the inclusion of
non-perturbative effects, in $\gamma N$ collisions.}
\end{figure}
%%%%%%%%%%%%%%%%%%%%%%%%%%%%%%%%%%%%%%%%%%%%%%%%%%%%%%%%%%%%%%%%%%%%
I start by presenting in fig.~\ref{phpt2} 
the $p_{\sss T}^2$ distributions measured in
photon--nucleon collision by the E687 \cite{Bellini94}
and by the E691~\cite{Anjos89,Anjos90} collaborations.
I also show the NLO QCD prediction for bare quarks (the solid histogram), 
which as you can see is significantly harder than the data\footnote{In this
subsection the relative normalization of experimental
distributions and theoretical curves has been fixed in order to give the same
total rate.}.
This is reasonable, since the non-perturbative fragmentation of the charm quark
into a charmed hadron will reduce the hadron momentum relative to the quark
one. As discussed in the previous lecture, we can describe this slow-down by
including a non-perturbative fragmentation function.
The dashed and dotted histograms in the figures include the convolution with a
Peterson fragmentation function (as well as with an intrinsic transverse
momentum, $k_{\sss T}$, as will be discussed in a moment). We chose here the
standard value of $\epsilon_c=0.06$, which was used in~\cite{Frixione97}. 
The agreement with
the data is good.

Let us consider now the case of hadroproduction, 
and focus on a
recent measurement of the single-inclusive differential cross-sections
performed at CERN by the WA92 collaboration~\cite{Adamovich96}, which
uses a $\pi^-$ beam of 350~GeV colliding with isosinglet nuclei. Consistent
results are obtained by using the 
E769 collaboration~\cite{Alves96a} data, using pion, proton and kaon
beams of 250~GeV on isosinglet targets, and more recent E706 
data~\cite{Apanasevich97} using 515 GeV pion beams.
                                                   
\begin{figure}
\centerline{\epsfig{figure=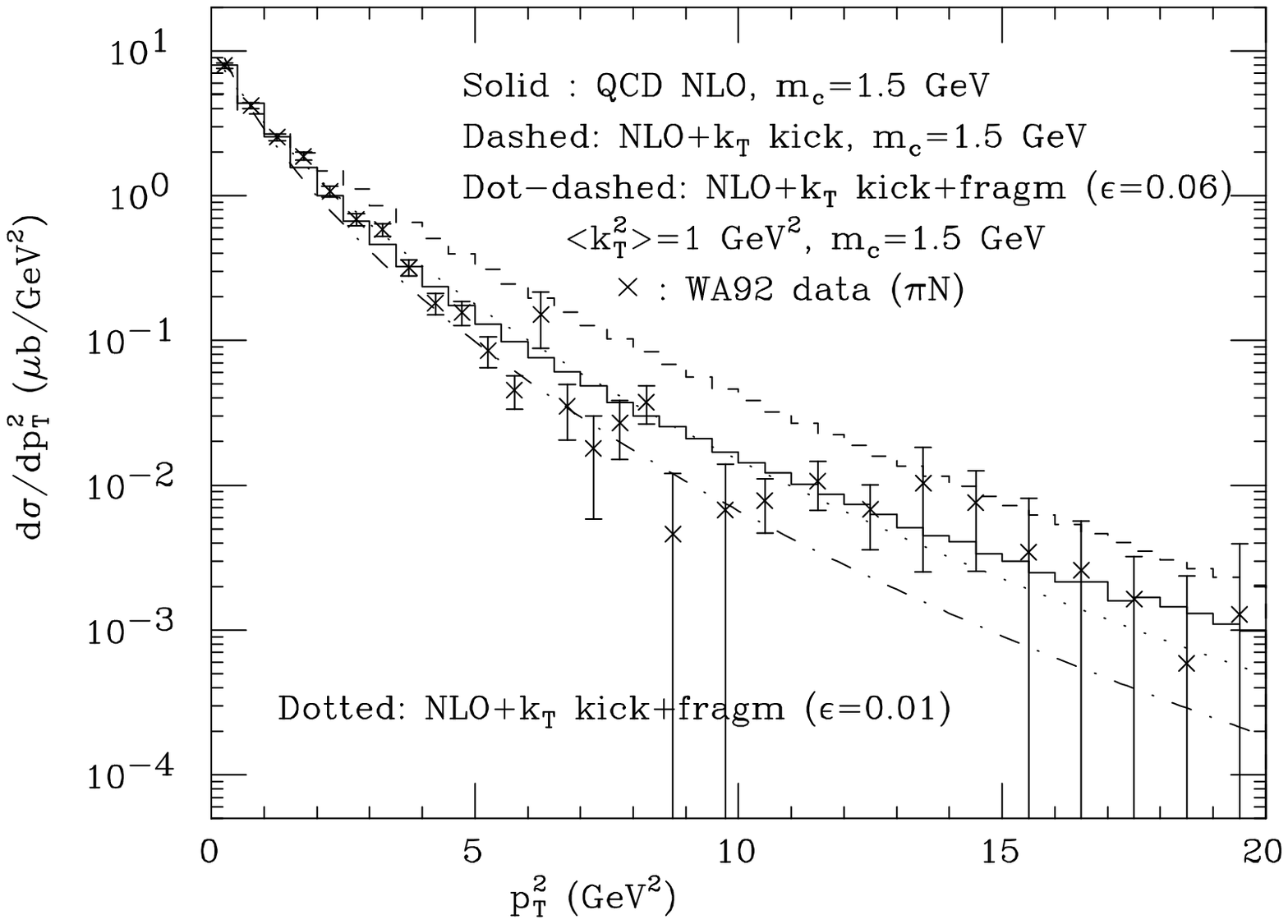,width=0.8\textwidth,clip=}}
\ccaption{}{ \label{hdrpt}                           
The single-inclusive $p_{\sss T}^2$ distribution measured by WA92,
compared to the NLO QCD predictions,                              
with and without the inclusion of non-perturbative effects.}
\end{figure}

In fig.~\ref{hdrpt} I show the comparison between the single-inclusive
$p_{\sss T}^2$ distributions measured by WA92 and the theoretical predictions.
%%%%%%%%%%%%%%%%%%%%%%%%%%%%%%%%%%%%%%%%%%%%%%%%%%%%%%%%%%%%%%%%%%%%
%%%%%%%%%%%%%%%%%%%%%%%%%%%%%%%%%%%%%%%%%%%%%%%%%%%%%%%%%%%%%%%%%%%%
The solid curve represents the pure NLO QCD prediction
for charm quarks, and as you see it almost exactly goes through the
experimental points. But this is not necessarily a good sign, since we just
learned in the study of photoproduction that we need to include an important
effect due to the non-perturbative fragmentation. The factorization theorem
discussed in the previous lecture tells us that these effects should be present
regardless of the nature of the beam, and once we find out that they must be
there in photoproduction, we should include them in hadroproduction as well. 
We would therefore expect that the final theoretical prediction, including the
fragmentation effects, will not agree with the data. Fortunately there is one
more effect of non-perturbative origin which comes to our rescue. That is the
famous {\em intrinsic} $k_{\sss T}$. Partons inside the proton cannot have
rigorously zero transverse momentum, because of the Fermi motion. It has been
known for a long time, mainly from the measurement of Drell-Yan events, that
quarks in the proton have an average intrinsic $k_{\sss T}$ of the order of
700~MeV. The dot-dashed line in fig.~\ref{hdrpt} is obtained by convoluting the
quark-level results with the same Peterson fragmentation function used in the
case of photoproduction, and with a gaussianly distributed intrinsic 
$k_{\sss T}$, with $\langle k_{\sss T}^2\rangle =1$~GeV$^2$.   This is a
reasonable choice for $\langle k_{\sss T}^2\rangle$, since gluons can be
expected to have a braoder $k_{\sss T}$ distribution than quarks.

As you can see                                                   
the effects of fragmentation and of the intrinsic $k_{\sss T}$ almost
completely cancel each other, and leave a $p^2_{\sss T}$ distribution which is
in fair agreement with the data. If we use the value of the Peterson
parameter $\epsilon_c$ extracted from the recent fits described in the first
lecture, $\epsilon_c=0.01$, we obtain an even better agreement (see the dotted
line in fig.~\ref{hdrpt}).                                               
Notice that in the photoproduction case
the $p_{\sss T}^2$ distribution is less sensitive to the choice of
$\langle k_{\sss T}^2\rangle$. This is because in photoproduction only one of
the two initial-state particles (i.e. the gluon) can have an intrinsic 
$k_{\sss T}$, while in the hadroproduction case 
both of them do. This reduced sensitivity is shown 
in fig.~\ref{phpt2}: the curves  with 
$\langle k_{\sss T}^2 \rangle =2$~GeV$^2$ 
and $\langle k_{\sss T}^2 \rangle =1$~GeV$^2$ are not significantly different
from one another.
                 
Having determined the values (or at least a range of values) for the
non-perturbative parameters which we expect should describe the production of
charm quarks at large $p_T$, we must now proceed and see whether the remaining
possible distributions that can be predicted do agree with the data or not. If
they did not, this would be an indication that there are extra phenomena at
play which we have not accounted for.

\subsubsection{Correlations in fixed-target production}
A distribution which is very sensitive to the presence on a non-perturbative 
$k_{\sss T}$-kick is the azimuthal correlation between the quark and the
antiquark. This is clear because already at LO, where the quarks are produced
back-to-back in the transverse plane and
the $\Delta\phi$ distribution is a sharp peak at $180^\circ$,
any transverse kick to the system will drastically change the topology of the
final state. 
Many experimental results on correlations between charmed particles
in hadro- and photoproduction have been obtained by different
experiments (see for example
refs.~\cite{Aguilar85b,Adamovich87,Aguilar88,Barlag91a,
Kodama91a,Aoki92a,Alvarez92,Frabetti93});  these reported
distributions of the azimuthal distance
between the charmed hadrons, the rapidity difference, the invariant
mass and the transverse momentum of the pair.
A detailed comparison of these results with QCD predictions is performed in
ref.~\cite{Frixione97}. More recently, new measurements of the azimuthal
distance and pair transverse momentum
for charmed mesons have been presented by the WA92
collaboration~\cite{Adamovich96a}.
In what follows we will focus on the distribution of $\Delta\phi$,
defined as the angle between the projections of the momenta of the pair
onto the transverse plane, and of the transverse
momentum of the pair $p_{\sss T}(Q\overline{Q})$. We will discuss whether NLO
QCD predictions can describe the available experimental data.

In leading-order QCD the heavy-quark pair is produced in the
back-to-back configuration, corresponding to $\Delta\phi=\pi$ and
$p_{\sss T}(Q\overline{Q})=0$. NLO corrections, as well as
non-perturbative effects, can cause a broadening of these
distributions, as illustrated in ref.~\cite{Mangano93} and \cite{Frixione97}.
                                                                           
\begin{figure}
\centerline{\epsfig{figure=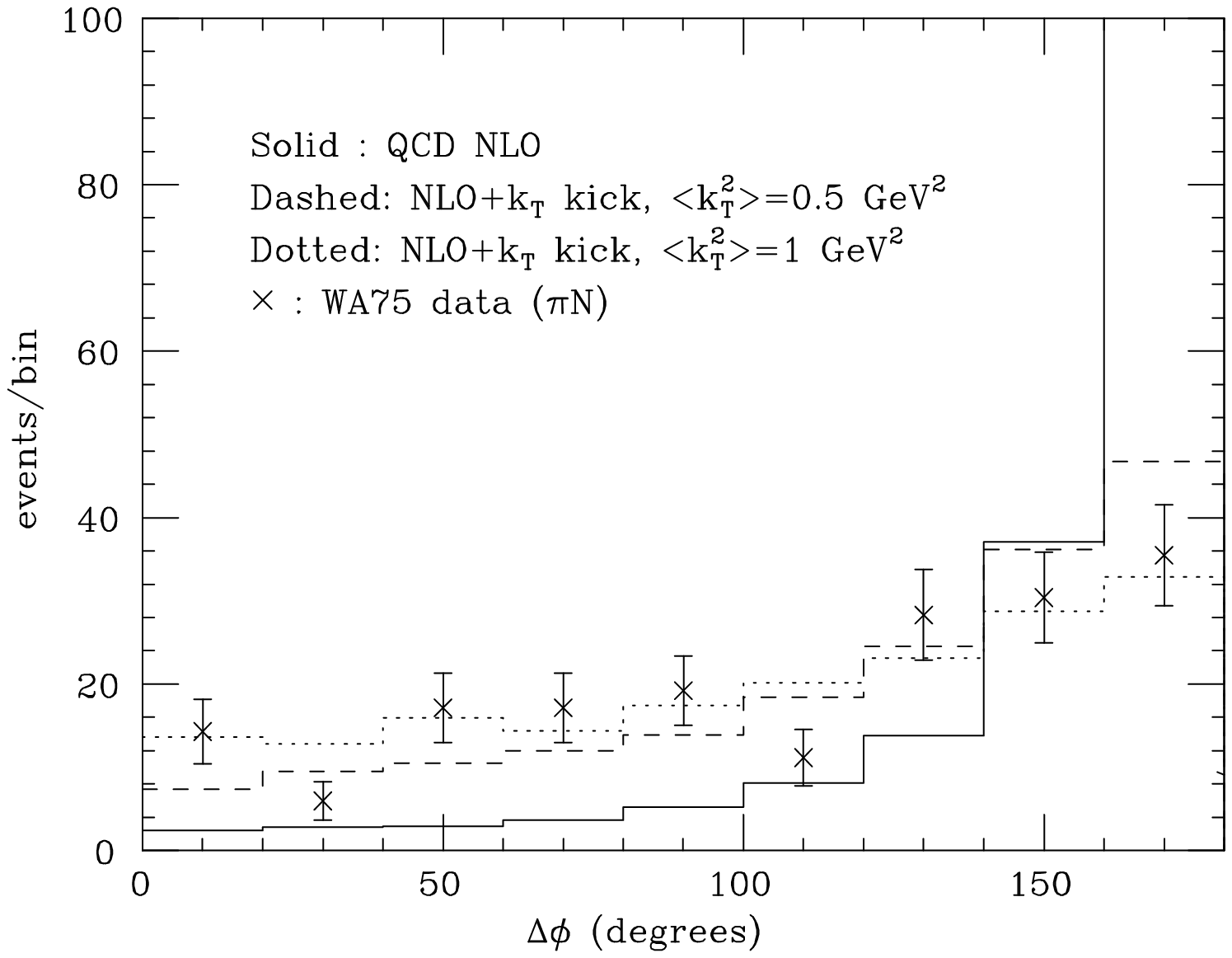,width=0.5\textwidth,clip=}
            \hspace{0.3cm}
            \epsfig{figure=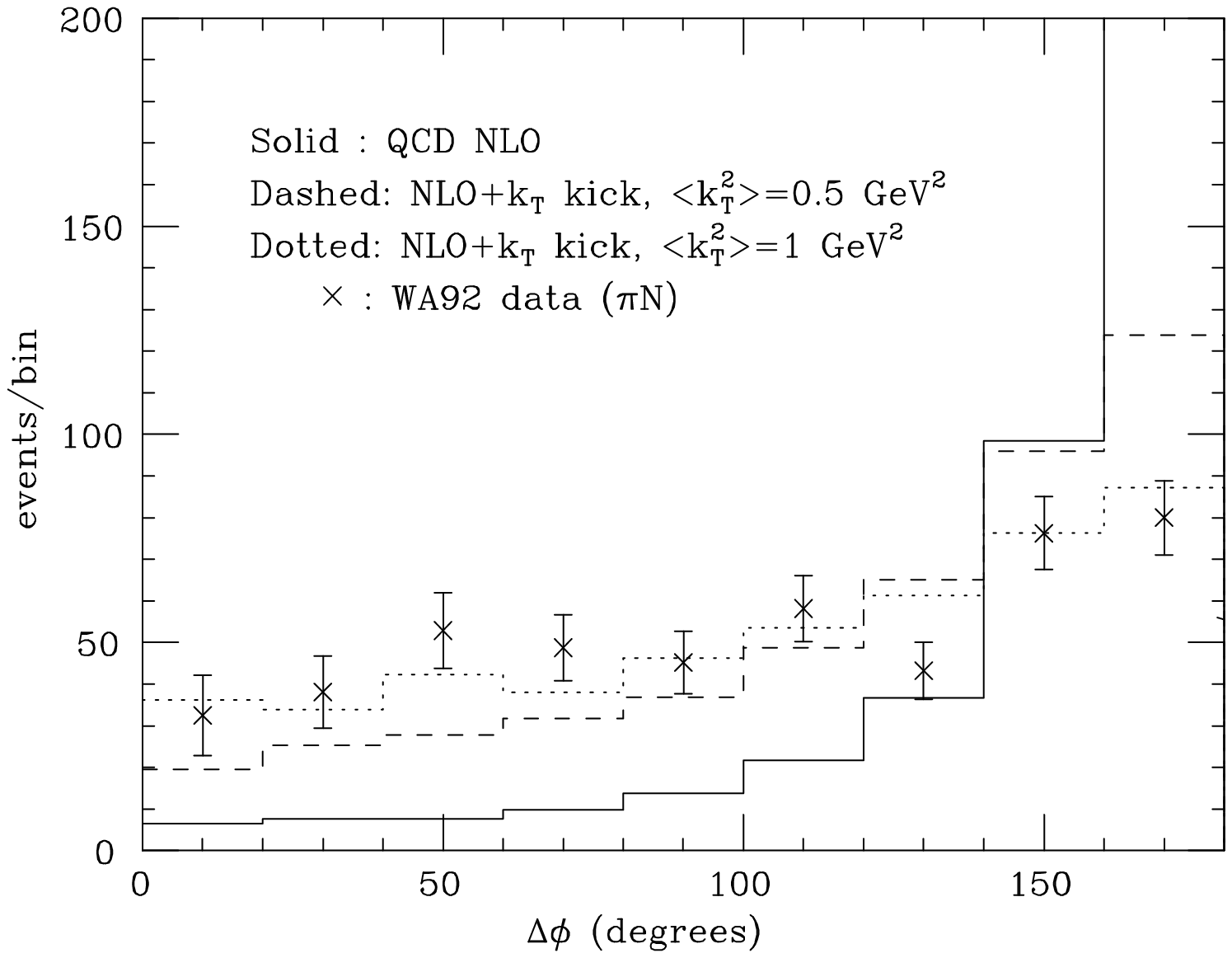,width=0.5\textwidth,clip=}}
\ccaption{}{\label{deltaphi}
Azimuthal correlation for charm production in $\pi N$ collisions:
NLO calculation versus the WA75 and WA92 data.
}
\end{figure}

We have chosen, as an illustration for hadroproduction, the cases of
the WA75 and the WA92 results, which have both been obtained in
$\pi^-N$ collisions at the same energy, $E_b=350$ GeV.
Let us first consider the $\Delta\phi$ distribution.
In fig.~\ref{deltaphi} we show (solid curves) the NLO
result superimposed on the data of the two experiments.
The charm mass was set to its default value, $m_c=1.5$~GeV.
In both cases, we see that the experimental data favour a much broader
distribution than the pure NLO QCD result for charm quarks.

One should, however, take into account also non-perturbative effects,
as in the case of single-inclusive distributions. 
The dashed and dotted curves in fig.~\ref{deltaphi} correspond to
the NLO prediction, supplemented
with the effect of an intrinsic transverse momentum
with $\langle k_{\sss T}^2\rangle=0.5$ GeV$^2$
and $\langle k_{\sss T}^2\rangle=1$ GeV$^2$, respectively.
We see that with $\langle k_{\sss T}^2 \rangle=0.5\;$GeV$^2$
it is impossible to describe the WA75 and WA92 data.
This conclusion differs from the one of ref.~\cite{Frixione94}
for the WA92 result. This is because in
ref.~\cite{Adamovich96a} the WA92 collaboration has improved
the study of correlations with respect to ref.~\cite{Adamovich95}
by considering a wider set of correlation variables and by improving
the statistics by a factor of 5.
WA92 and WA75 data now appear to be consistent with each other.
As is apparent from fig.~\ref{deltaphi}, the acceptable
value of $\langle k_{\sss T}^2\rangle=1$ GeV$^2$ is required to
describe the data. This is encouragingly consistent with the results we
obtained form the study of the inclusive-$p_T$ distributions!

%%%%%%%%%%%%%%%%%%%%%%%%%%%%%%%%%%%%%%%%%%%%%%%%%%%%%%%%%%%%%%%%%%%%%
%%%%%%%%%%%%%%%%%%%%%%%%%%%%%%%%%%%%%%%%%%%%%%%%%%%%%%%%%%%%%%%%%%%%%

\begin{figure}[htb]
\centerline{\epsfig{figure=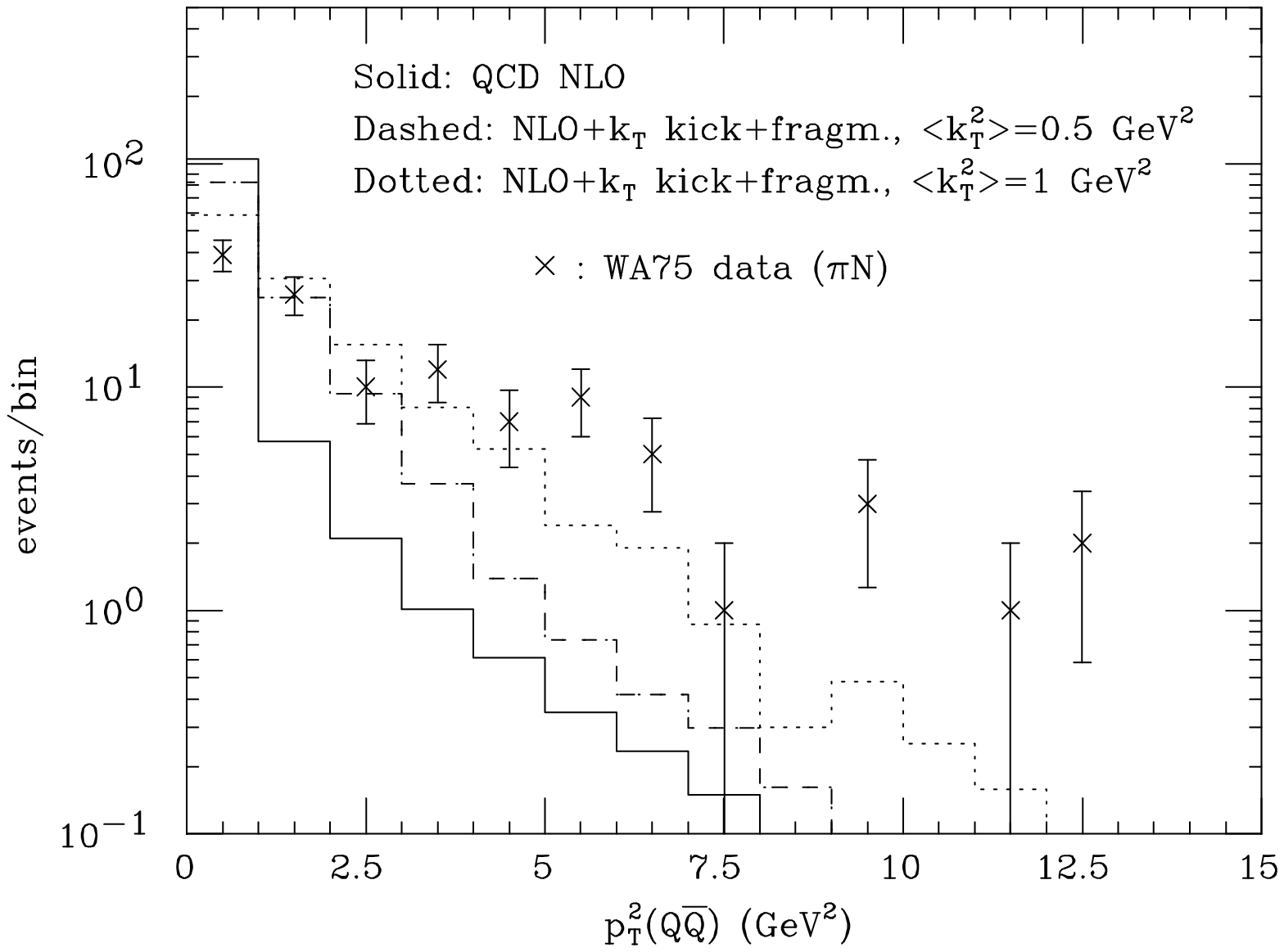,width=0.5\textwidth,clip=}
            \hspace{0.3cm}
            \epsfig{figure=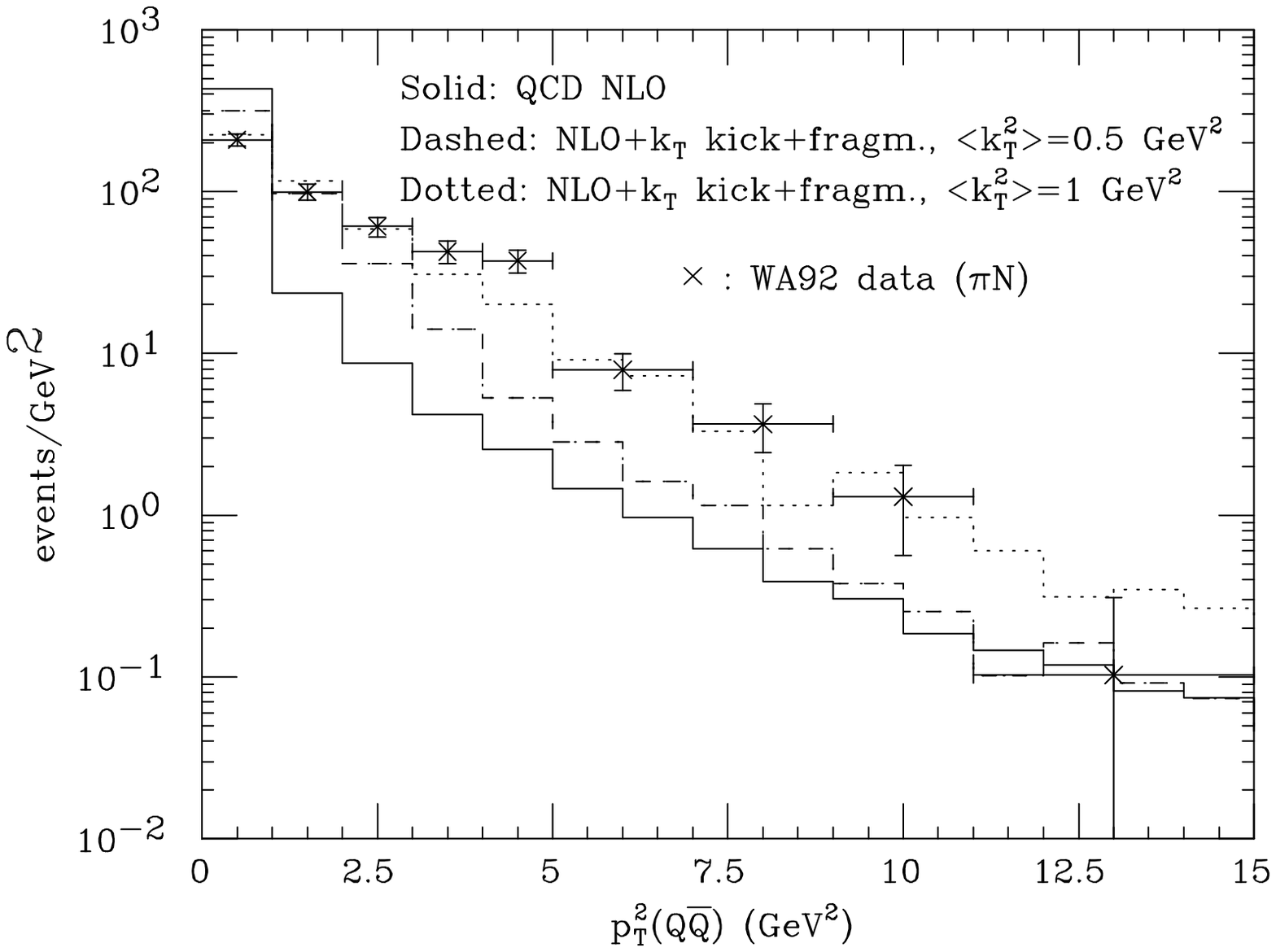,width=0.5\textwidth,clip=}}
\ccaption{}{\label{f:pt2qq}
NLO QCD result for the $p_{\sss T}^2(Q\overline{Q})$ supplemented
with an intrinsic transverse momentum for the incoming partons,
compared with the WA75 (left) and WA92 (right) data.
}
\end{figure}

\begin{figure}[htb]
\begin{center}
\mbox{\psfig{file=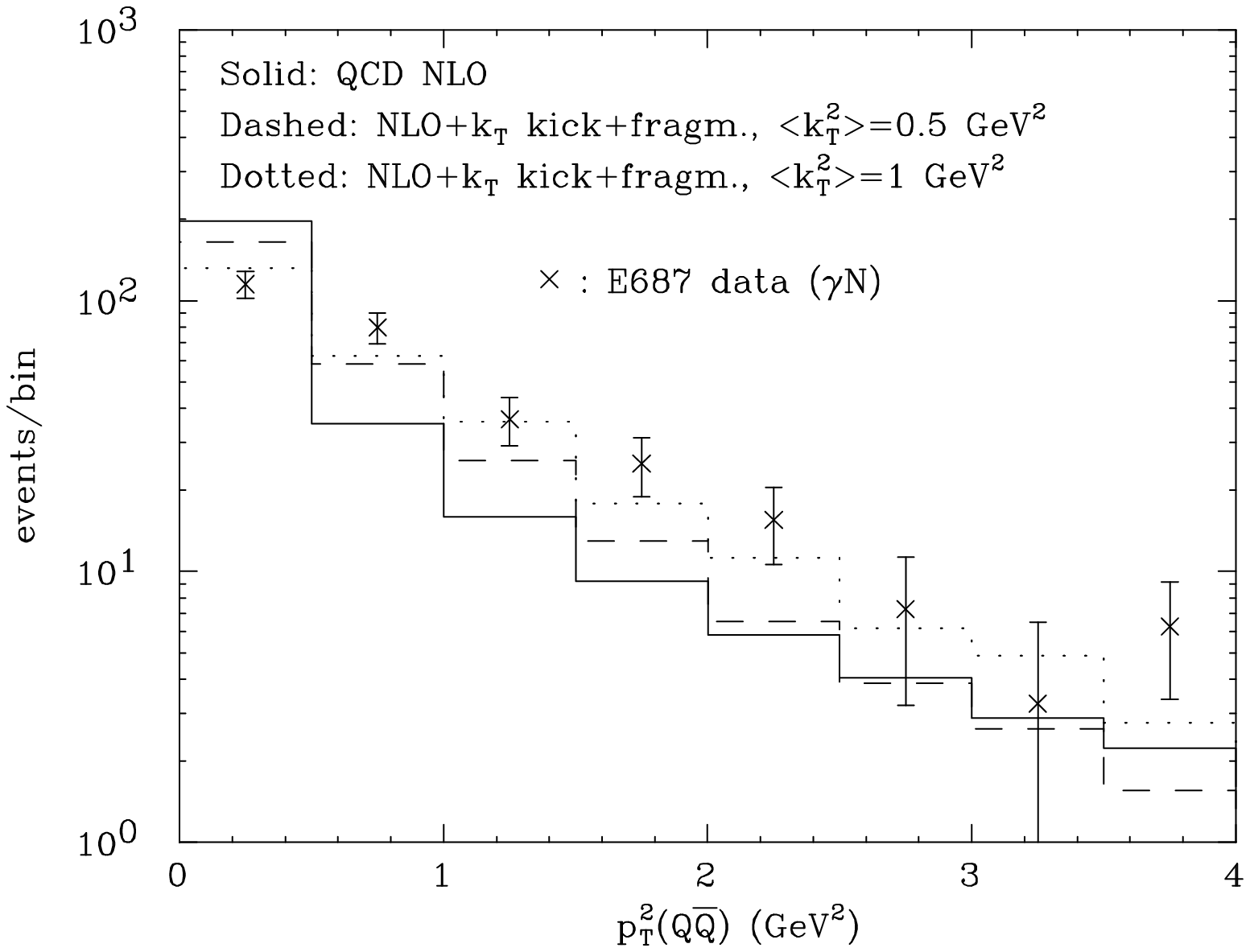,width=0.65\textwidth}}
\ccaption{}{\label{e687ptqq}
Transverse momentum distribution of the $D\bar{D}$ pair versus the
perturbative result for the E687 experiment.
}
\end{center}
\end{figure}

The WA75 collaboration, and recently the WA92 collaboration, published
in refs.~\cite{Aoki92a,Adamovich96a} the distribution of the transverse
momentum of the heavy-quark pair. This is yet another independent variable,
which is sensitive to both intrinsic $k_{\sss T}$ and fragmentation effects.
The theoretical prediction supplemented with a                              
$k_{\sss T}$-kick with $\langle k_{\sss T}^2\rangle = 1\;$GeV$^2$
cannot reproduce the WA75 data. This is a typical example of why, until
recently, no coherent overall picture of the mechanisms at play
in charm production was available. The most recent results from WA92, however, 
confirm the indications of the QCD calculations supplemented with 
$k_{\sss T}$-kick and fragmentation, 
as displayed in fig.~\ref{f:pt2qq}.
%%%%%%%%%%%%%%%%%%%%%%%%%%%%%%%%%%%%%%%%%%%%%%%%%%%%%%%%%%%%%%%%%%%%%
%%%%%%%%%%%%%%%%%%%%%%%%%%%%%%%%%%%%%%%%%%%%%%%%%%%%%%%%%%%%%%%%%%%%%
%%%%%%%%%%%%%%%%%%%%%%%%%%%%%%%%%%%%%%%%%%%%%%%%%%%%%%%%%%%%%%%%%%%%%
\begin{figure}[htb]
\centerline{\epsfig{figure=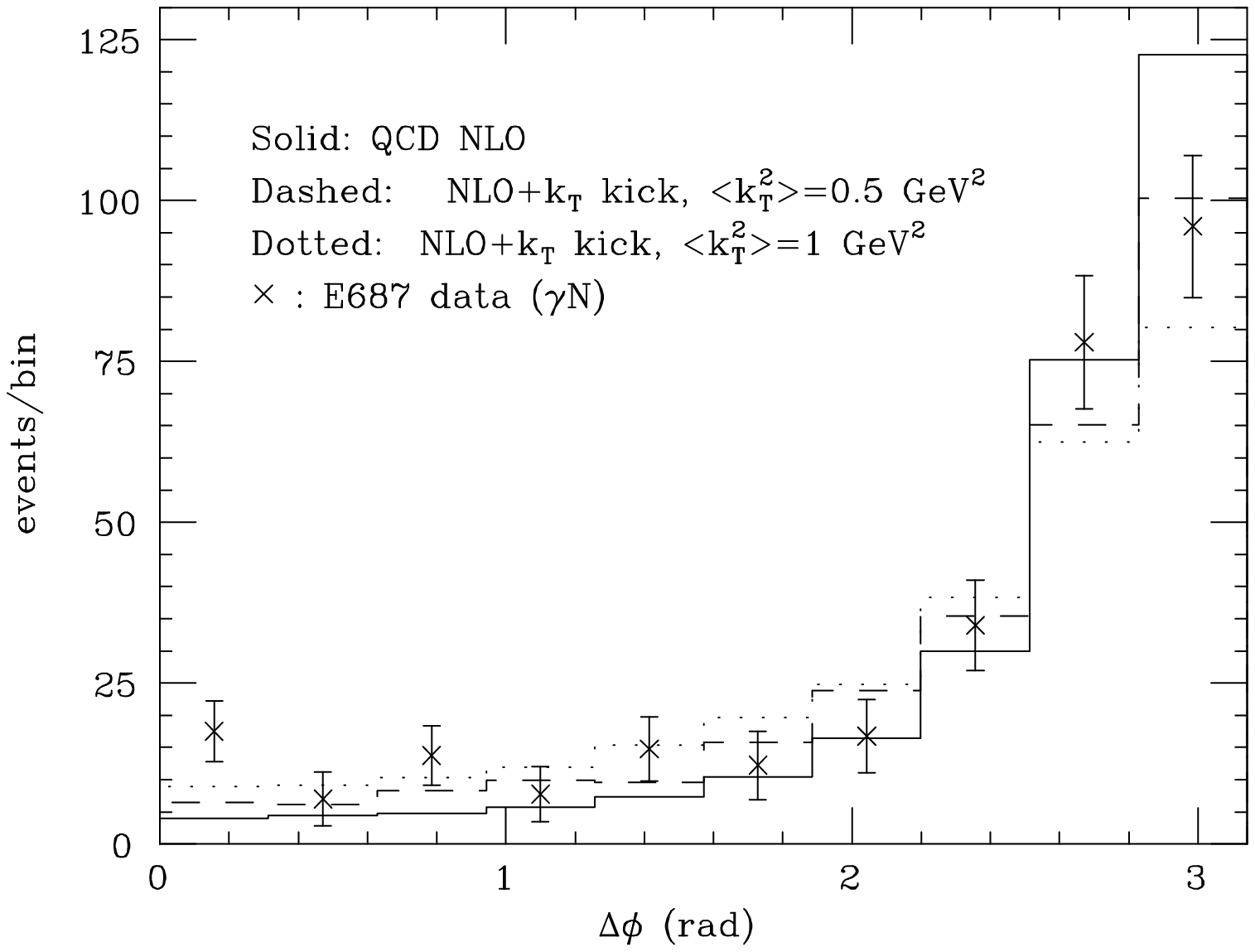,width=0.5\textwidth,clip=}
            \hspace{0.3cm}
            \epsfig{figure=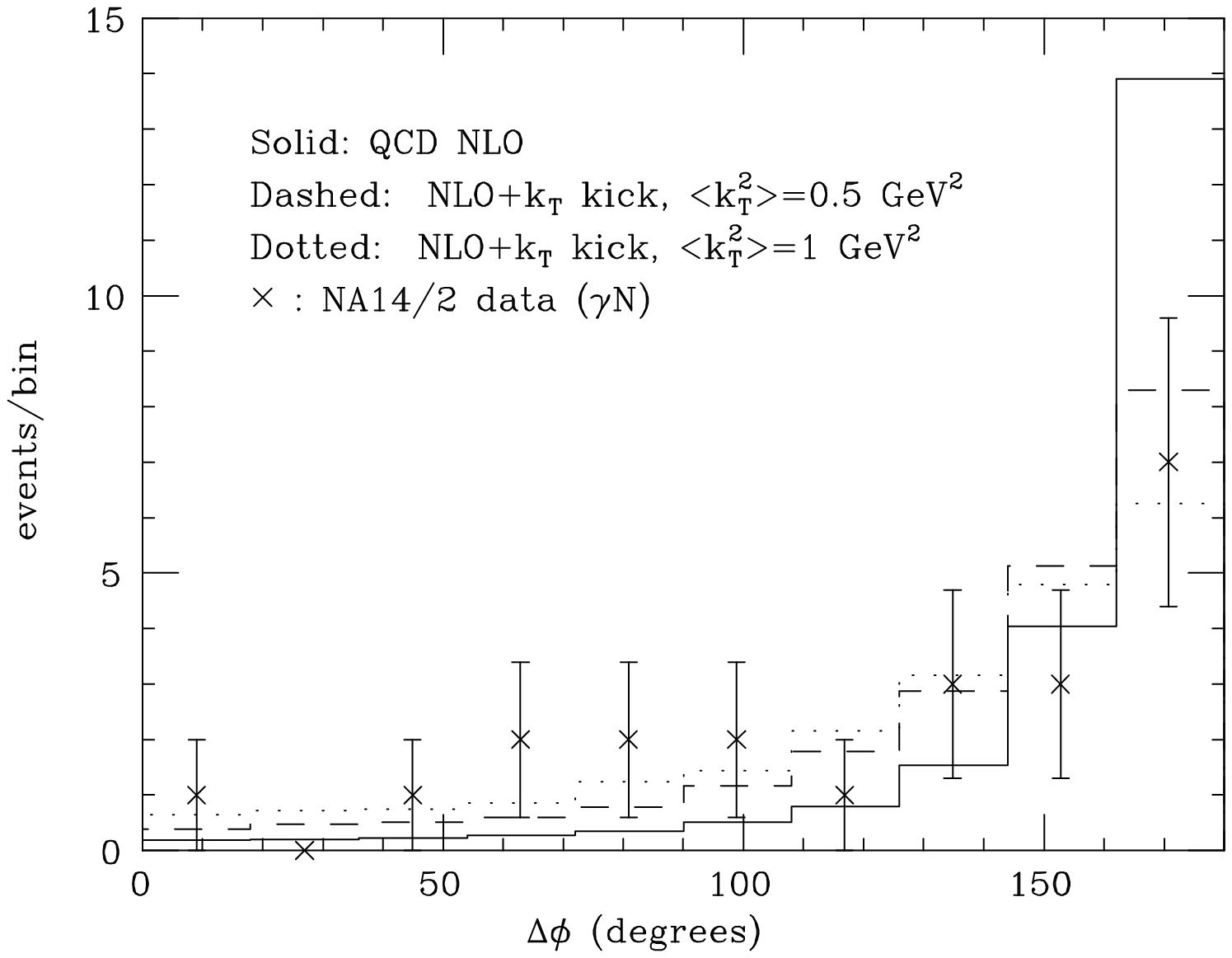,width=0.5\textwidth,clip=}}
\ccaption{}{\label{f:dphi_photo}
Azimuthal correlation of $D\bar{D}$ pair versus the perturbative
result in photoproduction for the E687 (left) and NA14/2 (right)
experiments.}
\end{figure}
%%%%%%%%%%%%%%%%%%%%%%%%%%%%%%%%%%%%%%%%%%%%%%%%%%%%%%%%%%%%%%%%%%%%%
Similar conclusions can be drawn from the study of photoproduction correlation
data (the $\Delta\phi$
distributions are shown in fig.~\ref{f:dphi_photo},
the distribution in the transverse momentum of the heavy-quark
pair in fig.~\ref{e687ptqq}).
In all cases $\langle k_{\sss T}^2 \rangle$ values of the order
of 1~GeV$^2$ are necessary (and sufficient!) to describe the data.
\\[0.3cm]
%%%%%%%%%%%%%%%%%%%%%%%%%%%%%%%%%%%%%%%%%%%%%%%%%%%%%%%%%%%%%%%%%%%%%
%%%%%%%%%%%%%%%%%%%%%%%%%%%%%%%%%%%%%%%%%%%%%%%%%%%%%%%%%%%%%%%%%%%%%
{\bf Conclusions:}
\begin{itemize}
\item For the first time since the beginning of charm production studies we
      have reached a compelling and coherent picture  of the production
dynamics in fixed-target experiments. Large data-sets and state-of-the-art
detector technology have been fundamental to make outstanding progress in the
quality of the data!
\item The production dynamics are well described by NLO QCD, supplemented with
      simple phenomenological descriptions of the NP effects:
      \begin{itemize}                                
        \item fragmentation at large $p_T$
        \item intrinsic $\langle k_T^2 \rangle = 1$~GeV$^2$
      \end{itemize}                            
\item The parametrization of these effects satisfies the expected universality
      properties: no dependence on
      \begin{itemize}             
        \item beam type and energy
        \item observable
      \end{itemize}
\item Additional fine tuning will be possible when more data will be available.
We can then 
expect to sort out the interplay between QCD and NP parameters in       
     a more accurate fashion. E.g.:
      \begin{itemize}                                        
        \item larger $\alpha_s$ $\to$ smaller $\langle k_T^2 \rangle$  (from
              $\Delta\phi$ spectrum)
        \item smaller $\langle k_T^2 \rangle$ $\to$ softer fragmentation (from
              $p_T$ spectra)
      \end{itemize}
\end{itemize}                                         

\subsection{Heavy-quark production at hadron colliders}\label{sec:colliders}
Hadron collisions are by far the largest source of heavy quarks available
today. While the environment of high-energy hadronic interactions does not
allow to trigger on the largest fraction of charm and bottom produced, the
production rates are so large that the number of recorded events
allows today $b$-physics studies that are competitive with those of $e^+e^-$
experiments. The introduction of new experimental
techniques, such as the use of silicon vertex detectors, which enable the
tagging of events containing bottom quarks~\cite{Bedeschi}, led in the recent
years to high-statistics and low-background measurements of the $b$-production
properties over a large range of transverse momenta. 

From the point of view of QCD studies, heavy-flavour production
in high-energy hadronic collisions has better potentials than in fixed-target
experiments. The $b$ quarks produced at large $p_{\sss T}$ can be studied in
perturbative
QCD with smaller contamination from non-perturbative effects.
For example, the impact of the initial-parton transverse momentum
is largely reduced with respect to fixed-target charm production.
Furthermore, the fragmentation properties of heavy flavours in high
transverse momentum jets can be directly studied, since the transverse
momenta involved are typically perturbative.

In spite of all of this, the comparison of theory with data collected by the
high-energy collider experiments (UA1, CDF and D0) shows some puzzling
discrepancies, which will be reviewed in the rest of this lecture.
In particular, we will concentrate on the inclusive $p_T$ distributions in the
central and forward rapidity regions, and on the azimuthal correlations between
$b$ and $\bar b$.                                                              

\subsubsection{Comparison of data and theory}
The distribution most commonly studied by the hadron-collider
experiments is
the $b$-quark differential \pt\ spectrum, integrated within a fixed
rapidity range and above a given \pt\ threshold (\ptmin~):
\be \label{eq:ptmin}
        \sigma(\ptmin) = \int_{\vert y \vert < y_{max}} d \, y
        \; \int_{\pt>\ptmin} d\,\pt \frac{d^2\sigma}{dy\, d \pt} \;.
\ee
The UA1 experiment at the CERN $p\bar{p}$
collider used $y_{max}=1.5$, while
the CDF and D0 experiments at the Tevatron use $y_{max}=1$.

\begin{figure}[htb]
\centerline{\epsfig{figure=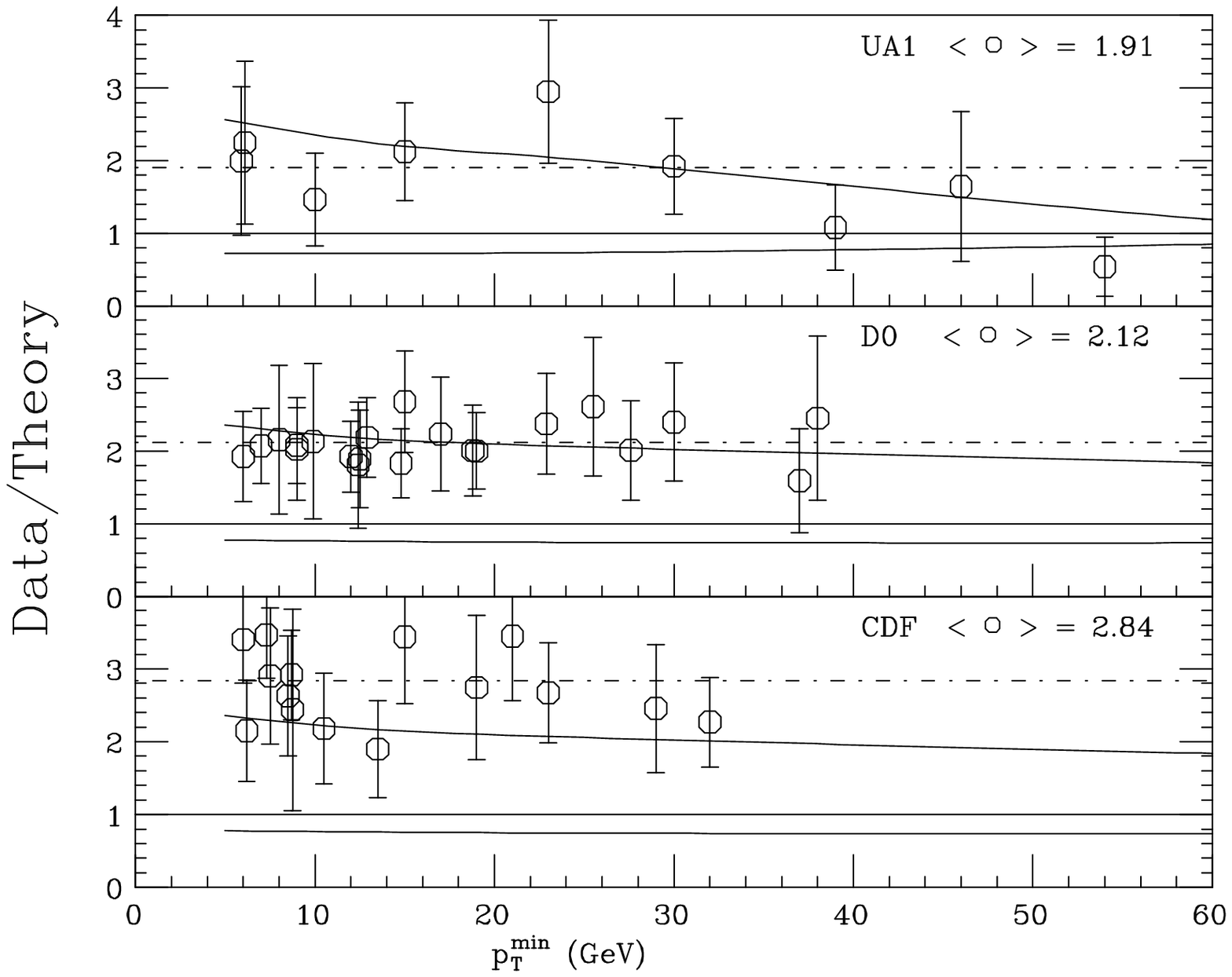,width=0.7\textwidth,clip=}}
\ccaption{}{ \label{fig:mrscomp}
Linear comparison between experimental data and theory for the
integrated $b$-quark \pt\ distribution. See the text for details.}
\end{figure}

The results obtained by CDF~\cite{Abe92} and D0~\cite{Abachi95a}
are shown in figs.~\ref{fig:mrscomp}, compared with the theoretical
predictions.             
For an easier evaluation of the results, we present them on a linear scale
in the form Data/Theory, and include the UA1 data~\cite{Albajar91} as well.
We divided the data by our central theoretical prediction. The dot-dashed  
straight lines are constant fits to the ratios, weighed by the inverse of
the experimental uncertainties.
The upper and lower solid lines correspond to the upper and lower
theoretical predictions divided by
the central prediction. The lower prediction is obtained
fixing the renormalization and factorizations scales to be equal to
$2\muo$ (with $\muo^2 \equiv \pt^2+m_b^2$), and using the 
MRSA$^\prime$ parton density set~\cite{Martin95}. 
The value of $\lambdamsb$ for this set ($152$~MeV)
yields a value of $\as$ significantly lower            
than that extracted from different observables \cite{Barnett96}. To estimate
the upper theoretical curve we use
the set MRS125~\cite{Martin95a}, for which 
$\as(M_{\rm Z})=0.125$, and a renormalizatoin scale equal to $\muo/2$.
Given the large values        
of \pt\ probed by the collider experiments, the $b$-mass dependence of the
theoretical result is small. We chose as a default the value
$\mb = 4.75$~GeV, for all our curves.
                                                                    
The first important thing to notice in fig.~\ref{fig:mrscomp} is that,
independently of the input parameters chosen, the shape of the theoretical
curves agrees  well with the data. Secondly, the results at
$\sqrt{S}=630$~GeV are by and large 
consistent with those at $\sqrt{S}=1800$~GeV. The
average ratio Data/Theory measured by UA1 differs by less than 10\% from the
ratio measured by D0, independently of the input parameters chosen.
The difference is however between 40 and 50\% if one uses the CDF data. If we
naively average the Tevatron data, we conclude that the relative
discrepancy between theory and data when changing the value of $\sqrt{S}$ from
630 to 1800~GeV is about 30\%, a number of the same order of
the estimate of $\log(1/x)$ effects discussed in the first lecture.

Independently of the beam energy, the data are higher by a factor
of about 2 than the default prediction based on $\mu=\muo$.
They are, however, well
reproduced by the upper theoretical curve. Therefore, while the
overall uncertainty of the theoretical prediction due to the scale choice
is large, there is currently no serious inconsistency between theory and data
for the inclusive \pt\ distribution of $b$ quarks at the Tevatron. The 30\%
discrepancy between the results of CDF and D0 is comparable to the discrepancy
between the extrapolation of the UA1 data to CDF, while UA1 and D0 data agree
at the level of 10\%.
                     
\begin{figure}[htb]                            
\centerline{\epsfig{figure=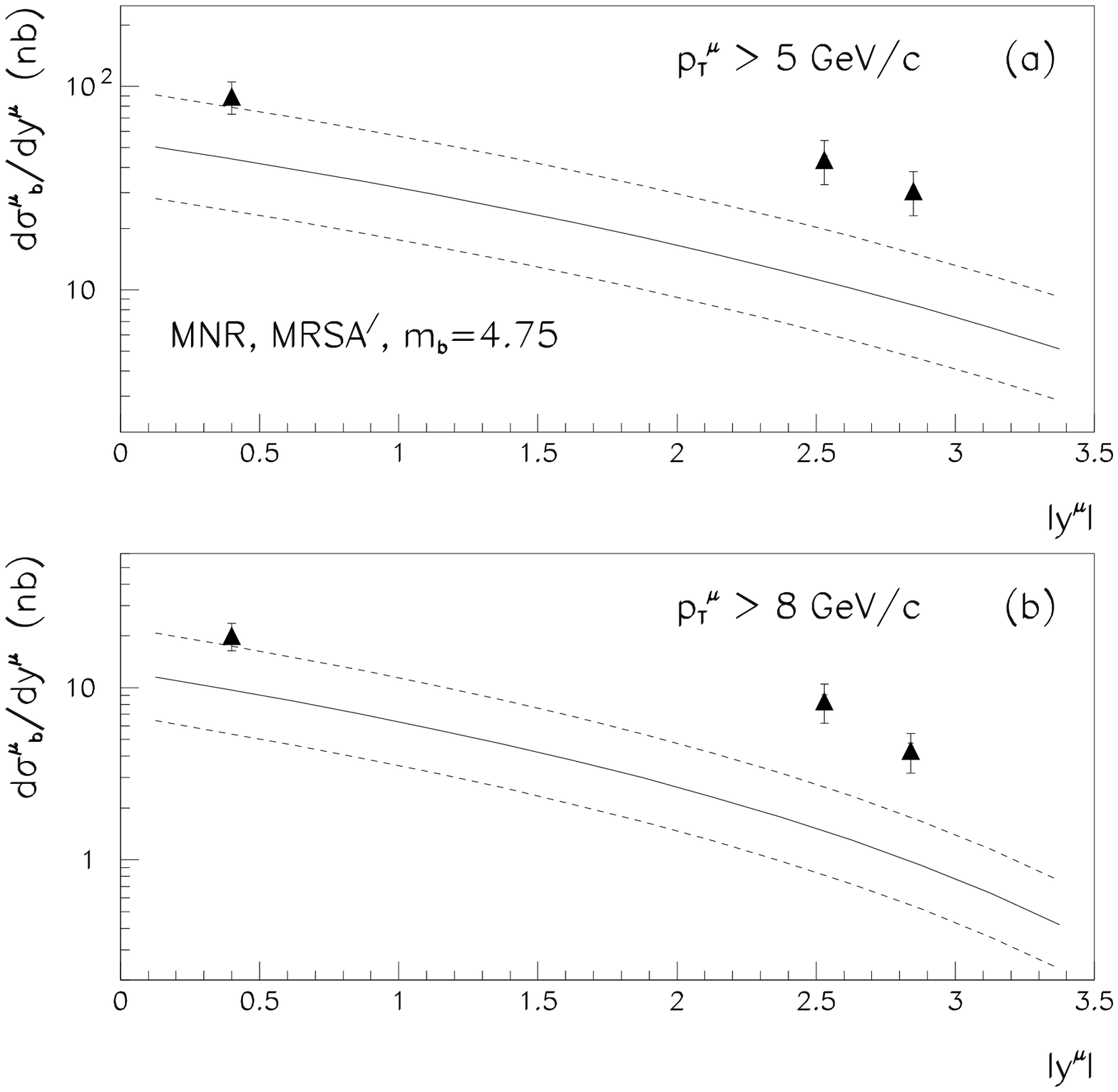,width=0.5\textwidth,clip=}}
\ccaption{}{ \label{fig:d0fmdy}                
Comparison between D0 data and theory for the $p\bar{p}\to (b\to\mu)+X$
cross-section at large rapidity.}
\end{figure}

Nevertheless, the discrepancy between CDF and D0 remains quite puzzling, as it
is not consistent with the quoted systematic and statistical errors.
In addition to this puzzle, a recent measurement of forward 
$b$ production by D0~\cite{Abachi96b}
(performed using muons in the rapidity range
$2.4<\vert \eta^{\mu} \vert<3.2$)
indicates a rate in excess by a factor of approximately 4 over the
central theoretical prediction.
The results are shown in fig.~\ref{fig:d0fmdy}.
Certainly no phenomenon of perturbative origin (such as higher-order
corrections) could explain such a discrepancy.

\begin{figure}[htb]
\centerline{\epsfig{figure=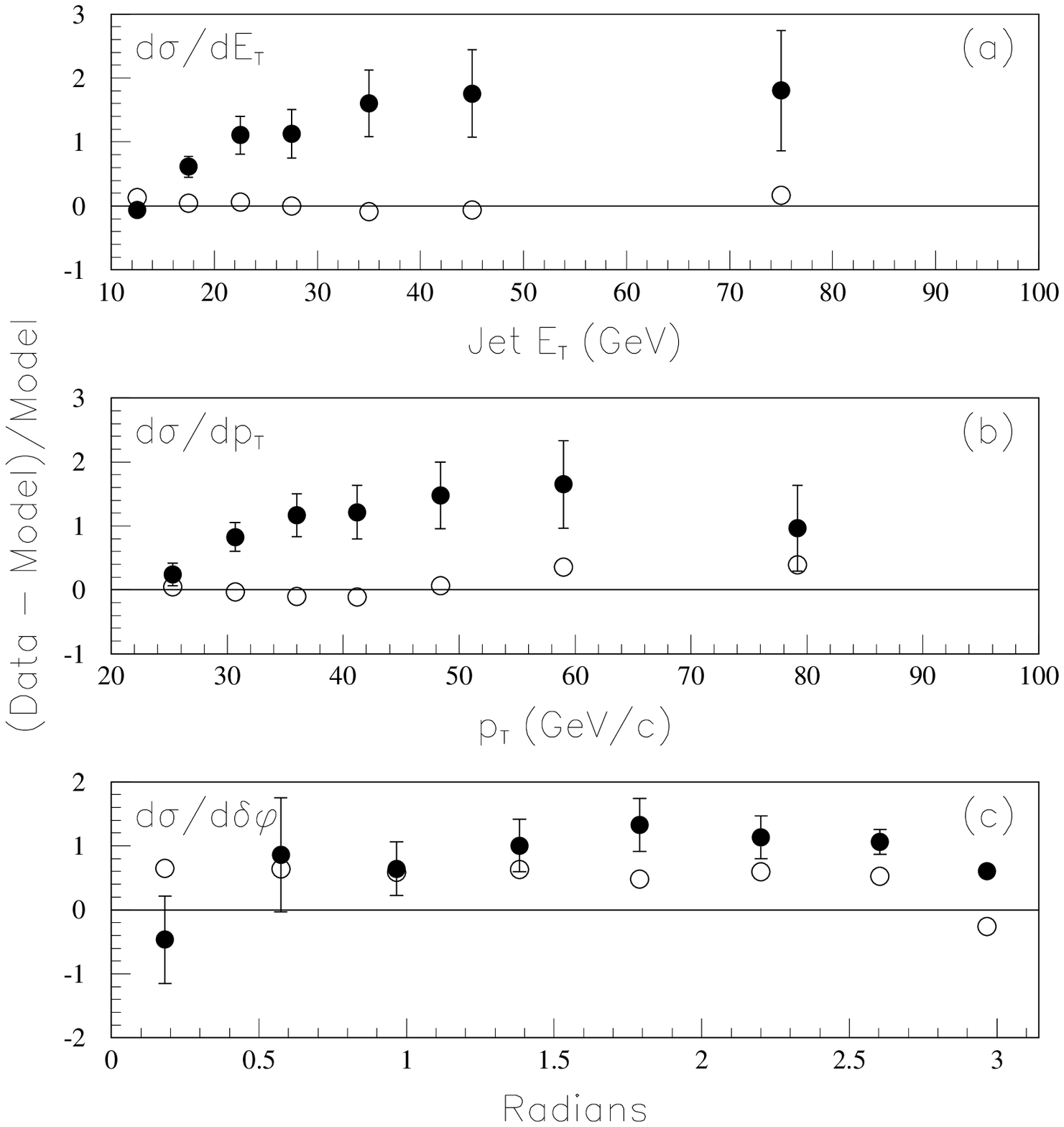,width=0.6\textwidth,clip=}}
\ccaption{}{ \label{fig:cdfmub}
CDF results on the $b\bar b$ correlations using $\mu$+$b$-tagged jet final
states~\cite{Abe96b}, compared to NLO QCD. Solid points correspond to the
default theoretical prediction, with scales $\mu=\muo$, empty circles
correspond to the difference between the choice $\mu=\muo/2$ and $\mu=\muo$.
Top figure: tagged-jet \et\ distribution.
Central: muon \pt\ spectrum. Bottom: azimuthal correlations.}
\end{figure}

\begin{figure}[htb]                      
\centerline{
   \epsfig{figure=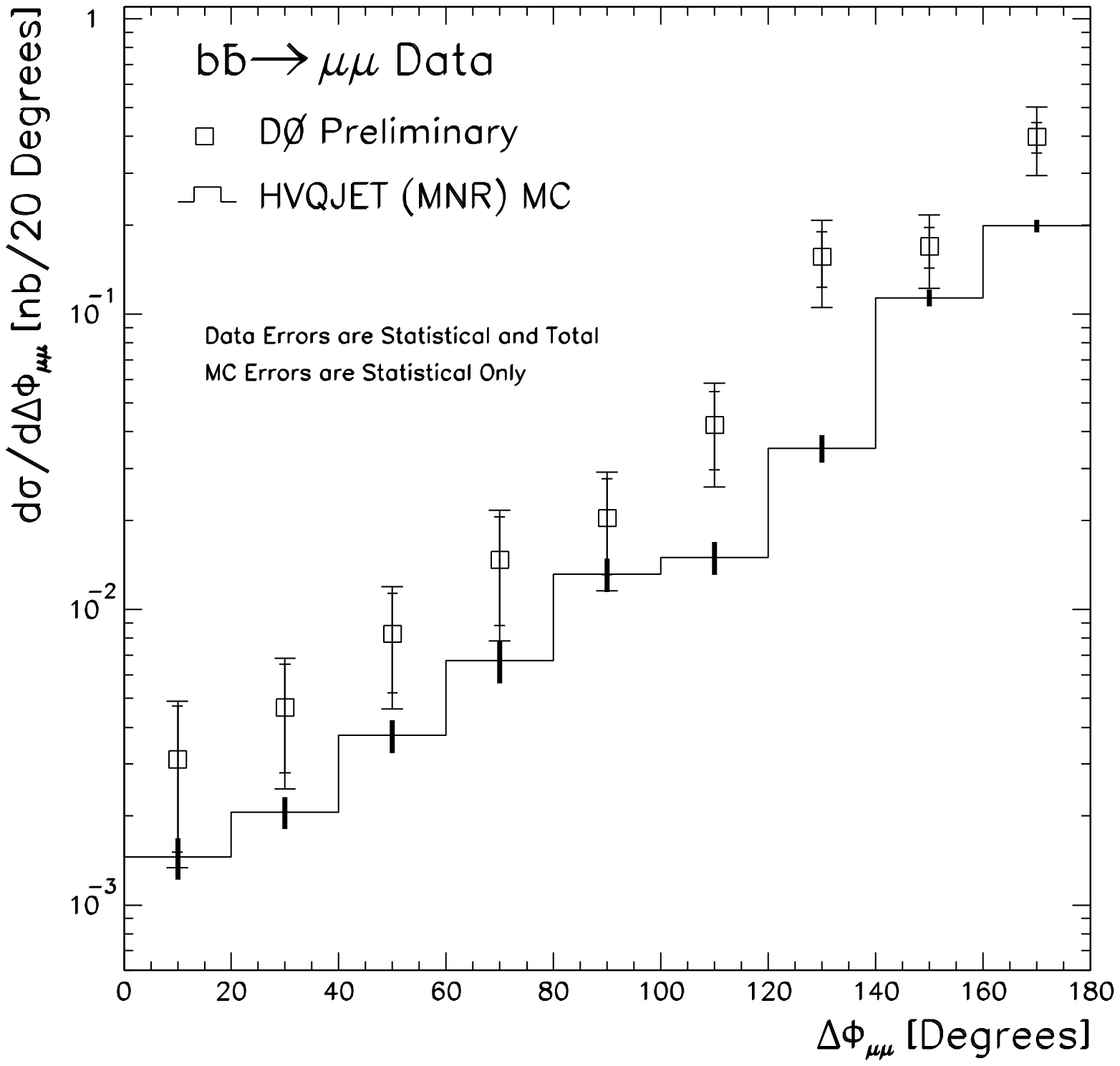,width=0.5\textwidth,clip=}
   \epsfig{figure=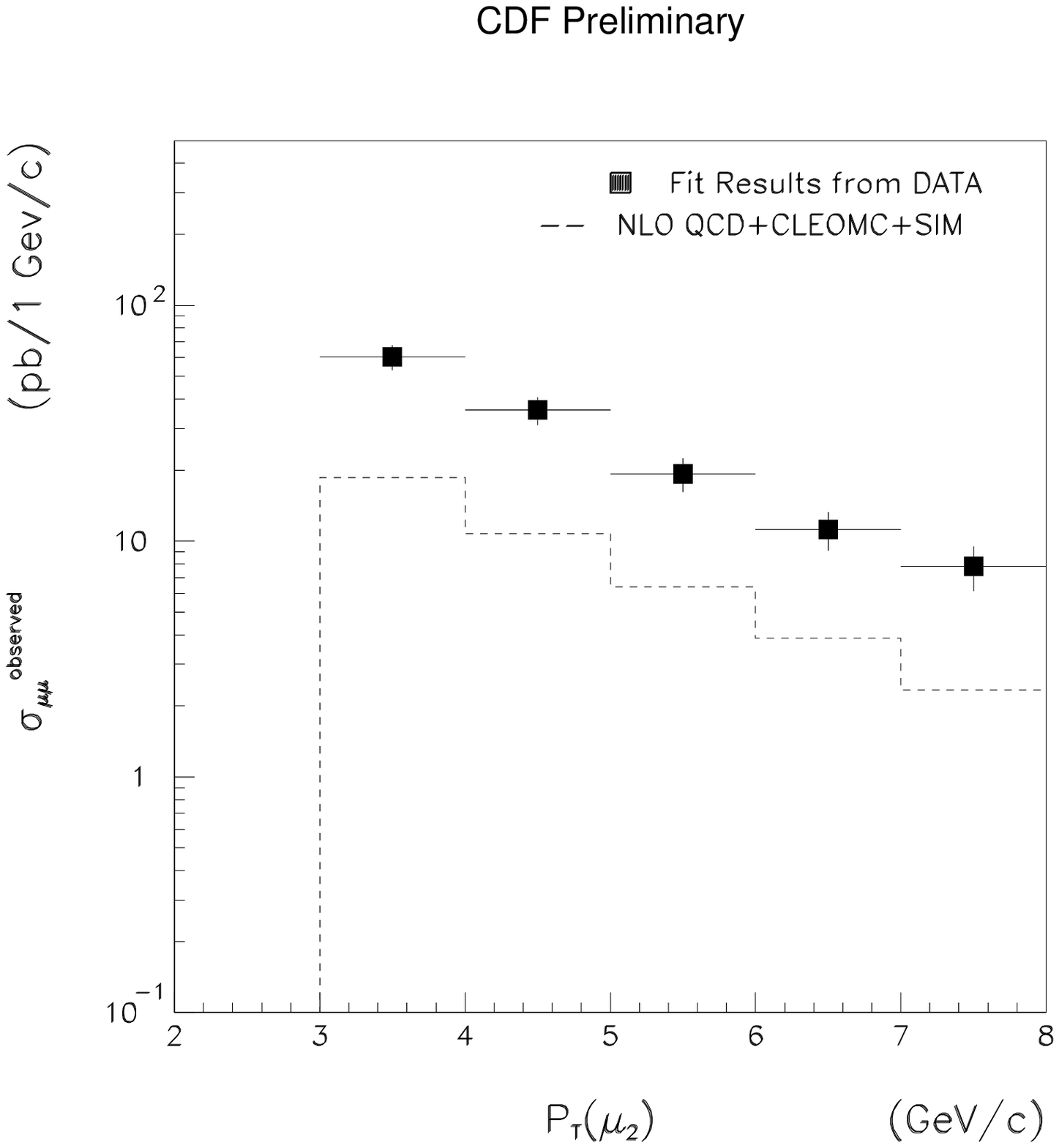,width=0.5\textwidth,clip=} }
\ccaption{}{ \label{fig:mmcorr}       
Results on $b\bar b$ correlations using dimuon final states,
compared to NLO QCD. Left: azimuthal-correlation results from D0.
Right: \pt\ distribution of the slowest muon from CDF.}
\end{figure}

Similar inconsistencies are found in the study of correlations.
A measurement has been recently reported by CDF,
using muons plus tagged $b$ jets~\cite{Abe96b}. The shape of the \dphi\
distribution is in reasonable agreement with QCD, while the distributions of
the jet \et\ and of the muon \pt\ are not (see fig.~\ref{fig:cdfmub}).
On the other hand, both CDF and D0 recently presented studies based on samples
of high-mass dimuons~\cite{Abe96d,Abachi96a}. The shapes for both $\Delta\phi$
and $p_{\sss T}^{b}$ for a given $p_{\sss T}^{\bar b}$ are in this case
well reproduced    
by theory, within the large uncertainties  (fig.~\ref{fig:mmcorr}).

Contrary to the case of $c\bar c$ correlations measured in fixed-target
experiments,
the measurement at hadron colliders is
sensitive to the modelling of the heavy-quark fragmentation, because of
possible trigger biases. A harder (softer) fragmentation function would enhance
(decrease) the efficiency for the detection of the softest of the two $b$
quarks. These effects could have an impact on the distributions reported by
CDF and D0.
CDF explored the effects of changes in the $\epsilon_b$
parameter within the Peterson fragmentation model, finding them negligible.
As we argued earlier, it cannot be excluded that a systematic study of other
possible parametrizations for the fragmentation modelling could lead to
significant effects.
Also the possible effects of the $k_{\sss T}$ kick have been studied by
CDF~\cite{Abe96b}, with the conclusion that not even an average $k_{\sss T}$
as large as 4~GeV could improve the agreement between theory and data in the
case of the muon plus $b$-tagged jet measurement.

In summary, while the overall features of $b$ production at the Tevatron are
well described by NLO QCD, there are still some apparent inconsistencies which
go beyond the quoted theoretical and experimental uncertainties. 
I will finish these lectures by presenting some studies which have been
reported recently, and others which are in progress, aimed at 
a better assessment of the theoretical systematics.
I will first present studies related to the resummation of large perturbative
logarithms, and then some considerations on the effects of non-perturbative
fragmentation.
                     
\begin{figure}[htb]
\centerline{\epsfig{figure=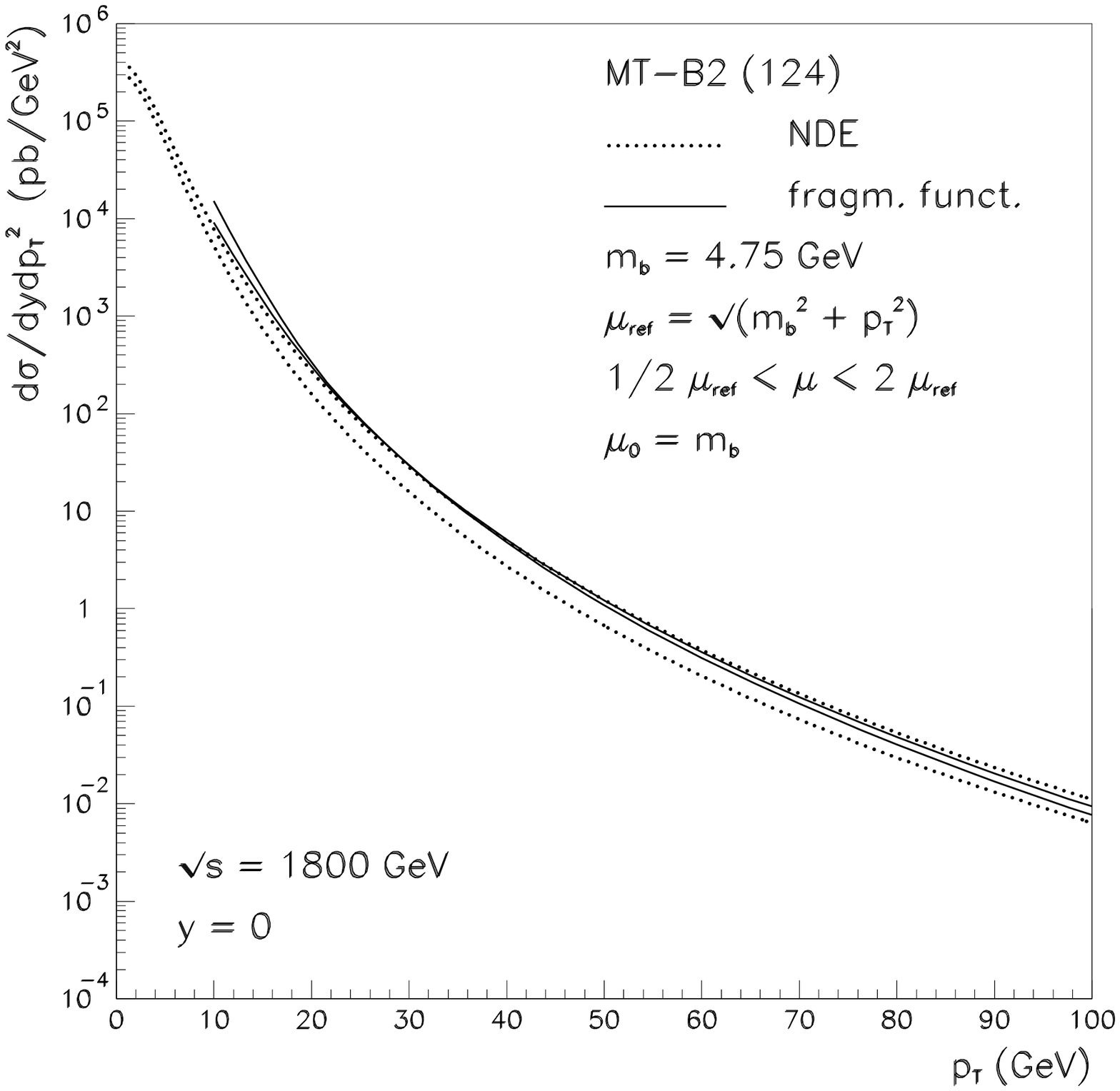,width=0.7\textwidth,clip=}}
\ccaption{}{ \label{fig:greco}
Differential $b$-quark \pt\ distribution: comparison of the fixed-order NLO
calculation (NDE stands for \cite{Nason88})
with the fragmentation-function approach \cite{Cacciari94};
$\mu_0$ is the scale at which the boundary conditions for the fragmentation
functions are set.}
\end{figure}

\subsubsection{Resummation of large-$p_T$ logarithms}
Two different groups have approached this problem in the recent past. Cacciari
and Greco \cite{Cacciari94} have folded the NLO cross-section for production of
a massless parton $i$ ($i=g,q$) \cite{Aversa89}
with the NLO fragmentation function for the
transition $i\to b$ \cite{Mele91}.
The evolution of the fragmentation functions resums all
terms of order $\as^n \log^n (\pt/m)$ and $\as^{n+1} \log^n (\pt/m)$.
All the dependence on the $b$-quark mass lies in the
boundary conditions for the fragmentation
functions. This approach ensures a full NLO accuracy, up to corrections of
order $m^2/(m^2+p_{\scriptscriptstyle T}^2)$. In particular, this formalism
describes NLO corrections to the gluon
splitting process, which in the \oacube\ calculation is only included
at the leading order.
One can verify, by looking at the Born-level production process as a
function of the quark mass, that,
in order for the mass corrections not to exceed the 10\% level, it is
however necessary to limit the applications of this formalism to the region of
$\pt \gsim 20$~GeV.

Figure~\ref{fig:greco} shows the differential $b$-quark \pt\ distribution
obtained in the fragmentation-function approach, compared to the standard
fixed-order NLO result. Several features of this figure should be
noticed. To start with, the scale
dependence is significantly reduced with respect to the fixed-order calculation.
Furthermore, in the range of
applicability of this calculation (i.e. $\pt \gsim 20$~GeV) the result of the
fragmentation-function approach lies on the upper side of the fixed-order NLO
calculation. The resummed calculation is, however, always
within the uncertainty band of the fixed-order one. Finally,
notice that the overall effect of the
inclusion of higher-order logarithms is a steepening of the \pt\ spectrum,
as is natural to expect.

Another calculation has recently appeared, by Scalise, Olness and Tung
\cite{Scalise96}. In this approach the authors employ a strategy developed in
the case of DIS in refs. \cite{Aivazis94,Aivazis94a}.
Here initial- and final-state mass
singularities are resummed as in the fragmentation-function approach, and
the result is then matched in the
low-\pt\ region to the fixed-order NLO massive calculation.
At large \pt\ this calculation does not include as yet, however, the full set
of NLO corrections to the hard-process matrix elements.

The preliminary numerical results of this study \cite{Scalise96}
are consistent with those of
the approach by Cacciari and Greco; in particular, they support the conclusion
that in the \pt\ range explored by the Tevatron experiments the resummed
results are consistent with the fixed-order ones, provided a scale
$\mu$ of the order of $\muo/2$ or smaller is selected.

\subsubsection{Non-perturbative fragmentation}
We start with a remark: the use
of the Peterson fragmentation function might not be justified in the
context of hadroproduction of heavy quarks.
As a simple observation, we point out here that while the measurement of
heavy-quark spectra in $e^+e^-$ data is mostly sensitive to the first moment
of the fragmentation function, corresponding to the average jet
energy carried by the heavy hadron, the \pt\ distributions in hadronic
collisions are sensitive to higher moments of the non-perturbative
fragmentation function. In
fact, assuming for simplicity a perturbative \pt\ spectrum of the form:
\be
        \frac{d\sigma}{d\pt} = \frac{A}{\pt^n} \; ,
\ee
it is easy to prove that the resulting hadron spectrum, after convolution with
a given fragmentation function $f(z)$, will be given by:
\be
        \frac{d\sigma}{d\pt} = \frac{A}{\pt^n} \;
        \times \; \int \, dz \, z^{n-1} \, f(z) \; .
\ee
In the case of the Tevatron, $n$ turns out to be approximately 5.
It is not unlikely, therefore, that  alternative models for the
non-perturbative fragmentation of heavy quarks, which could equally well fit
the $e^+e^-$ data, could give rise to significant differences when applied to
production in hadronic collisions. We also remark that
the gluon component of the fragmentation function is not important
in $e^+e^-$ physics, while it may be crucial in hadroproduction.

\begin{figure}
\centerline{\epsfig{figure=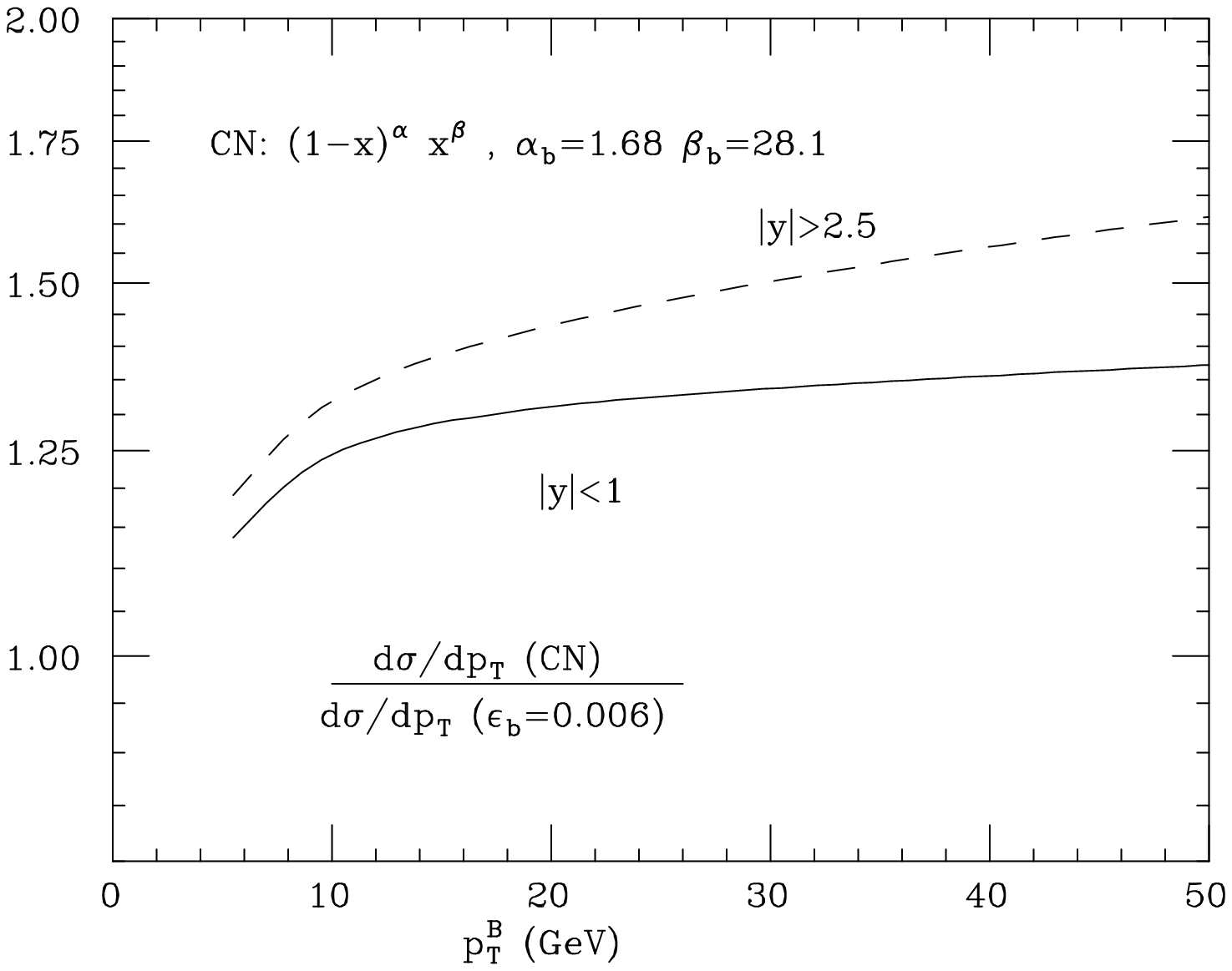,width=0.9\textwidth,clip=}}
\ccaption{}{ \label{fig:bptCN}                
Ratio of the $B$-meson $p_T$ distribution using the
Colangelo-Nason fragmentation function (with the
parameters $\alpha$ and $\beta$ fitted at NLO), relative to that 
obtained by using the Peterson fragmentation function and $\epsilon_c=0.06$.
Central production (solid) and forward production (dashes).}
\end{figure}                                                
Another important point concerns the correlation which exists between the
chosen value of $\as$, the accuracy at which the perturbative fragmentation
function is evaluated, and the value of the parameters describing the
non-perturbative fragmentation extracted from the comparison with the
$e^+e^-$ data. A larger value of $\as$, and a NLO calculation, give rise to a
softer perturbative fragmentation function than would be obtained from a
smaller value of $\as$ or a LO accuracy. As a result, the non-perturbative
fragmentation function will be harder. This was already noticed in the first
lecture, where we compared the values of the Peterson parameter $\epsilon$
extracted from a LO and a NLO fit to the Argus and Opal $D$-production data
($\epsilon_c=0.06$ at LO, $\epsilon_c=0.01$ at NLO). 

A complete study of the effect of these differences for $b$ production at the
Tevatron has not been carried out as yet. 
As a crude estimate of the effects, however, we can just compare the $p_T$
spectra obtained by convoluting the fixed-order NLO calculation of the
$b$-$p_T$ distribution with NP fragmentation functions fitted at the LO and at
the NLO.
In fig.~\ref{fig:bptCN} I plot the ratio of the $B$-meson $p_T$ distribution
obtained by folding with the Colangelo-Nason fragmentation function (with the
parameters $\alpha$ and $\beta$ fitted at NLO), relative to that 
obtained by using the Peterson fragmentation function and $\epsilon_c=0.06$,
the default choice of CDF and D0. This is done for two different rapidity
intervals, one describing {\em central} production, the other describing {\em
forward} production.

In the central region we obtain differences of up to 30\%, which are consistent
with the excess in rate reported by CDF. In the forward region the effect is
larger, up to a factor of 50\%, but not large enough to explain the
deviation seen by D0 using their forward-muon data.

We summarize the main conclusions of the studies presented in this
section:
\begin{enumerate}
\item There is good agreement between the shape of the $b$-quark \pt\
distribution predicted by NLO QCD and that observed in the data for central
rapidities.
\item Although the data are higher by a factor of approximately 2
with respect to the theoretical prediction with the default choice
of parameters, extreme (although acceptable)
choices of $\lambdamsb$ and of renormalization and factorization scales
bring the theory in perfect agreement with the data of UA1
and D0, and within 30\% of the CDF measurements.
\item The choice of low values of $\muR$ and $\muF$, which helps the agreement
between theory and data, is favoured by studies of
higher-order logarithmic corrections.
\item The 30\% difference between CDF and D0 cannot be explained by theory, and
will need to be understood before further progress can be made.
\item Forward production of $b$ quarks at D0 
indicates a larger discrepancy between theory and data, which cannot be
accomodated by higher-order perturbative corrections, nor by the
systematic uncertainties induced by the choice of non-perturbative fragmentation
functions.                                                     
\end{enumerate}
As should be clear from the list above, more work is necessary both on the
experimental and on the theoretical sides to improve our understanding of
bottom production at high energy. To which extent the current inconsistencies
will affect future physics plans at the Tevatron and at the LHC it is not
clear. It is unlikely however that measurements such as CP asymmetries will be
affected by these issues, since they rely on ratios of rates. Nevertheless the
study of $b$ production remains an important test of our ability to perform
solid perturbative predictions, and any progress which can be made
in the coming future will certainly benefit our understanding of QCD at high
energies.                                                  
\\[0.5cm]
{\bf Acknowledgements:} I am very grateful to L. Moroni and I. Bigi for the
invitation to attend this School. I also wish to
thank my collaborators, S. Frixione, P. Nason and G. Ridolfi, with whom most of
the work presented here was done over the years.

\end{document}